\documentclass[]{raa}

\def\kms  {km~s$^{-1}$}
\def\R    {$\mathcal{R}$~}

\usepackage{graphicx,times,amsmath,amssymb,longtable}
\usepackage[labelsep=space,labelfont=bf,font=normalsize]{caption}
\usepackage{natbib}
\usepackage{hyperref}
\bibpunct{(}{)}{;}{a}{}{,}

\hypersetup{pdftitle = The title of my PDF, pdfauthor = My name, pdfsubject= The subject, pdfkeywords = keyword1 keyword2 keyword3}
\hypersetup{colorlinks = true, linkcolor = green, anchorcolor = red, citecolor = blue, filecolor = red, pagecolor = red, urlcolor = red}

\begin{document}

   \title{CO observations of a sample towards nearby galaxies
}

   \volnopage{Vol.0 (20xx) No.0, 000--000}      
   \setcounter{page}{1}          

   \author{Fa-Cheng Li
      \inst{1,2}
   \and Yuan-Wei Wu
      \inst{1}
   \and Ye Xu
      \inst{1}
   }

   \institute{Purple Mountain Observatory, \& Key Laboratory for Radio Astronomy, Chinese Academy of Sciences,
             Nanjing 210008, China; {\it lifc@pmo.ac.cn}\\
        \and
             University of Chinese Academy of Sciences, Beijing 100039, China\\
   }

   \date{Received~~20xx month day; accepted~~20xx~~month day}

\abstract{We have simultaneously observed $^{12}$CO, $^{13}$CO, and C$^{18}$O (J=1$-$0) rotational transitions in the centers of a sample of 58 nearby spiral galaxies using the 13.7-m millimeter-wave telescope of the Purple Mountain Observatory. Forty-two galaxies were detected in $^{13}$CO emission, but there was a null detection for C$^{18}$O emission with a sigma upper limit of 2 mK. Central beam ratios, \R, of $^{12}$CO and $^{13}$CO range mostly from 5 to 13, with an average value of 8.1$\pm$4.2, which is slightly lower than previous estimates for normal galaxies. Clear correlations are found between $^{12}$CO and $^{13}$CO luminosities. An average $X$-factor of $1.44\pm0.84\times10^{20}\ \mathrm{cm}^{-2}~(\mathrm{K}~\mathrm{km}~\mathrm{s}^{-1})^{-1}$ is slightly lower than that in the Milky Way.
\keywords{galaxies:ISM--molecules:galaxies--millimeter lines:ISM--star formation:ISM
}
}

   \authorrunning{Fa-Cheng Li, Yuan-Wei Wu \& Ye Xu}           
   \titlerunning{CO observations of nearby galaxies}  

   \maketitle

%
%
\section{Introduction}           
\label{sect:intro}

Molecular hydrogen, H$_2$, constitutes a dominant part of molecular clouds in the interstellar medium (ISM) in galaxies and is most closely related to star formation. The current method of studying molecular clouds in external galaxies involves the  observation of rotational transitions of carbon dioxide, CO. H$_2$ lacks a dipole moment and therefore, quadrupole or vibrational transitions cannot be excited under typical cold temperature conditions in giant molecular clouds (GMCs). Rotational transitions of CO can easily be generated by collisions with H$_2$, particularly the line from the first excited level to ground, J=1$-$0, can be excited under the condition of very low temperatures and density of only 10 K and 300 cm$^{-3}$ respectively. Thus, $^{12}$CO, as well as its isotopic variants, remain the most straightforward and reliable tracer of H$_2$ in molecular clouds. In addition, there is a well-known CO$-$H$_2$ conversion factor, called the X factor, and it is defined as
\begin{equation}
X_{\mathrm{CO}} = \frac{N(\mathrm{H}_2)}{I_{\mathrm{CO}}} ~ [\mathrm{cm}^{-2}~(\mathrm{K}~\mathrm{km}~\mathrm{s}^{-1})^{-1}],
\label{eq:x}
\end{equation}
where $N(\mathrm{H}_2)$ is the column density of H$_2$ in cm$^{-2}$ and $I_{\mathrm{CO}}$ is the integrated line intensity of $^{12}$CO.

The first CO detections in external galaxies were carried out by \citet{1975ApJ...199L..75R} and \citet{1975ApJ...199L..79S}. Later, \citet{1995ApJS...98..219Y} published the Five Colleges Radio Astronomy Observatory (FCRAO) extragalactic CO survey at $\lambda=2.6$ mm of a large sample of 300 galaxies with 1412 positions using the 14 m telescope at 45$''$ resolution. The detection rate is 79\% and 193 galaxies were observed in multiple positions. \citet{1993A&AS...97..887B} observed both $^{12}$CO(1$-$0) and $^{12}$CO(2$-$1) emission from the centers of 81 nearby spiral galaxies using the 30-m telescope at the Institut de Radioastronomie Millimetrque (IRAM) at a resolution of 23$''$ and 12$''$ for $^{12}$CO(1$-$0) and $^{12}$CO(2$-$1), respectively, and finding an average (and median) $^{12}$CO(2$-$1) to $^{12}$CO(1$-$0) line ratio to be 0.89$\pm$0.06. \citet{1997ApJ...478..144S} also used the IRAM 30-m telescope to observe $^{12}$CO(1$-$0) transitions in 37 ultraluminous infrared galaxies and discovered that interacting galaxies also have relatively high CO luminosity. There are plenty of other CO surveys of nearby galaxies using either single-dishes or even interferometer telescopes that have aimed to map the molecular gas distribution or kinematics within galaxies and have worked out properties of molecular gas clouds from galaxy to galaxy \citep{1999ApJS..124..403S, 2001PASJ...53..757N, 2003ApJS..145..259H, 2009AJ....137.4670L}. However, most of these surveys are based on $^{12}$CO J=1$-$0, J=2$-$1 or even higher rotational transitions, while there are few systematic studies of CO isotopologue transitions of a larger sample of galaxies has yet to be published. This is probably because $^{12}$CO J=1$-$0 is found to not be an accurate measure of the amount of molecular gas, however, $^{13}$CO emission with lower opacity may give more reliable constraints on the H$_2$ column density.

\citet{1991A&A...247..320S} presented observations of $^{13}$CO (1$-$0) emission from 16 nearby spiral galaxies using the 12-m telescope at the National Radio Astronomy Observatory. They found the ratio of $^{12}$CO (1$-$0) and $^{13}$CO (1$-$0) emittance to be insensitive to variations in global parameters such as inclination angle and Hubble type. The detection revealed a range of central beam ratios from 5 to 16.6, mostly from 7 to 11. \citet{1995A&A...300..369A} studied molecular gas in 32 infrared-bright galaxies, which consists mostly of starbursts. They presented several line ratios, among which they suggested that the ratio of $^{12}$CO/$^{13}$CO (1$-$0) can be a measurement of the cloud environment in galaxies. \citet{2001ApJS..135..183P} performed a mapping survey of $^{12}$CO and $^{13}$CO J=1$-$0 emissions along the major axes of 17 nearby galaxies. Their work resulted in an average central $^{12}$CO/$^{13}$CO intensity ratio of 11.6$\pm$1.9 implying that the X-factor is probably lower in most galactic nuclei. A non-liner correlation between CO and far-infrared luminosity exists in galaxies because luminous galaxies have a higher star formation efficiency (SFE) \citep{1988ApJ...334..613S, 1991ARA&A..29..581Y, 2004ApJ...606..271G}. \citet{1998ApJ...507L.121T} collected previous observational results of $^{12}$CO and $^{13}$CO emissions and compared far-infrared luminosity with that of $^{12}$CO and $^{13}$CO, respectively. They found that the $^{13}$CO depression in luminous starburst mergers may account for a higher abundance ratio of $^{12}$CO and $^{13}$CO than that in normal galaxies. \citet{2005ARA&A..43..677S} and \citet{2010ApJ...714L.118D} further confirmed the validity of this correlation when studying high-redshift star forming galaxies.

In order to systemically study the physical properties of external galaxies, we present simultaneous observations of $^{12}$CO, $^{13}$CO, and C$^{18}$O J=1$-$0 emissions from the centers of 58 nearby galaxies, mostly spiral, using the Purple Mountain Observatory (PMO) 13.7-m millimeter radio telescope. And it is the second time we carry out observations towards nearby galaxies using this telescope after \citet{2011RAA....11..787T}. The observations and sample selection are described in Section~\ref{sect:Obs}. We then present results with detected CO spectra and derived parameters in Section~\ref{sect:data}. The analysis and discussions are in Section~\ref{sect:analysis} and finally the summary is in Section~\ref{sect:summary}.


\section{Sample and Observations}
\label{sect:Obs}

\subsection{Sample selection}

We selected a sample of 58 nearby galaxies from the FCRAO Extragalactic CO Survey \citep{1995ApJS...98..219Y}. The selection criteria were as follows: (1) $I(^{12}\mathrm{CO})\geq$~3 K \kms, where K is in antenna temperature. Strong $^{12}$CO emission usually indicates a relatively high detection rate of isotopic variants. (2) 10h$\leq$R.A$\leq$13h and Decl.$\geq$-10$^{\circ}$, in order to not conflict with the galactic time in the northern sky. Physical properties of the Infrared Astronomical Satellite (IRAS) data are summarized in Table~\ref{Tab:basic}, which are obtained from the IRAS Revised Bright Galaxy Sample (RGBS) \citep{2003AJ....126.1607S} and the SIMBAD database.

\begin{center}
\tabcolsep=3pt
\renewcommand{\arraystretch}{1.2}
\small
 \begin{longtable}[c]{ccccrr@{$\times$}lr}
  \caption{Basic Properties of the Sample}\label{Tab:basic} \\
  \hline
  \hline\noalign{\smallskip}
   \multicolumn{1}{c}{Galaxy} &  &  &  &  & \multicolumn{2}{c}{IR size} & \\
   \noalign{\smallskip}\cline{6-7}\noalign{\smallskip}
   \multicolumn{1}{c}{Name} & Type & R.A. DECL. & \multicolumn{1}{c}{$cz$} & \multicolumn{1}{c}{$D$} & \multicolumn{2}{c}{major$\times$minor} & $T_{\mathrm{dust}}$ \\
   &  &(J2000) & \multicolumn{1}{c}{(\kms)} & \multicolumn{1}{c}{(Mpc)} & \multicolumn{2}{c}{(arcmin)} &  \multicolumn{1}{c}{(K)} \\
   \multicolumn{1}{c}{(1)} & (2) & (3) & \multicolumn{1}{c}{(4)} & \multicolumn{1}{c}{(5)} & \multicolumn{1}{c}{(6)} & \multicolumn{1}{c}{(7)} & \multicolumn{1}{c}{(8)} \\
 \noalign{\smallskip}\hline\noalign{\smallskip}
  \endfirsthead
   \caption[]{-- \textit{Continued}} \\
\hline
  \hline\noalign{\smallskip}
   \multicolumn{1}{c}{Galaxy} &  &  &  &  & \multicolumn{2}{c}{IR size} & \\
   \noalign{\smallskip}\cline{6-7}\noalign{\smallskip}
   \multicolumn{1}{c}{Name} & Type & R.A. DECL. & \multicolumn{1}{c}{$cz$} & \multicolumn{1}{c}{$D$} & \multicolumn{2}{c}{major$\times$minor} & $T_{\mathrm{dust}}$ \\
   &  &(J2000) & \multicolumn{1}{c}{(\kms)} & \multicolumn{1}{c}{(Mpc)} & \multicolumn{2}{c}{(arcmin)} &  \multicolumn{1}{c}{(K)} \\
   \multicolumn{1}{c}{(1)} & (2) & (3) & \multicolumn{1}{c}{(4)} & \multicolumn{1}{c}{(5)} & \multicolumn{1}{c}{(6)} & \multicolumn{1}{c}{(7)} & \multicolumn{1}{c}{(8)} \\
 \noalign{\smallskip}\hline\noalign{\smallskip}
  \endhead
  \noalign{\smallskip}\hline\endfoot
M 066   &  Sb     &  11:20:15.026 $+$12:59:28.64 &   740  & 10.04  &  6.64 	&   3.65  &  34.71  \\
M 108   &  Sc     &  11:11:30.967 $+$55:40:26.84 &   698  & 13.85  &  7.91 	&   1.74  &  33.07  \\
NGC3079 &  S      &  10:01:57.924 $+$55:40:48.00 &  1142  & 18.19  &  4.49 	&   1.08  &  34.67  \\
NGC3169 &  Sb     &  10:14:15.099 $+$03:27:58.03 &  1305  & 20.61  &  2.73 	&   2.02  &  31.24  \\
NGC3184 &  Sc     &  10:18:16.985 $+$41:25:27.77 &   593  & 12.58  &  6.72 	&   5.72  &  29.71  \\
NGC3593 &  S0     &  11:14:37.002 $+$12:49:04.87 &   578  &  5.04  &  3.48 	&   1.43  &  35.25  \\
NGC3628 &  Sbc    &  11:20:17.018 $+$13:35:22.16 &   825  & 10.04  & 10.58  &   2.54  &  35.54  \\
NGC3631 &  Sc     &  11:21:02.944 $+$53:10:09.95 &  1158  & 21.58  &  4.19 	&   4.02  &  31.07  \\
NGC3672 &  Sc     &  11:25:02.476 $-$09:47:43.44 &  1899  & 27.70  &  2.97 	&   1.13  &  31.30  \\
NGC3675 &  Sb     &  11:26:08.584 $+$43:35:09.30 &   804  & 12.69  &  4.45 	&   2.00  &  29.15  \\
NGC3690 &  Sm     &  11:28:31.600 $+$58:33:44.00 &  3159  & 47.74  &  1.61 	&   1.41  &  47.31  \\
NGC3810 &  Sc     &  11:40:58.737 $+$11:28:16.07 &  1001  & 15.36  &  3.05 	&   2.08  &  32.12  \\
NGC3893 &  Sc     &  11:48:38.207 $+$48:42:38.84 &   892  & 16.35  &  2.88 	&   2.13  &  33.06  \\
NGC3938 &  Sc     &  11:52:49.453 $+$44:07:14.63 &   800  & 14.75  &  4.00 	&   3.80  &  30.57  \\
NGC4030 &  Scdr   &  12:00:23.643 $-$01:05:59.87 &  1427  & 24.50  &  2.67 	&   2.35  &  31.41  \\
NGC4038 &  Sc     &  12:01:52.480 $-$18:52:02.90 &  1563  & 21.54  &  5.20 	&   3.10  &  35.55  \\
NGC4039 &  Sc     &  12:01:53.700 $-$18:53:08.00 &  1563  & 21.54  &  3.10 	&   1.60  &  35.55  \\
NGC4041 &  Sc     &  12:02:12.173 $+$62:08:14.23 &  1243  & 22.78  &  1.70 	&   1.39  &  33.67  \\
NGC4051 &  SBab   &  12:03:09.686 $+$44:31:52.54 &   728  & 13.11  &  4.73 	&   2.60  &  33.04  \\
NGC4088 &  Sc     &  12:05:34.189 $+$50:32:20.50 &   696  & 13.37  &  4.39 	&   2.11  &  33.35  \\
NGC4096 &  Sc     &  12:06:01.161 $+$47:28:42.09 &   523  &  9.63  &  5.76 	&   1.73  &  30.21  \\
NGC4102 &  Sb     &  12:06:23.115 $+$52:42:39.42 &   859  & 16.89  &  1.78 	&   0.98  &  39.20  \\
NGC4157 &  Sb     &  12:11:04.365 $+$50:29:04.85 &   790  & 13.30  &  4.94 	&   0.89  &  31.02  \\
NGC4194 &  I      &  12:14:09.573 $+$54:31:36.03 &  2555  & 40.33  &  0.67 	&   0.46  &  45.18  \\
NGC4212 &  Sc     &  12:15:39.375 $+$13:54:05.30 &   -81  & 15.29  &  2.97 	&   1.42  &  32.15  \\
NGC4254 &  Sc     &  12:18:49.625 $+$14:24:59.36 &  2403  & 15.29  &  4.50 	&   4.27  &  32.65  \\
NGC4258 &  Sbc    &  12:18:57.620 $+$47:18:13.39 &   448  &  7.10  &  11.14 &   5.46  &      -  \\
NGC4273 &  SBc    &  12:19:56.063 $+$05:20:36.12 &  2400  & 15.29  &  1.71 	&   1.20  &  33.27  \\
NGC4293 &  Sap    &  12:21:12.891 $+$18:22:56.64 &   893  & 16.50  &  5.29 	&   1.80  &      -  \\
NGC4298 &  Sc     &  12:21:32.790 $+$14:36:21.78 &  1141  & 15.29  &  2.97 	&   1.78  &  28.71  \\
NGC4302 &  Sc     &  12:21:42.477 $+$14:35:51.94 &  1149  & 19.20  &  5.37 	&   0.64  &      -  \\
NGC4303 &  SABbc  &  12:21:54.950 $+$04:28:24.92 &  1570  & 15.29  &  4.64 	&   3.48  &  34.39  \\
NGC4312 &  Sa     &  12:22:31.359 $+$15:32:16.51 &   153  & 16.50  &  3.39 	&   0.88  &  30.75  \\
NGC4321 &  Sc     &  12:22:54.899 $+$15:49:20.57 &  1571  & 15.20  &  7.23 	&   5.64  &  31.89  \\
NGC4402 &  Sb     &  12:26:07.566 $+$13:06:46.06 &   237  & 15.29  &  2.97 	&   0.59  &  30.04  \\
NGC4414 &  Sc     &  12:26:27.089 $+$31:13:24.76 &   720  & 17.68  &  2.86 	&   1.60  &  32.92  \\
NGC4419 &  SBa    &  12:26:56.433 $+$15:02:50.72 &  -261  & 15.29  &  2.54 	&   0.81  &  34.26  \\
NGC4433 &  Sbc    &  12:27:38.610 $-$08:16:42.42 &  2913  & 41.68  &  1.52 	&   0.55  &  36.60  \\
NGC4457 &  SB0/Sa &  12:28:59.011 $+$03:34:14.19 &   882  & 12.40  &  2.00 	&   1.36  &  34.84  \\
NGC4490 &  Sd     &  12:30:36.710 $+$41:38:26.60 &   641  & 10.48  &  5.32 	&   2.29  &  35.83  \\
NGC4501 &  Sbc    &  12:31:59.216 $+$14:25:13.48 &  2284  & 15.29  &  5.47 	&   2.41  &  29.94  \\
NGC4527 &  Sb     &  12:34:08.496 $+$02:39:13.72 &  1771  & 15.29  &  4.62 	&   1.57  &  34.51  \\
NGC4535 &  SBc    &  12:34:20.310 $+$08:11:51.94 &  1957  & 15.77  &  5.84 	&   2.92  &  31.09  \\
NGC4536 &  Sc     &  12:34:27.129 $+$02:11:16.37 &  1802  & 14.92  &  4.61 	&   2.40  &  39.51  \\
NGC4567 &  Sc     &  12:36:32.703 $+$11:15:28.33 &  2262  & 15.29  &  3.72 	&   2.19  &  31.50  \\
NGC4568 &  Sc     &  12:36:34.292 $+$11:14:19.07 &  2262  & 15.29  &  4.06 	&   1.59  &  31.50  \\
NGC4569 &  Sab    &  12:36:49.816 $+$13:09:46.33 &  -235  & 15.29  &  6.92 	&   2.77  &  31.57  \\
NGC4579 &  Sab    &  12:37:43.527 $+$11:49:05.46 &  1519  &  7.73  &  4.60 	&   3.22  &  28.86  \\
NGC4631 &  Sc     &  12:42:08.009 $+$32:32:29.44 &   630  & 15.29  &  9.25 	&   2.78  &  35.93  \\
NGC4647 &  Sc     &  12:43:32.542 $+$11:34:56.89 &  1415  & 15.29  &  2.58 	&   2.04  &  30.97  \\
NGC4654 &  SBcd   &  12:43:56.638 $+$13:07:34.86 &  1037  & 12.82  &  4.47 	&   2.01  &  31.16  \\
NGC4666 &  Sbc    &  12:45:08.676 $-$00:27:42.88 &  1495  & 20.65  &  3.32 	&   0.93  &  33.29  \\
NGC4691 &  S0a    &  12:48:13.600 $-$03:19:57.70 &  1124  & 21.71  &  2.80 	&   2.30  &  38.37  \\
NGC4710 &  S0a    &  12:49:38.958 $+$15:09:55.76 &  1125  & 15.29  &  3.73 	&   0.90  &  33.67  \\
NGC4736 &  Sb     &  12:50:53.148 $+$41:07:12.55 &   323  &  4.83  &  4.13 	&   3.27  &  37.40  \\
NGC4818 &  Sab    &  12:56:48.907 $-$08:31:31.08 &  1051  &  9.37  &  3.34 	&   1.60  &  41.33  \\
NGC4826 &  Sb     &  12:56:43.696 $+$21:40:57.57 &   349  &  3.09  &  6.82 	&   3.89  &  33.76  \\
NGC4845 &  Sb     &  12:58:01.242 $+$01:34:32.09 &  1224  & 15.09  &  3.85 	&   1.08  &  32.93  \\

\end{longtable}
\tablecomments{0.81\textwidth}{Column (1): Names of the sample galaxies.~Column (2): Morphological types taken from the SIMBAD database (\url{http://simbad.u-strasbg.fr/simbad/}).~Column (3): Adopted tracking center of observed galaxies; Units of right ascension are hours, minutes, and seconds, and units of declination are degrees, arcminutes, and arcseconds. The data in Column (4)-(5) except for NGC4258, NGC4293, NGC4302, NGC4312 and NGC4457 are taken from \citet{2003AJ....126.1607S}.~Column (4): The heliocentric radial velocity computed as c times the redshift z. Column (5): Distances including luminosity ones and metric ones. Column (6)-(7): Infrared angular sizes taken from mostly 2MASS data using SIMBAD.~Column (8): Dust temperature derived from the RBGS IRAS 60 $\mu$m/100 $\mu$m color assuming an emissivity that is proportional to the frequency $\nu$; those without RBGS\citep{2003AJ....126.1607S} data are calculated following \citet{1996ARA&A..34..749S} using \textit{IRAS Point Source Catalog} (PSC:1988). }
\end{center}

\subsection{Observations}

We made observations between February and June 2011 and supplementary observations between September and October 2011 and also December 2012, using the PMO 13.7-m millimeter-wave telescope located at Delingha, Qinghai, China. The observations were made after the newly developed 3$\times$3 multi-beam sideband separation superconducting spectroscopic array receiver (SSAR) was added. The receiver employs a two-sideband superconductor-insulator-superconductor (SIS) mixer, which allowed us to simultaneously observe $^{12}$CO J=1$-$0 emission in the upper sideband (USB) and $^{13}$CO and C$^{18}$O J=1$-$0 emissions in the lower sideband (LSB). A high definition Fast Fourier Transition Spectrometer (FFTS) as backend enabled a bandwidth of 1 GHz and a velocity resolution of 0.16 \kms at 115.271 GHz. Single-point observations using beam 5 of SSAR were done in the "ON-OFF" position switching mode, with a pointing accuracy of nearly 5$''$. The Half Power Beam Width was 52$''$ at 115.271 GHz and the main beam efficiency, $\eta_{\textrm{mb}}$, for USB and LSB were 0.46 and 0.5 respectively between February and June 2011 and were 0.44 and 0.48 during October 2011\footnote{See the Status Report \url{http://www.radioast.csdb.cn/zhuangtaibaogao.php}}. Typical system temperatures were 220 K at 115.271 GHz and 130 K at 110.201 GHz during our observations.


\begin{figure}[p]
  \centering
\includegraphics[width=0.33\textwidth, angle=-90]{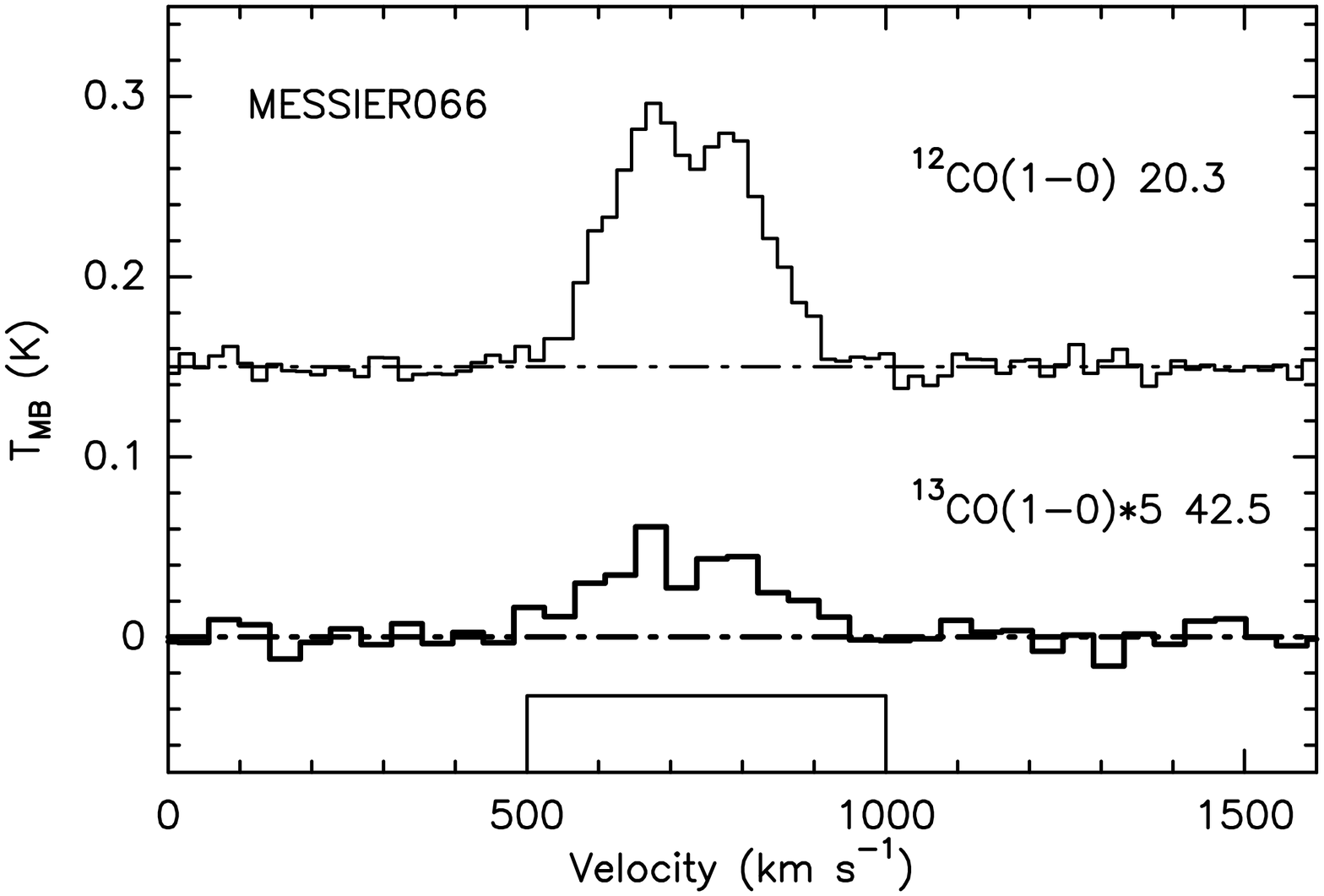}
\includegraphics[width=0.33\textwidth, angle=-90]{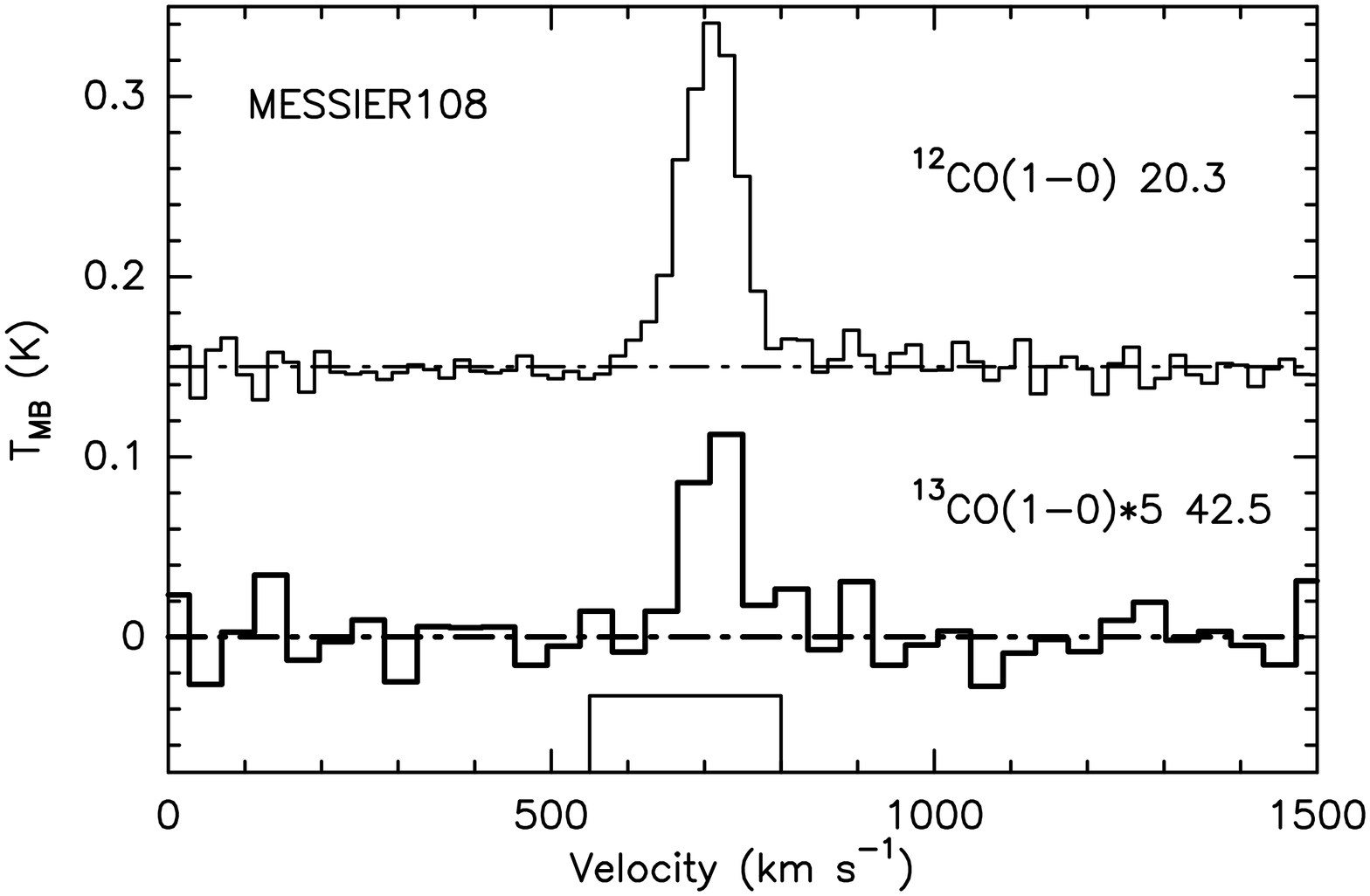}
\includegraphics[width=0.33\textwidth, angle=-90]{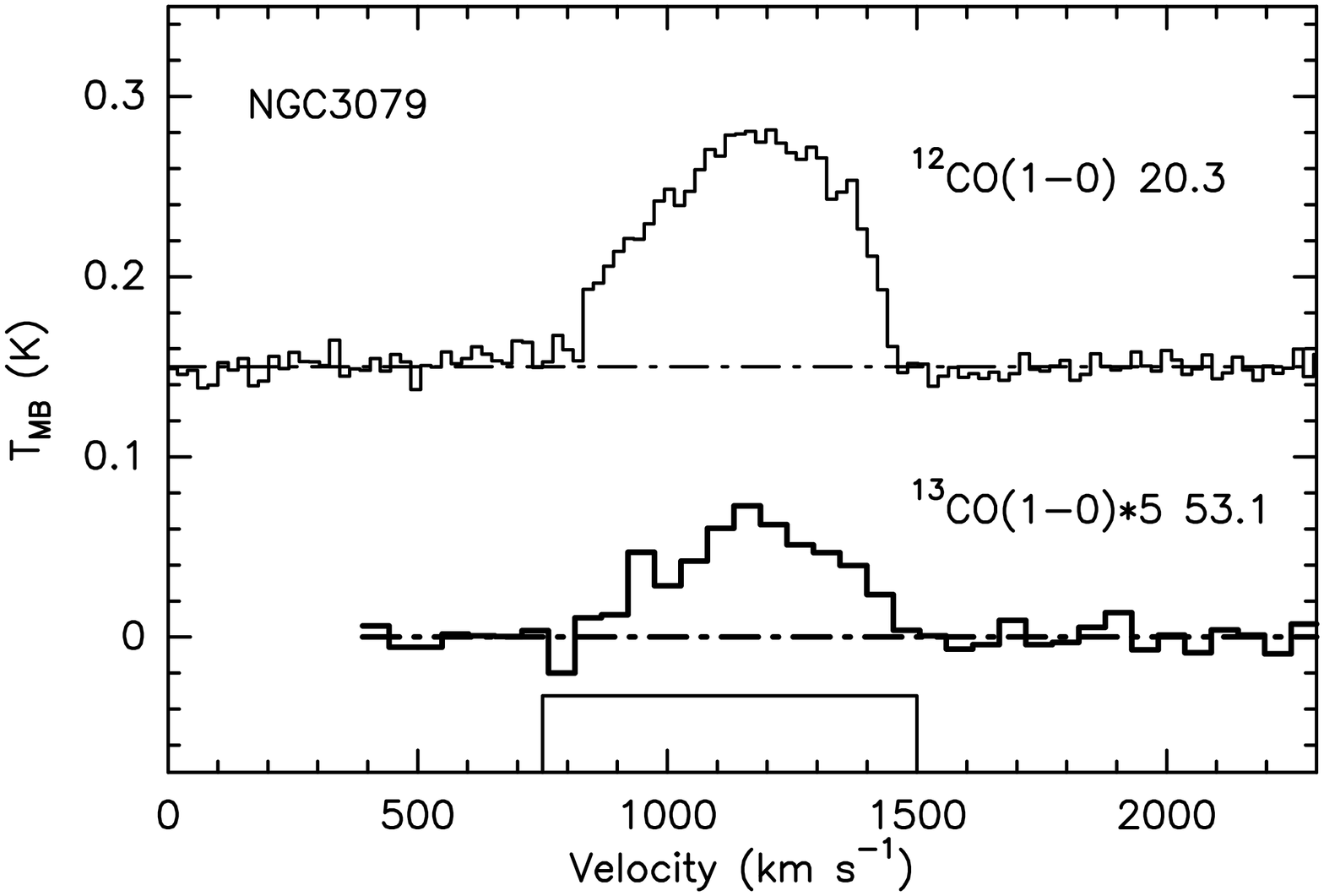}
\includegraphics[width=0.33\textwidth, angle=-90]{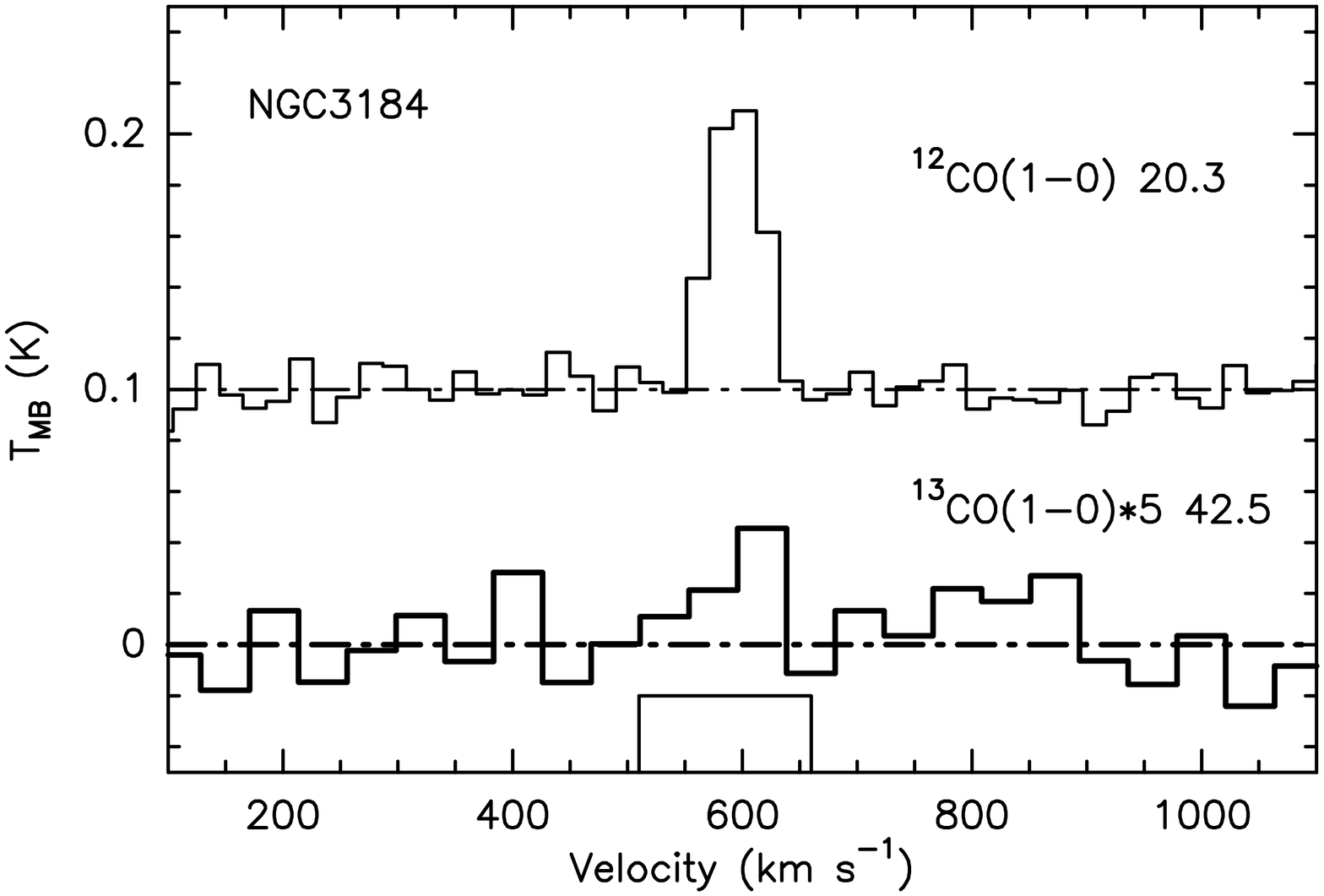}
\includegraphics[width=0.33\textwidth, angle=-90]{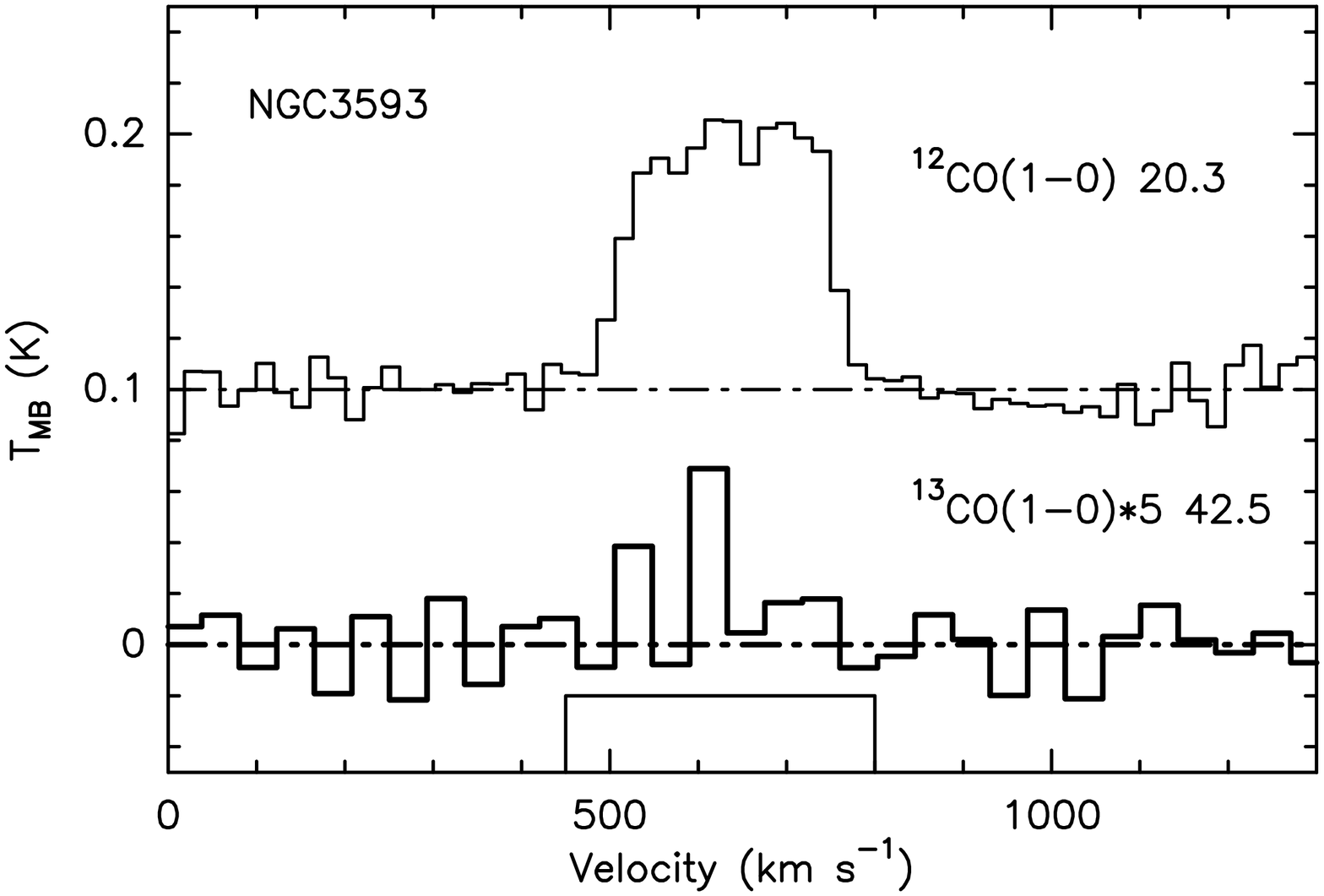}
\includegraphics[width=0.33\textwidth, angle=-90]{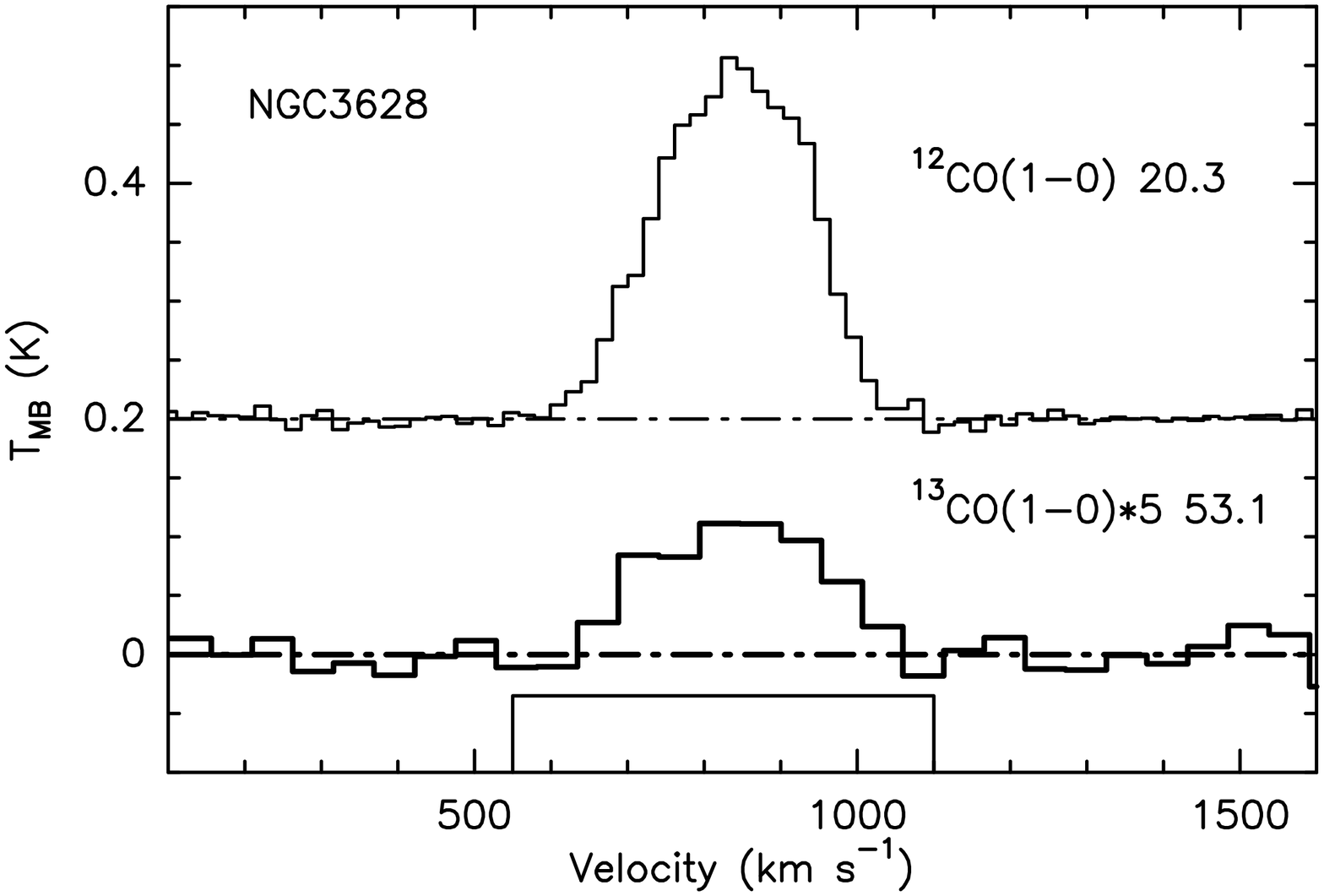}
\includegraphics[width=0.33\textwidth, angle=-90]{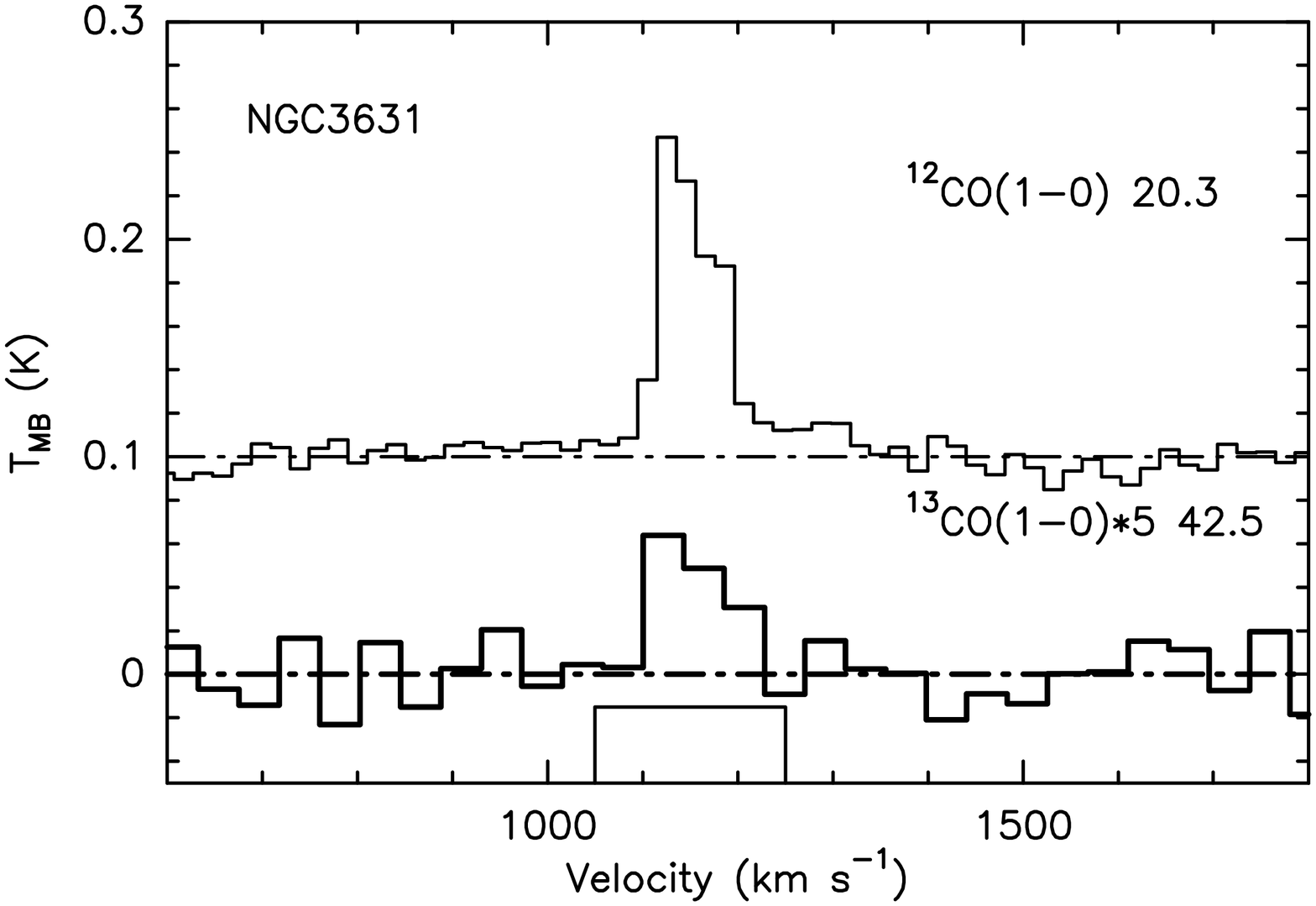}
\includegraphics[width=0.33\textwidth, angle=-90]{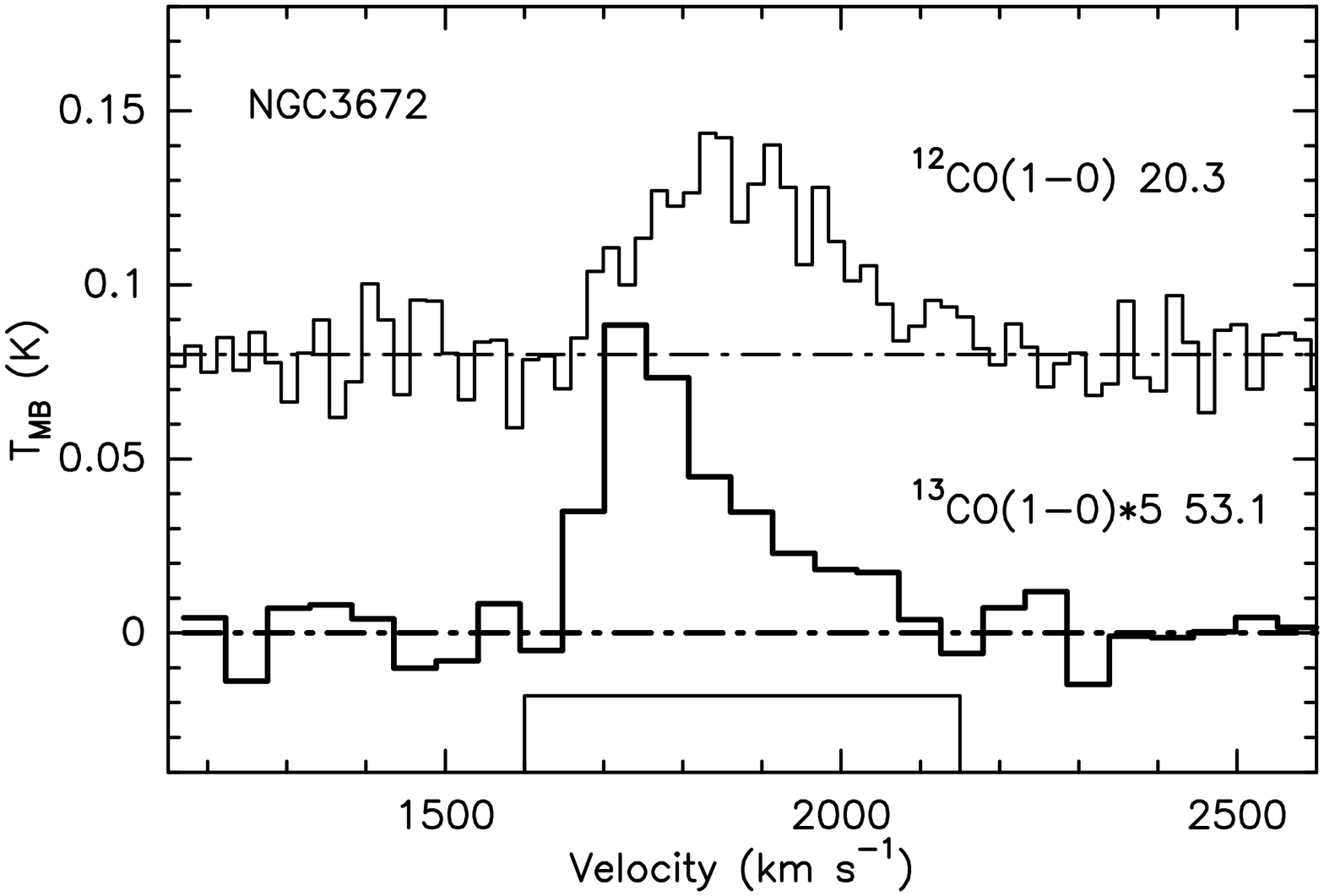}
\caption{Observed spectra of $^{12}$CO (thin lines) and $^{13}$CO (thick lines) in the central regions of $^{13}$CO detected galaxies. Velocities are the radio velocities with respect to LSR. The spectra of $^{13}$CO emissions are multiplied by 5 for comparison. The spectra are on the scale of main beam temperature. All the spectra were smoothed to a velocity resolution of about 20 \kms for $^{12}$CO and about 40 \kms for $^{13}$CO in order to limit the rms noise. The number right after the spectrum label is specified velocity resolution for individual source. The window for emission feature is drawn on the bottom of the axis box in each plot. }
\label{Fig:spectra}
\end{figure}

\begin{figure}[!htp]
  \centering
\includegraphics[width=0.33\textwidth, angle=-90]{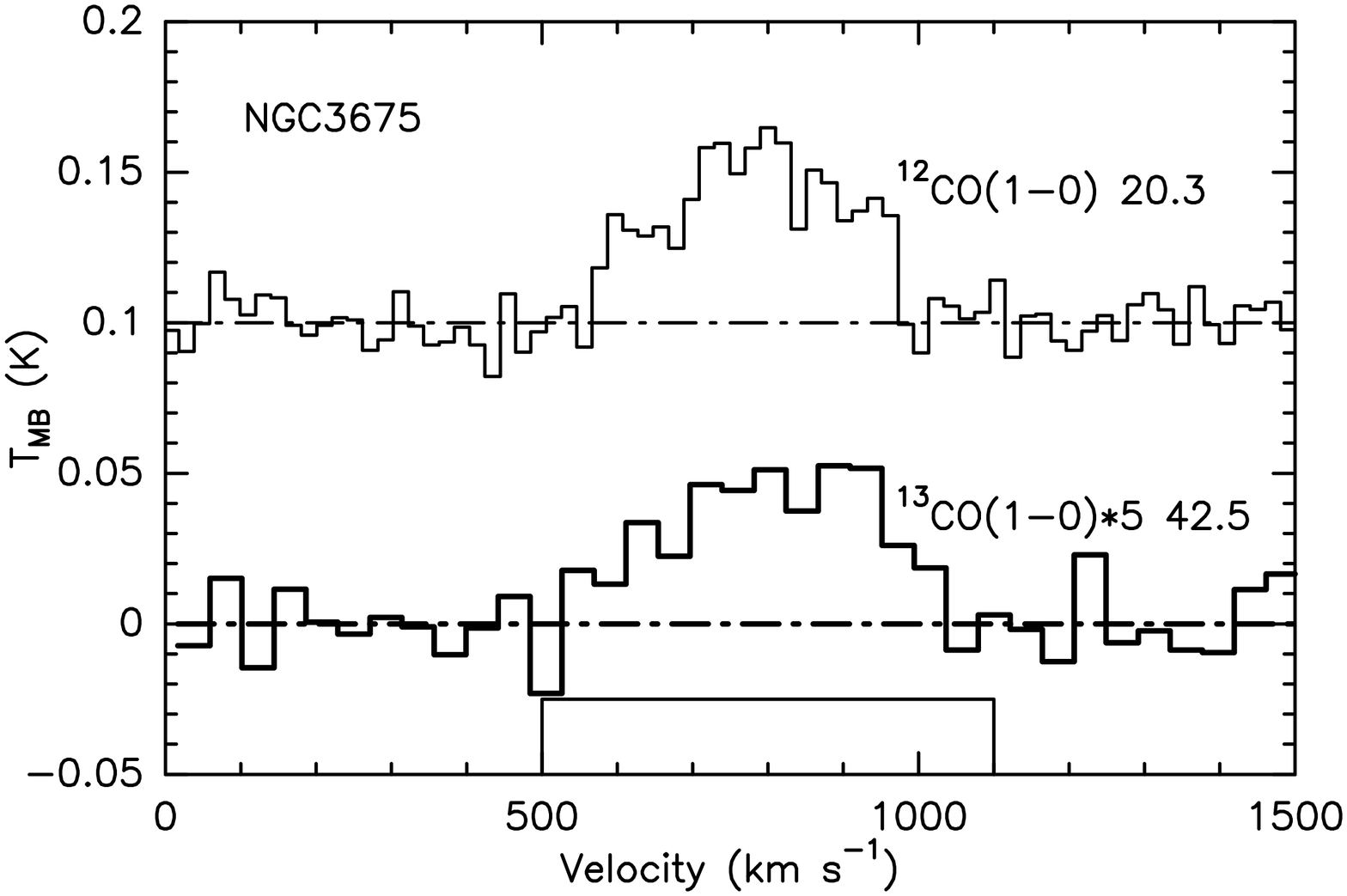}
\includegraphics[width=0.33\textwidth, angle=-90]{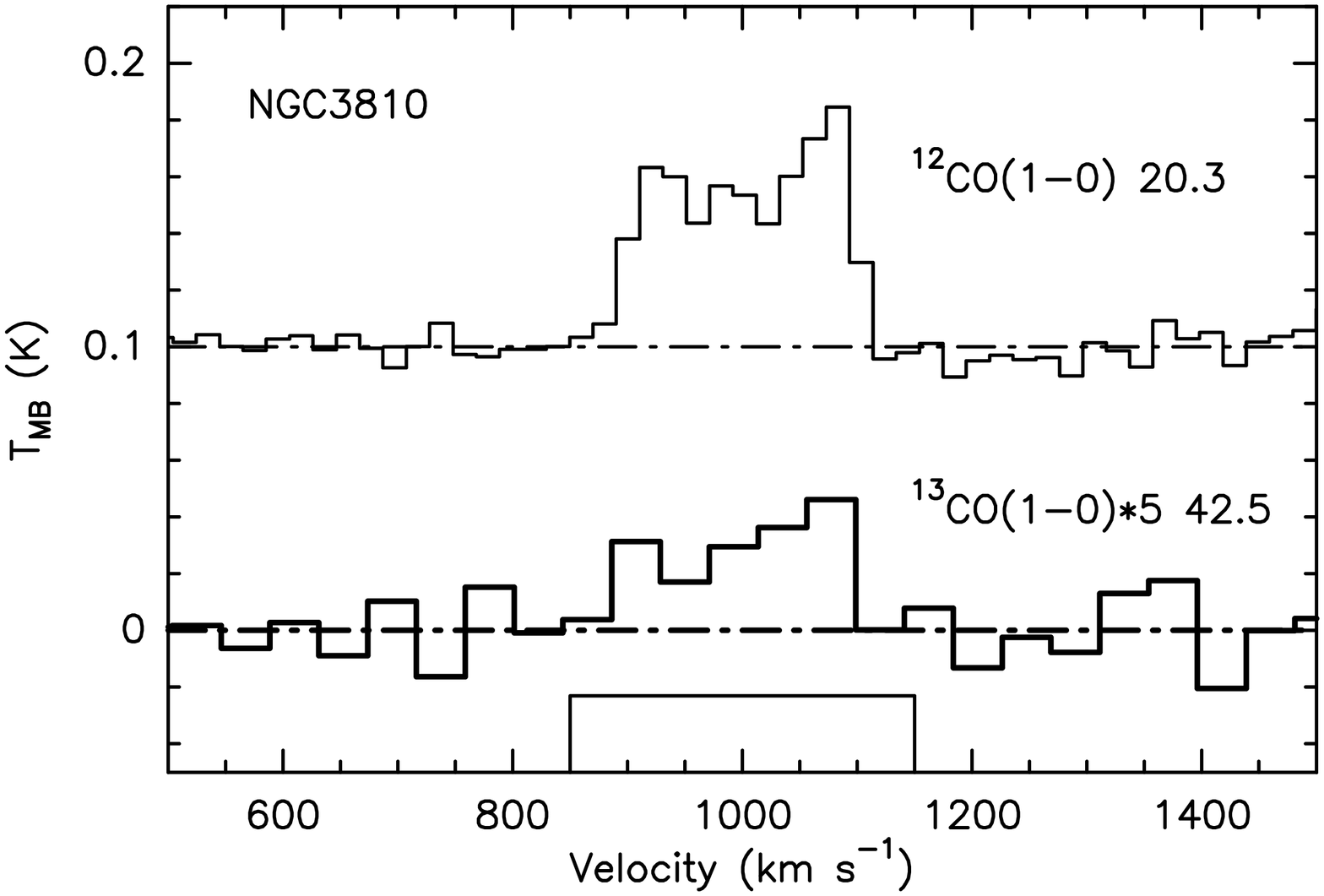}
\includegraphics[width=0.33\textwidth, angle=-90]{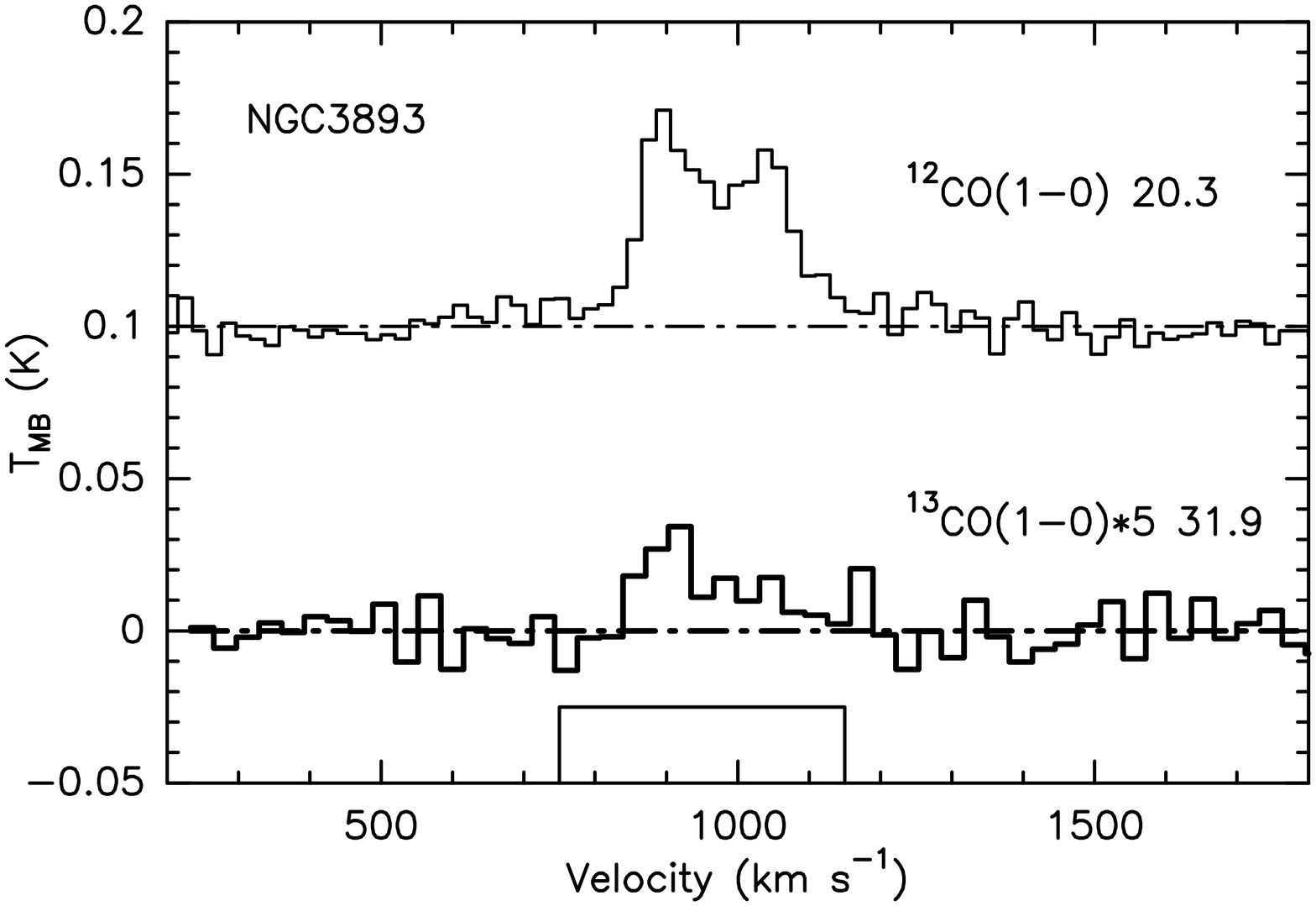}
\includegraphics[width=0.33\textwidth, angle=-90]{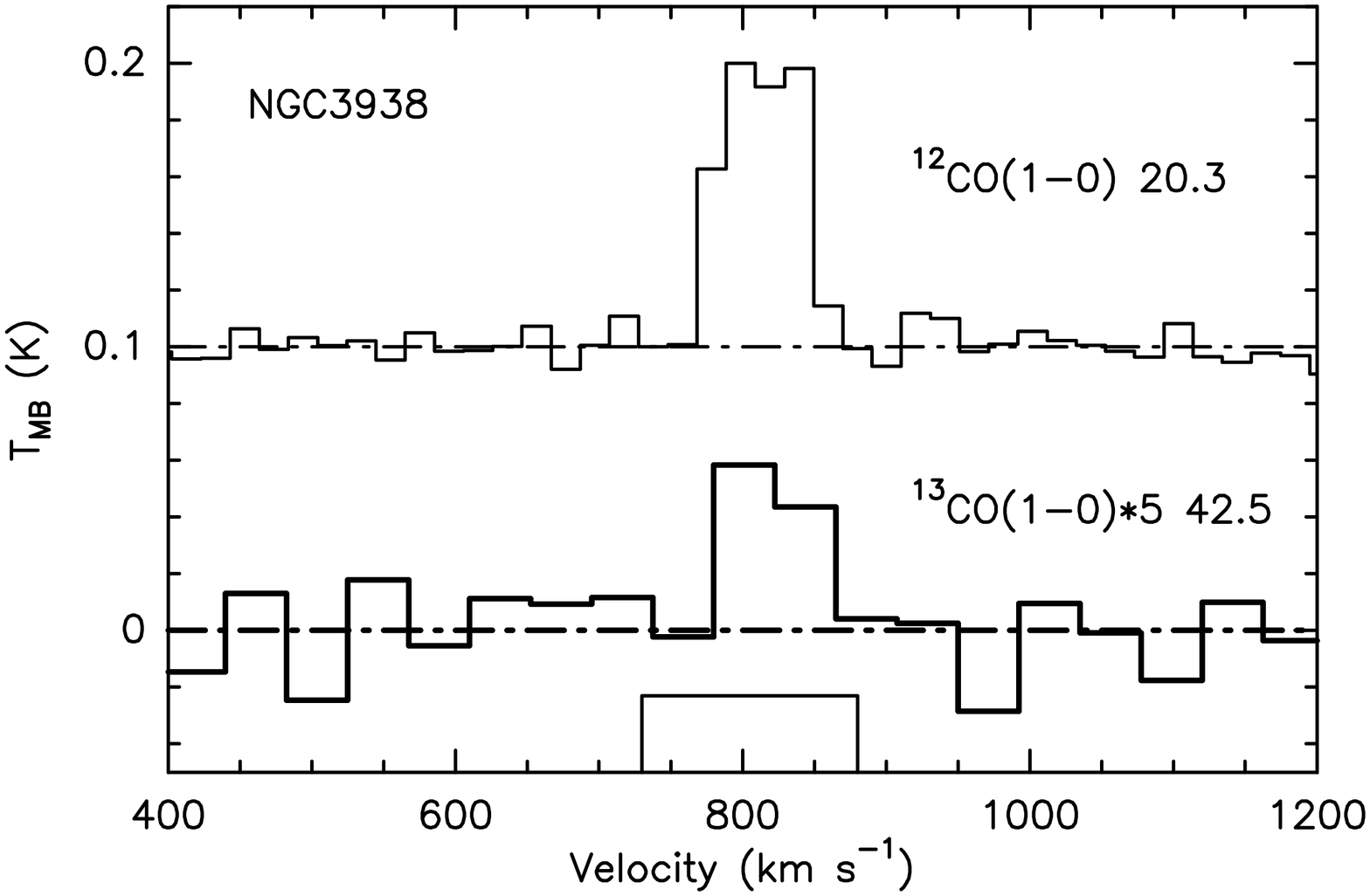}
\includegraphics[width=0.33\textwidth, angle=-90]{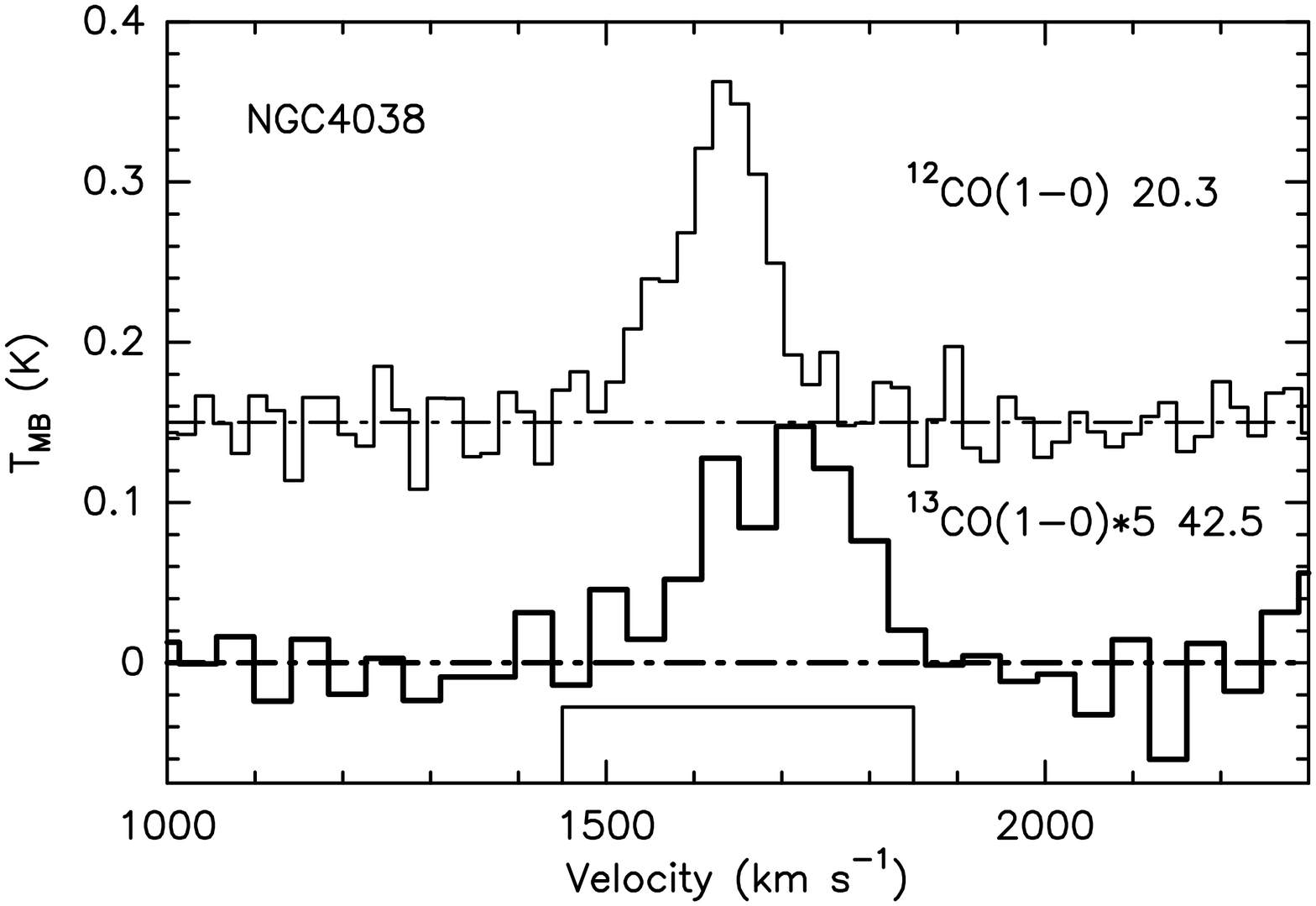}
\includegraphics[width=0.33\textwidth, angle=-90]{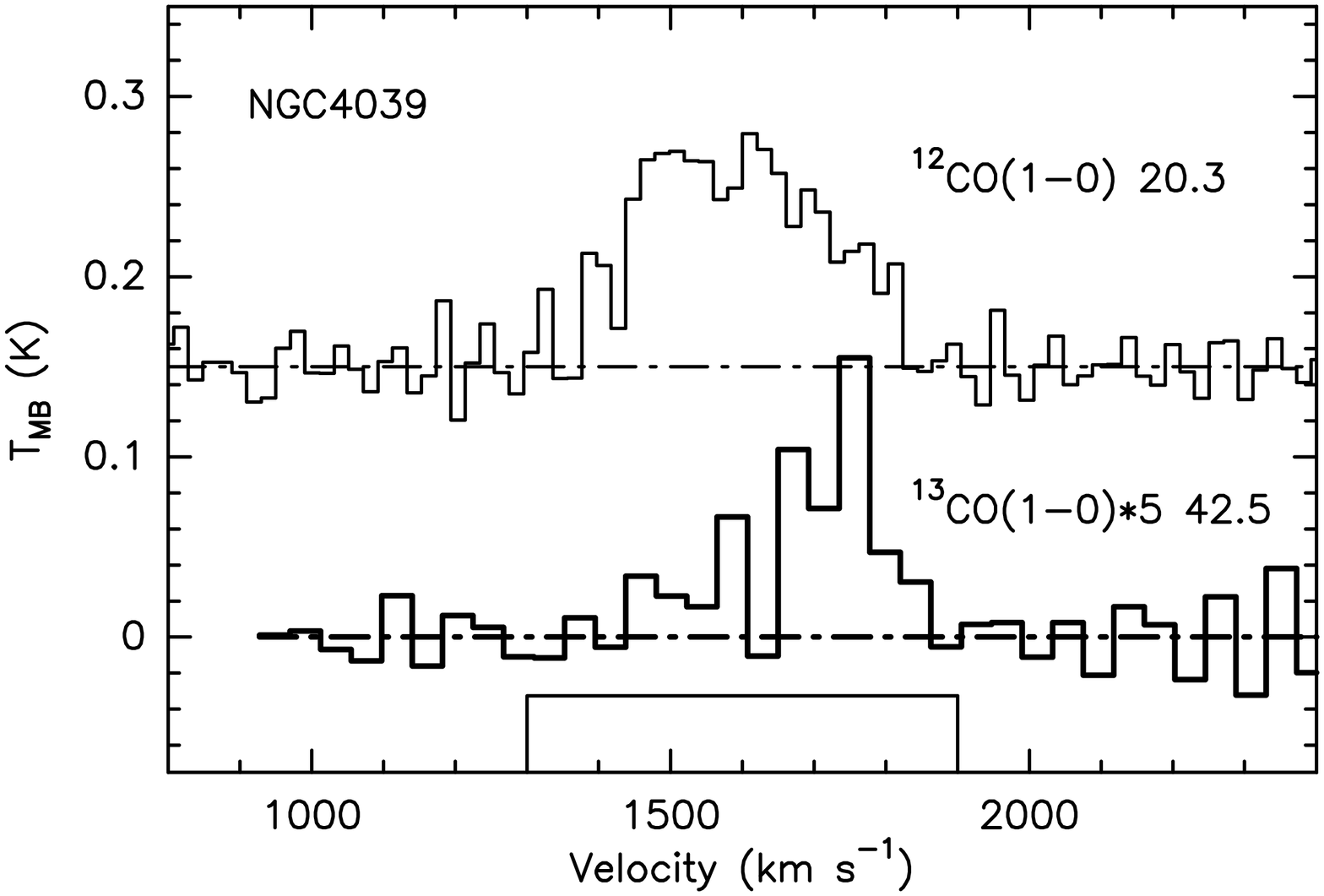}
\includegraphics[width=0.33\textwidth, angle=-90]{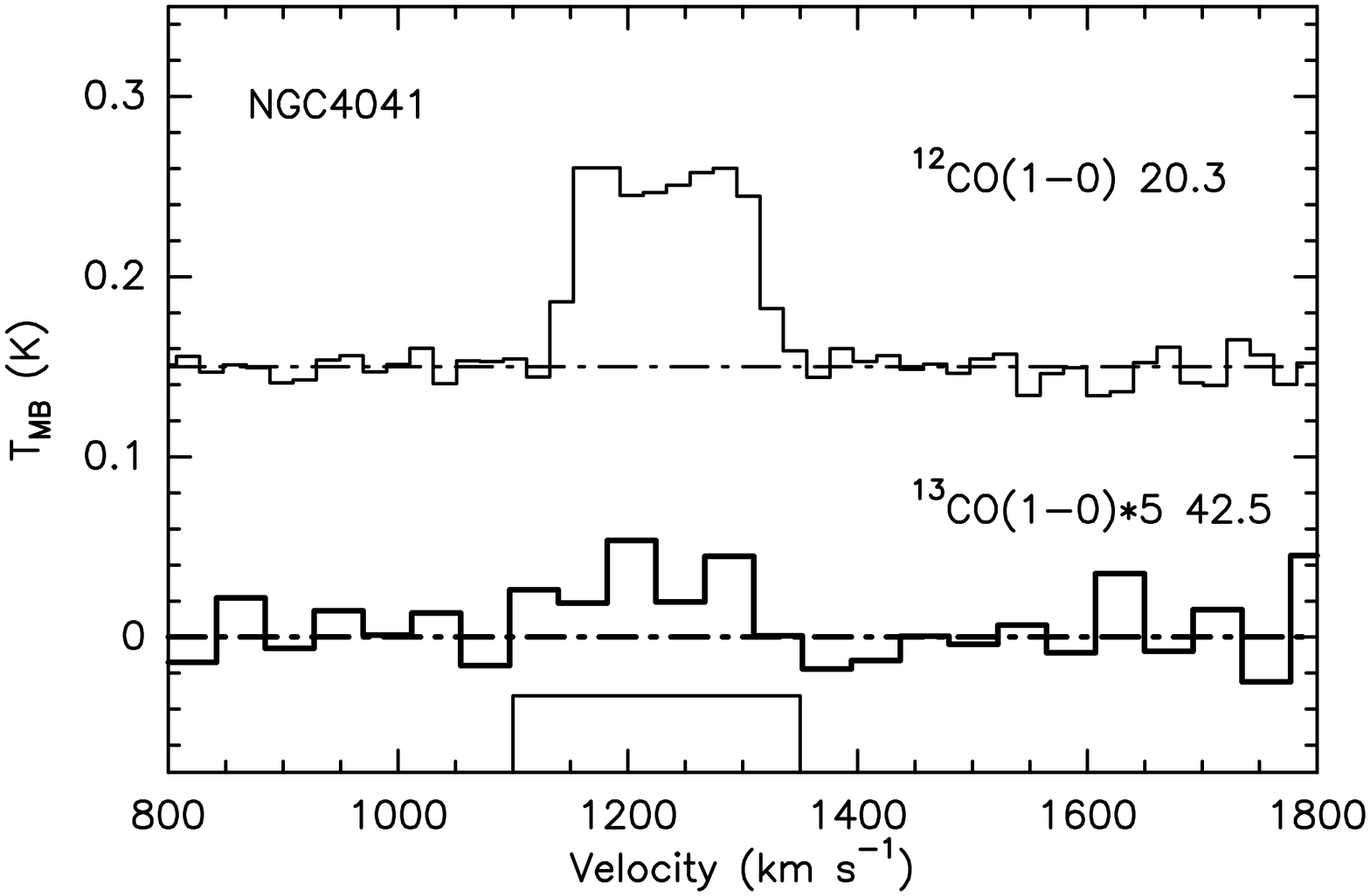}
\includegraphics[width=0.33\textwidth, angle=-90]{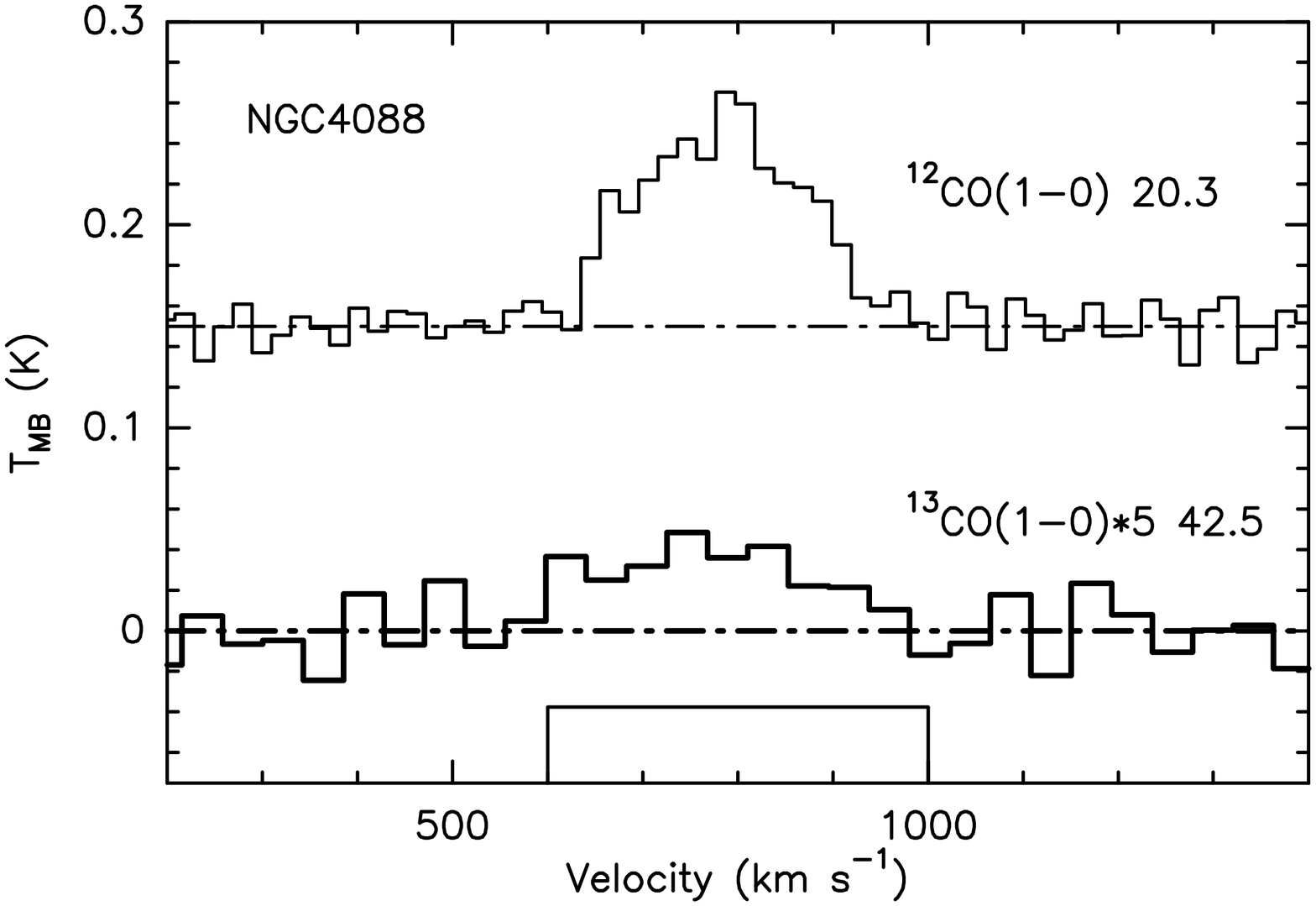}
\\[1cm]
\textbf{Fig. 1}\quad\textit{-- Continued}
\vspace{1cm}
\end{figure}

\begin{figure}[!htp]
  \centering
\includegraphics[width=0.33\textwidth, angle=-90]{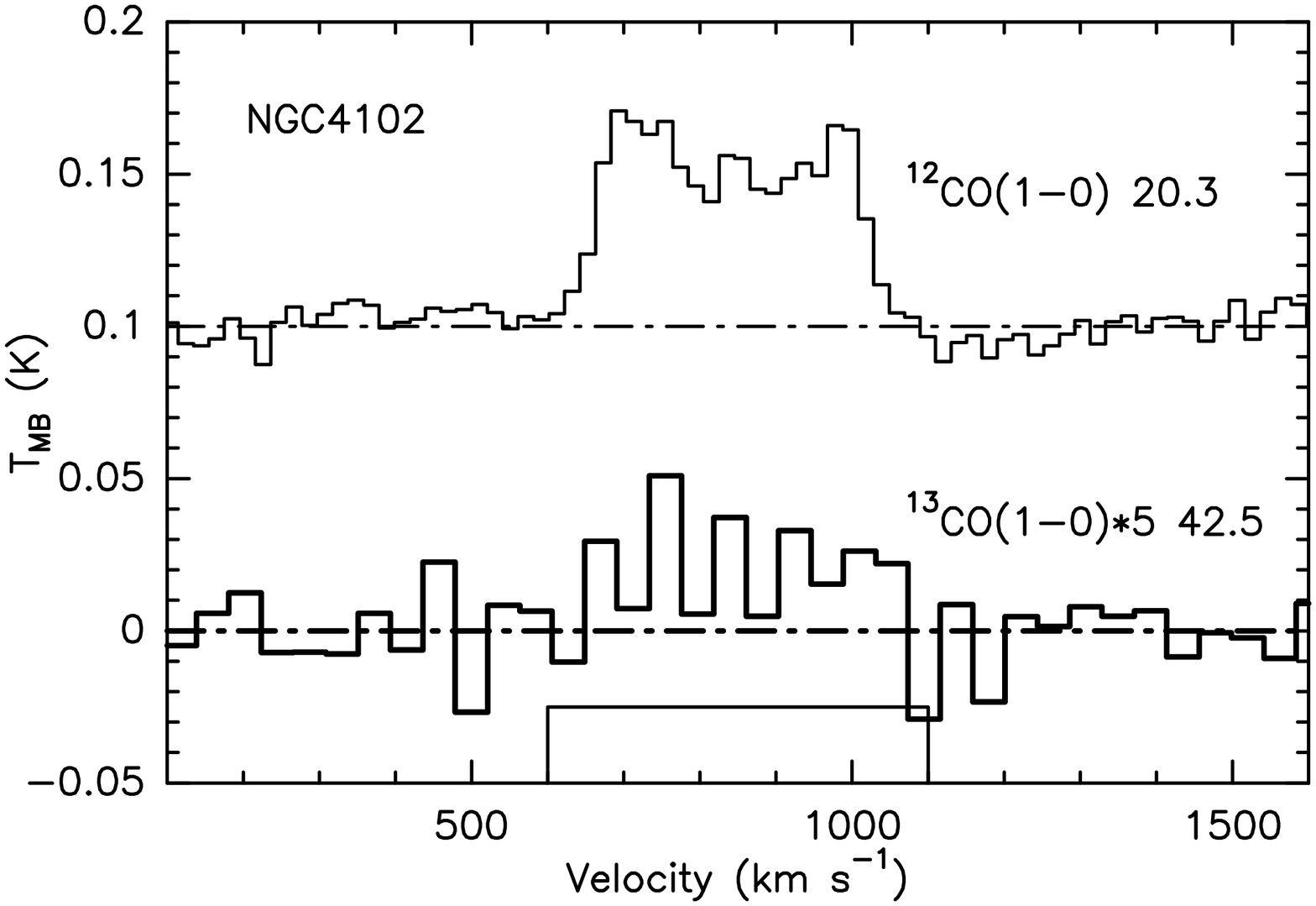}
\includegraphics[width=0.33\textwidth, angle=-90]{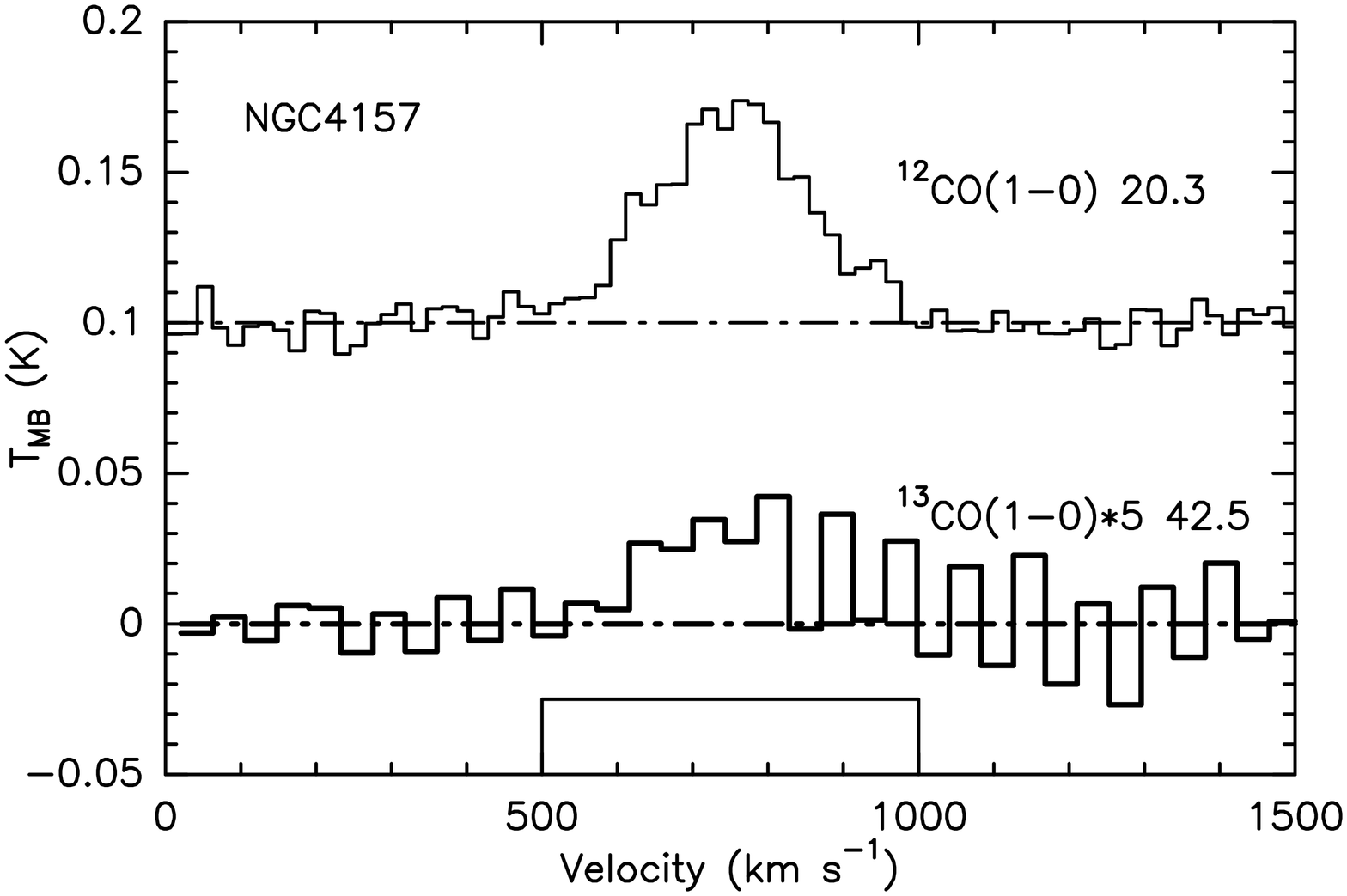}
\includegraphics[width=0.33\textwidth, angle=-90]{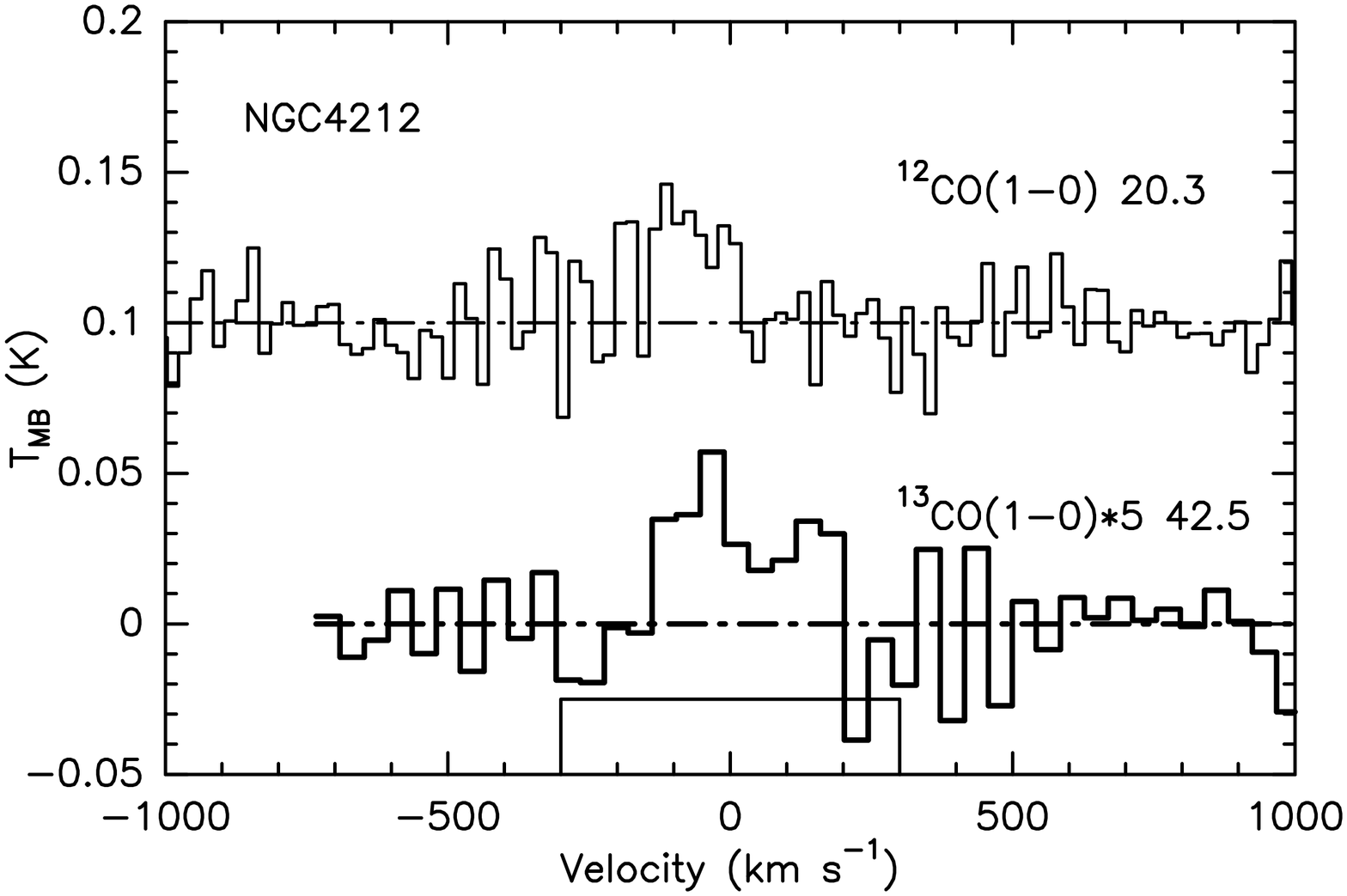}
\includegraphics[width=0.33\textwidth, angle=-90]{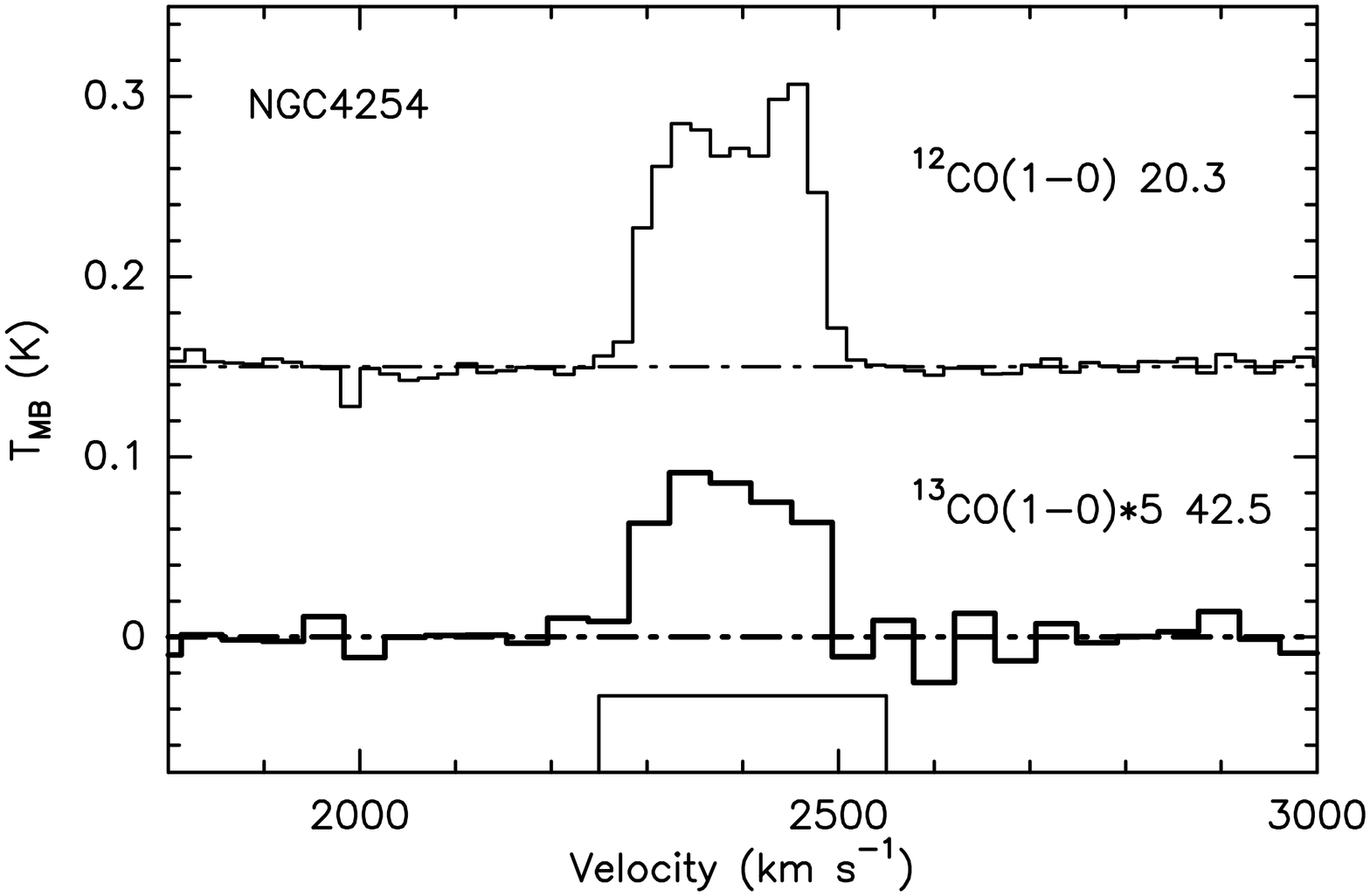}
\includegraphics[width=0.33\textwidth, angle=-90]{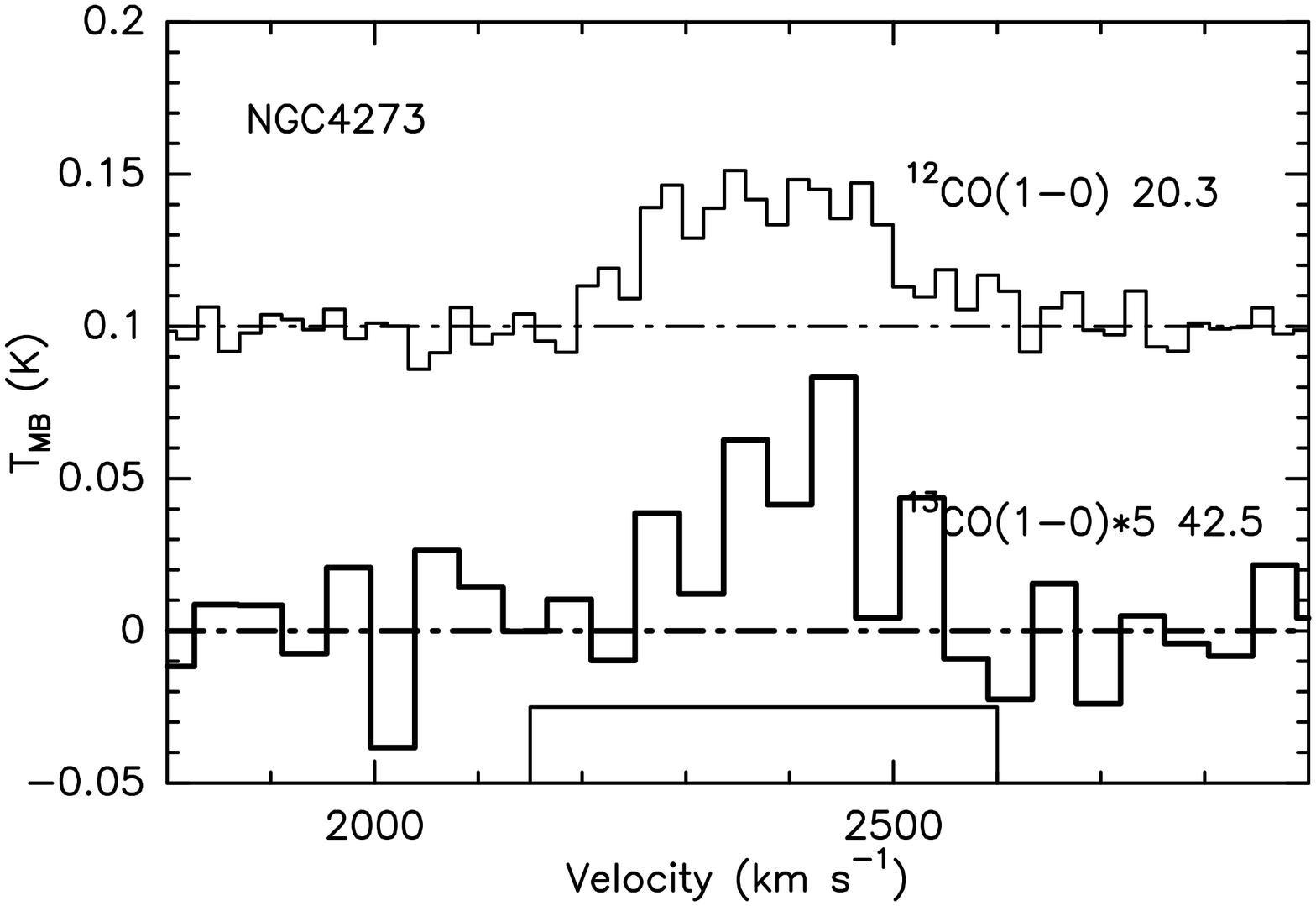}
\includegraphics[width=0.33\textwidth, angle=-90]{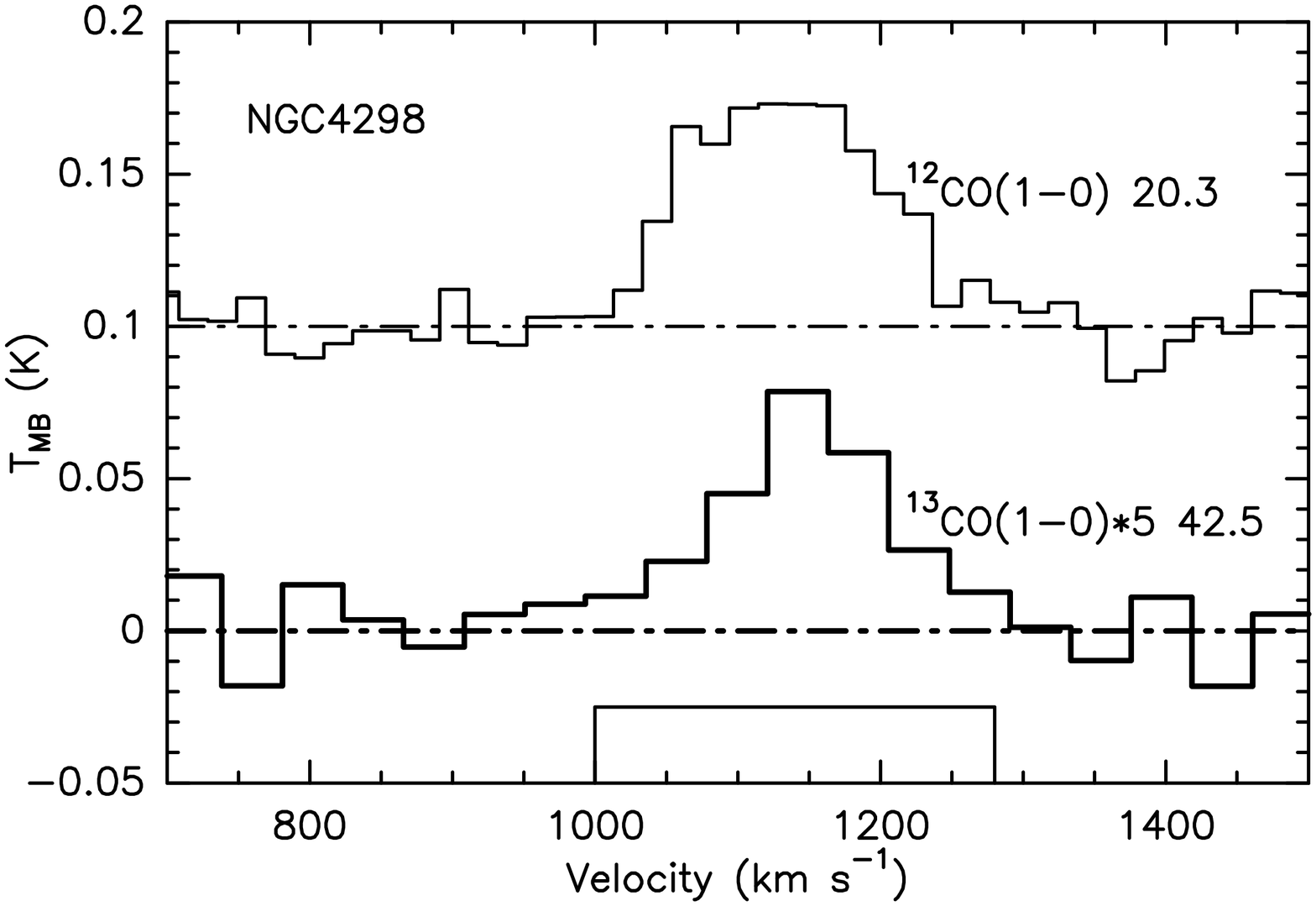}
\includegraphics[width=0.33\textwidth, angle=-90]{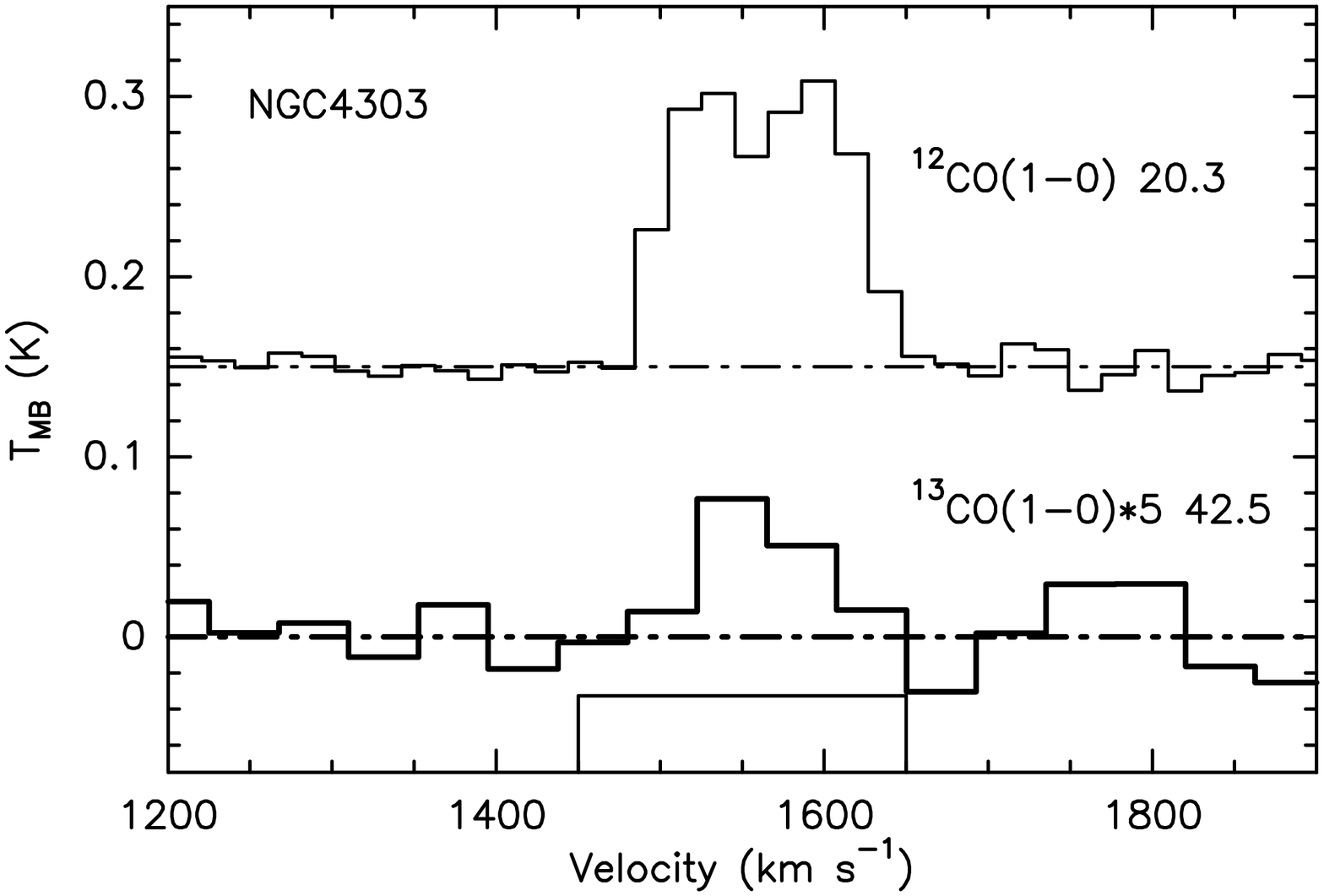}
\includegraphics[width=0.33\textwidth, angle=-90]{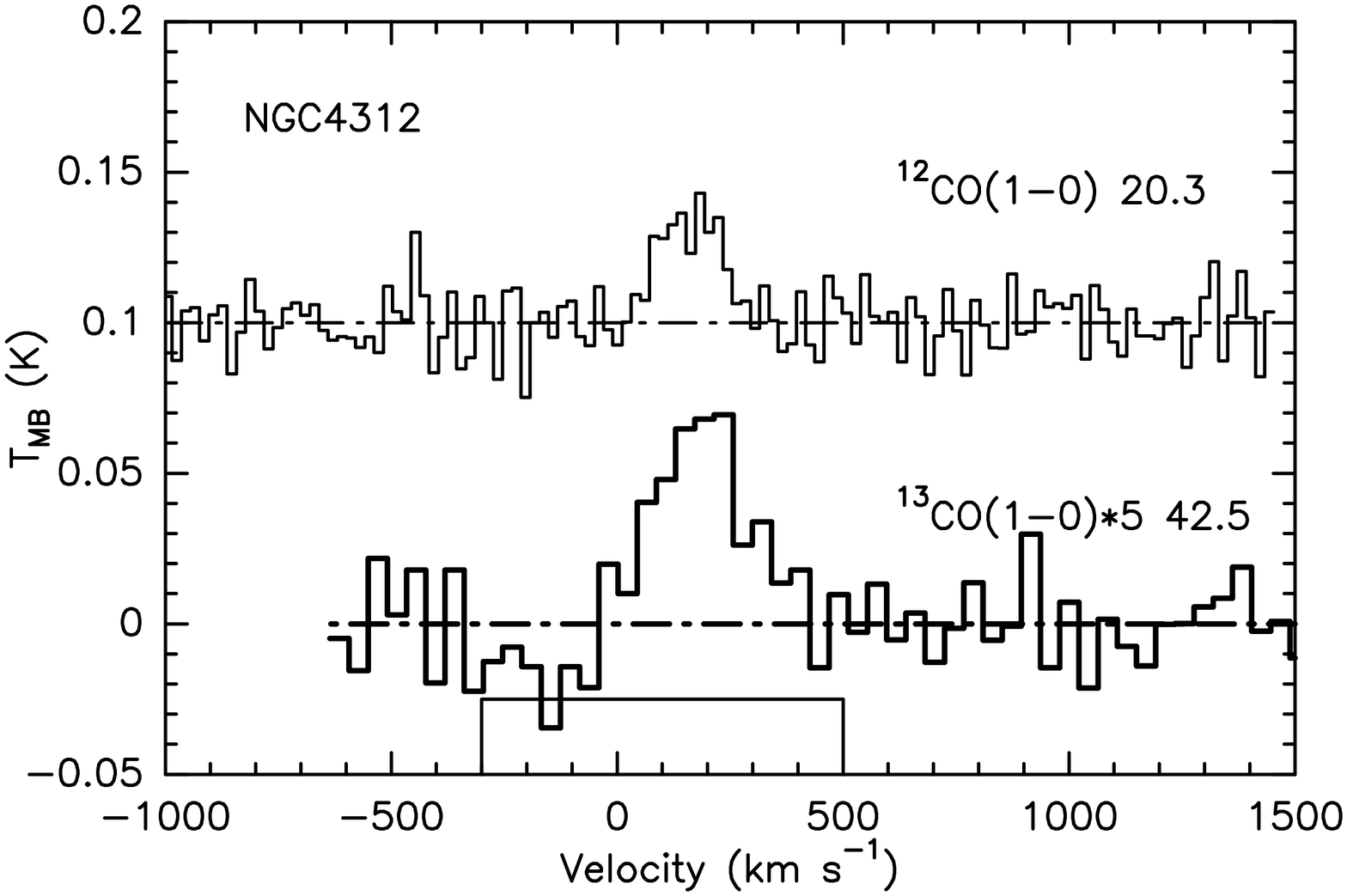}
\\[1cm]
\textbf{Fig. 1}\quad\textit{-- Continued}
\vspace{1cm}
\end{figure}

\begin{figure}[!htp]
  \centering
\includegraphics[width=0.33\textwidth, angle=-90]{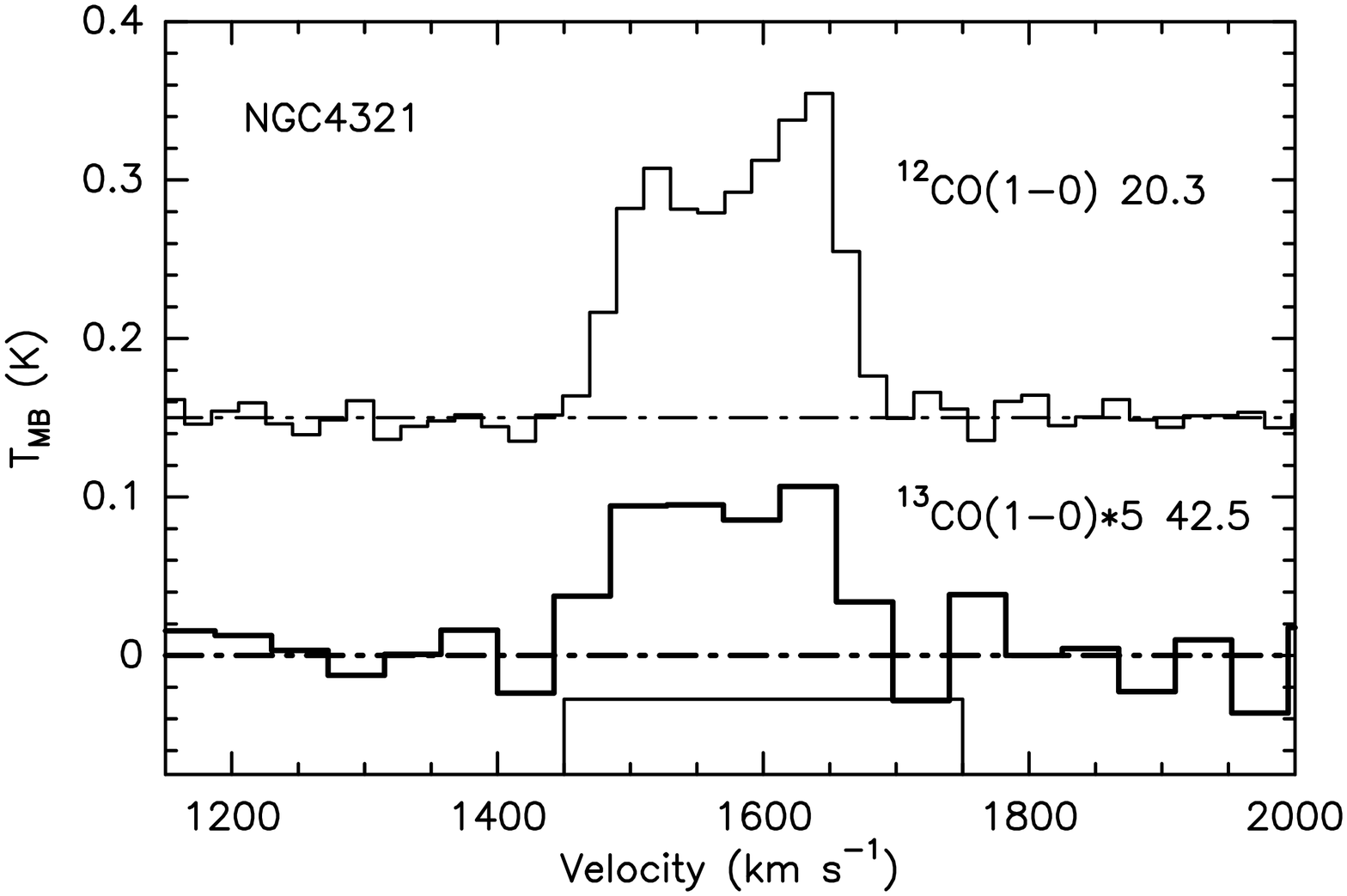}
\includegraphics[width=0.33\textwidth, angle=-90]{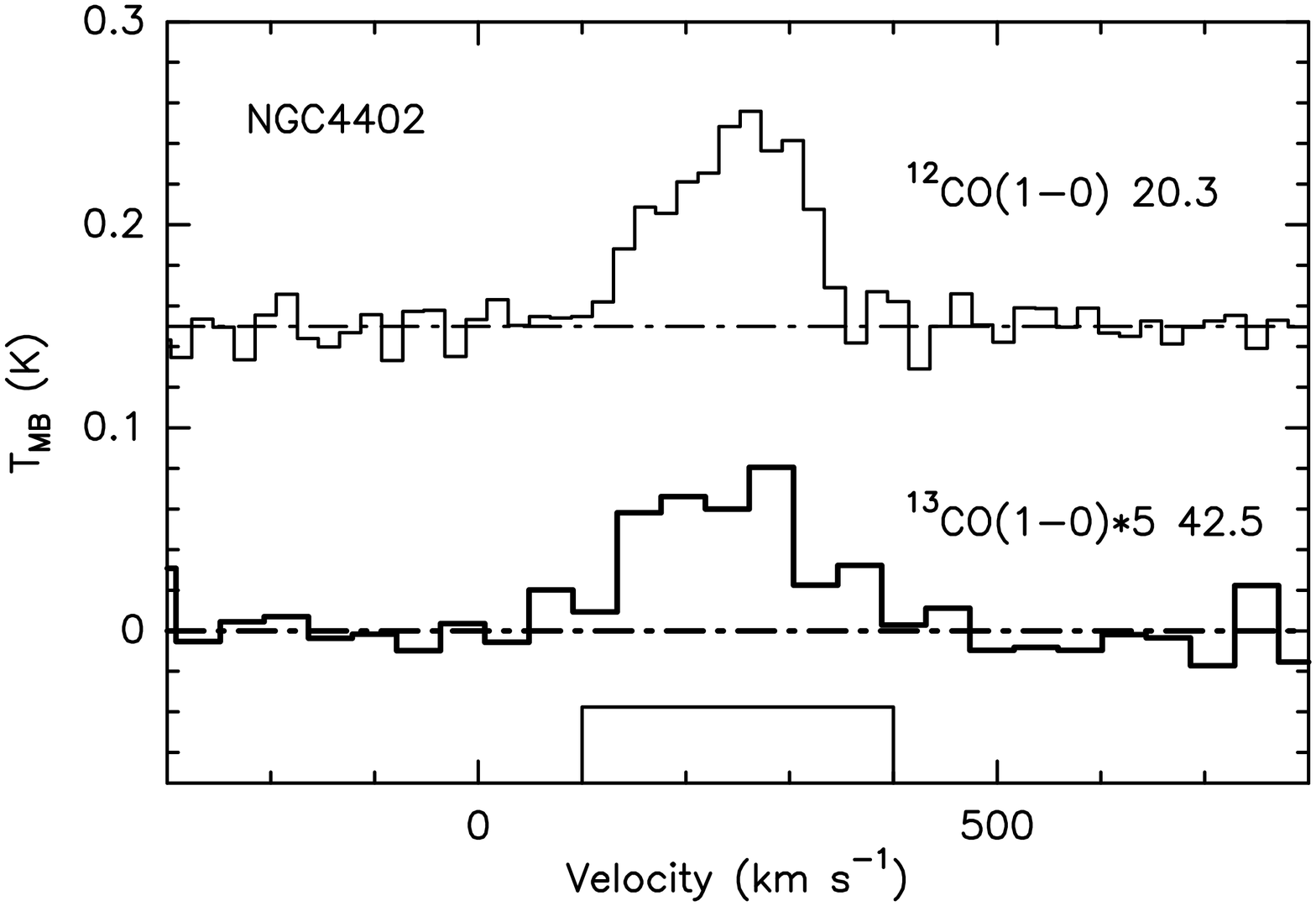}
\includegraphics[width=0.33\textwidth, angle=-90]{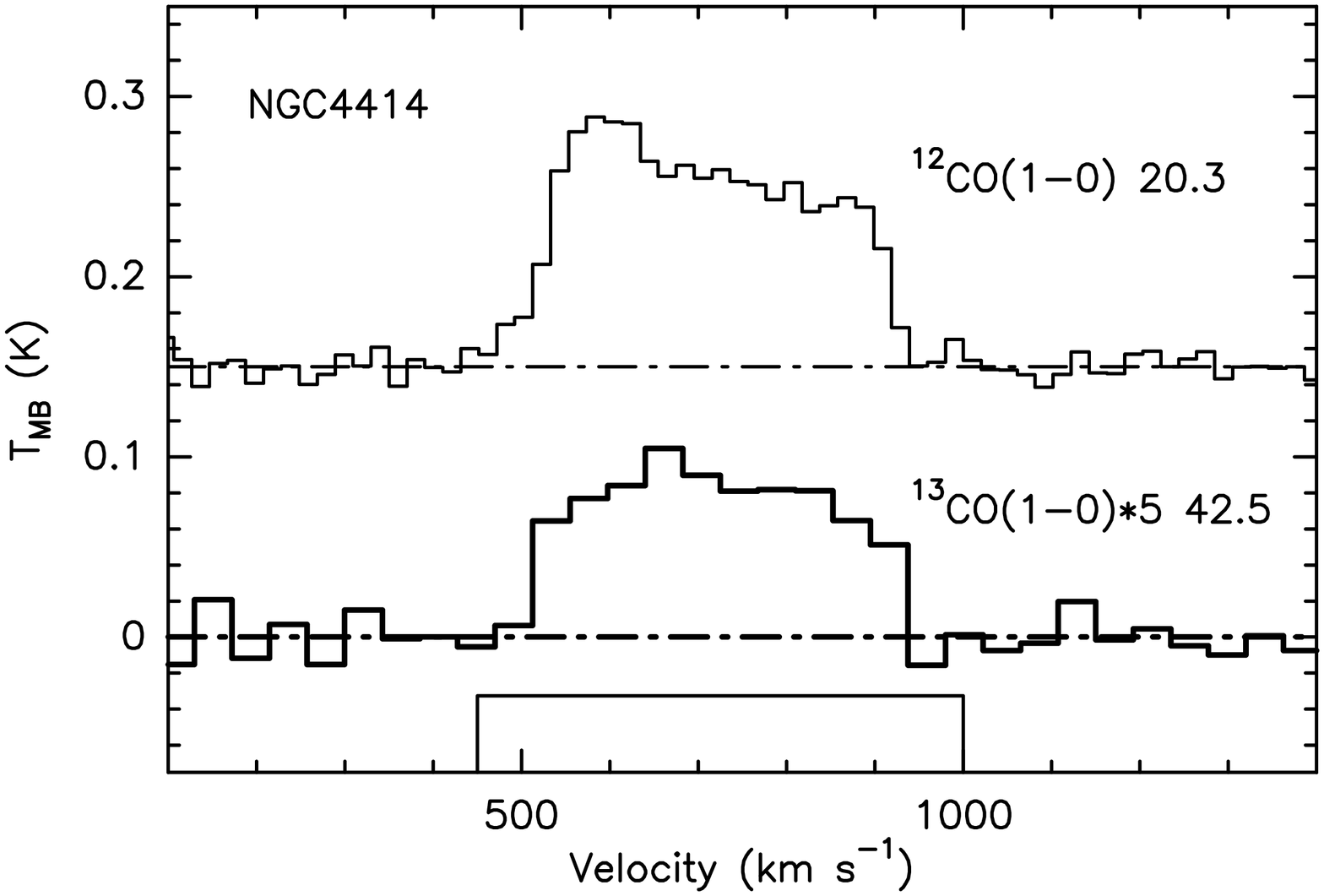}
\includegraphics[width=0.33\textwidth, angle=-90]{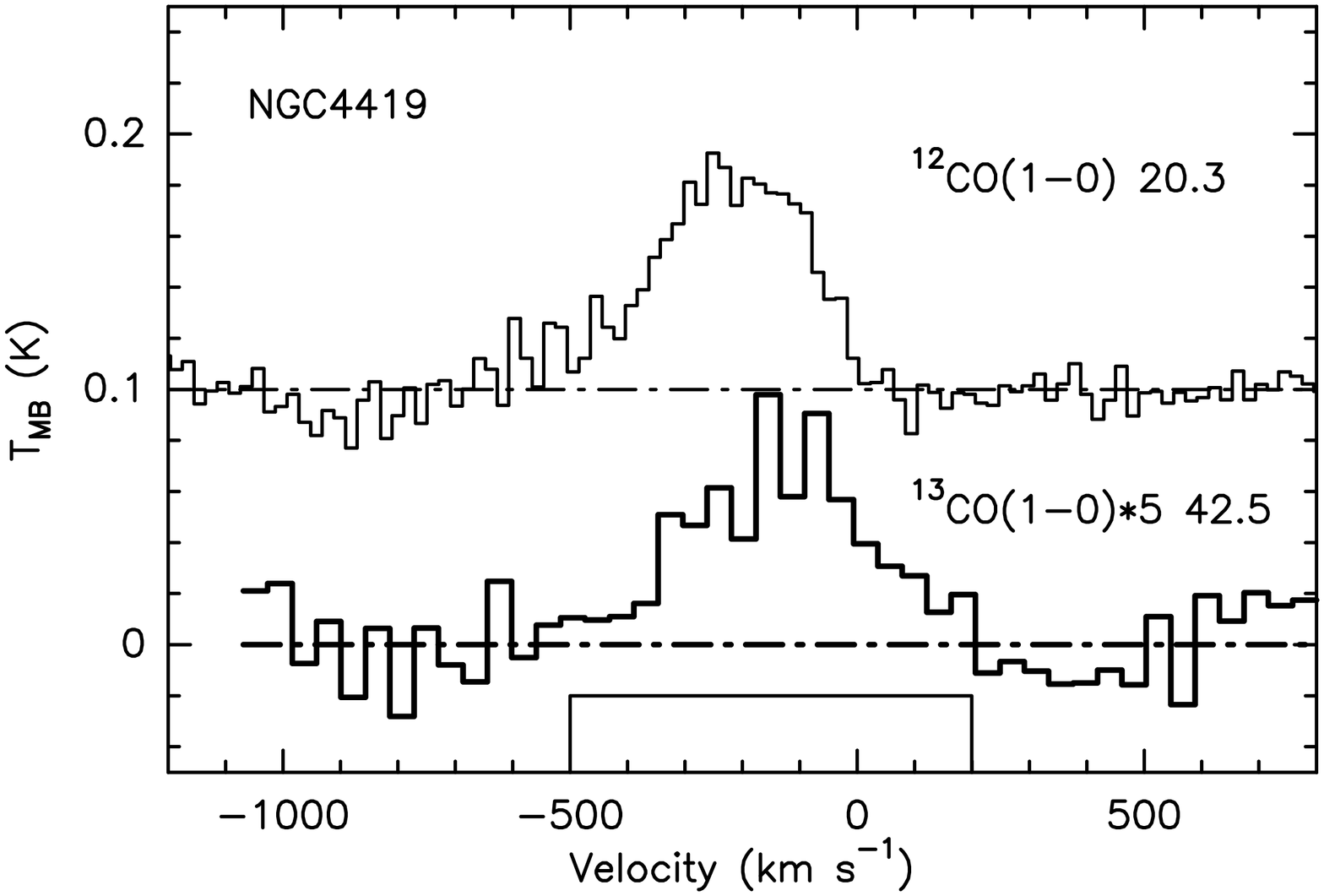}
\includegraphics[width=0.33\textwidth, angle=-90]{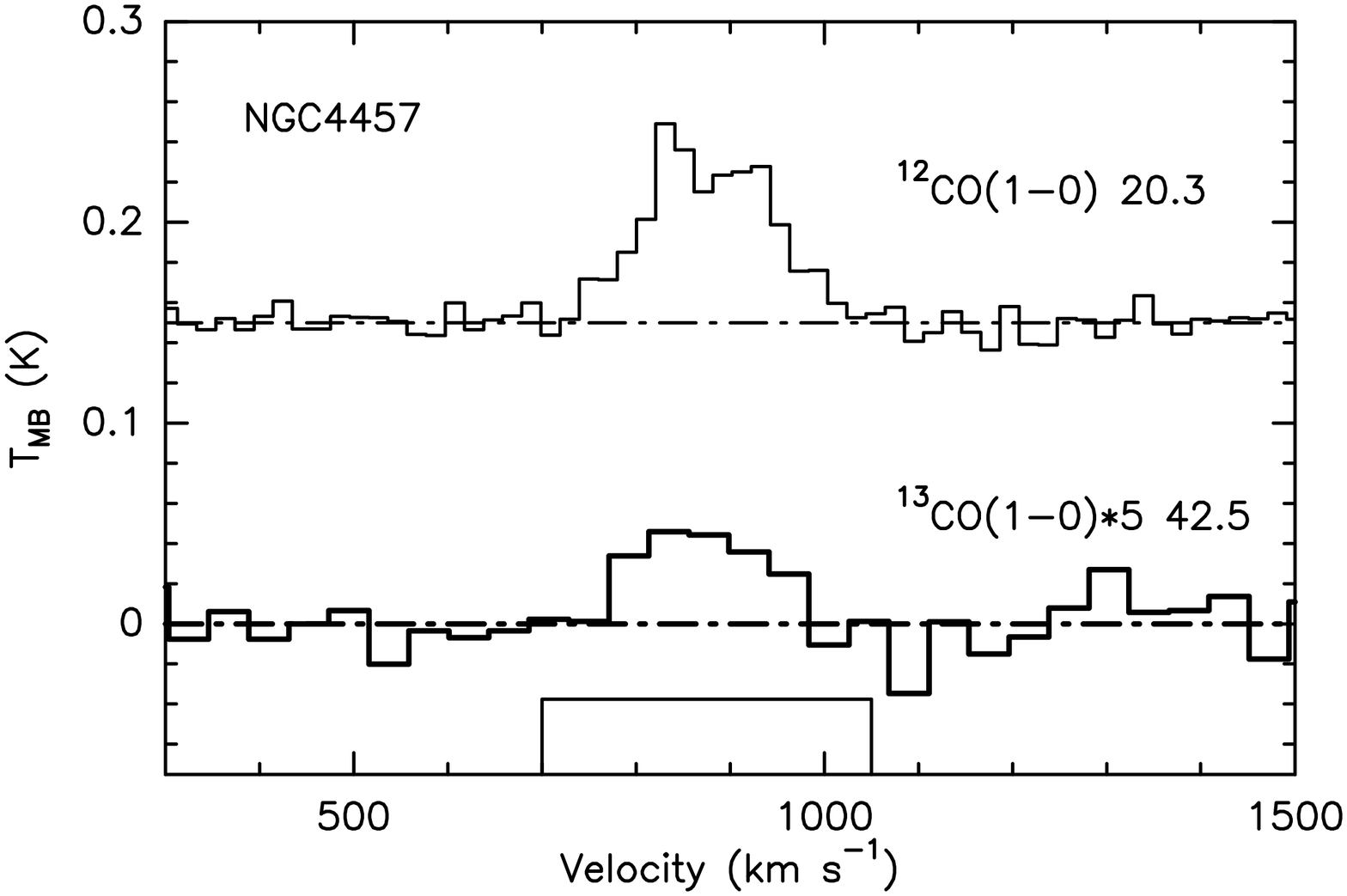}
\includegraphics[width=0.33\textwidth, angle=-90]{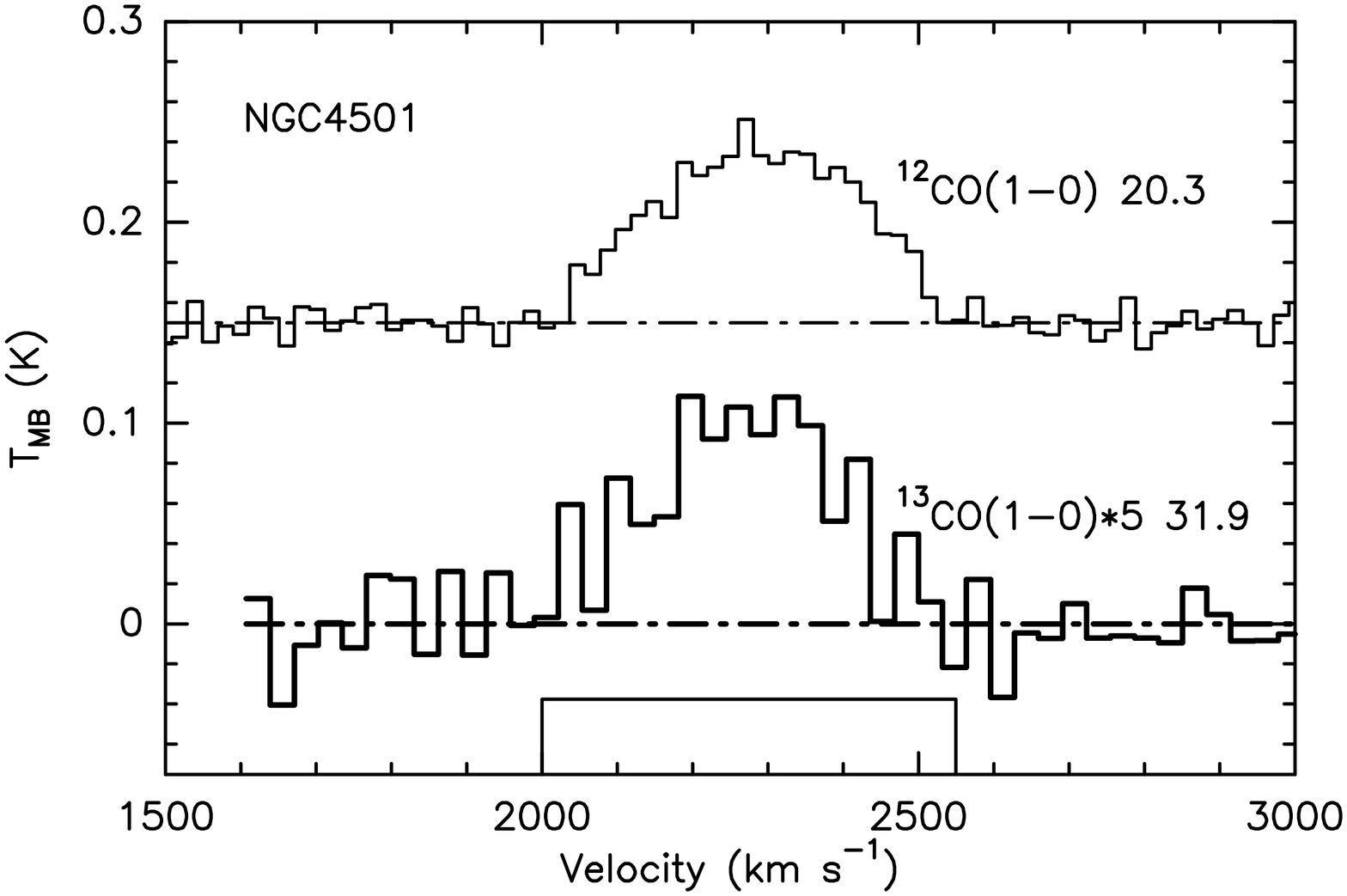}
\includegraphics[width=0.33\textwidth, angle=-90]{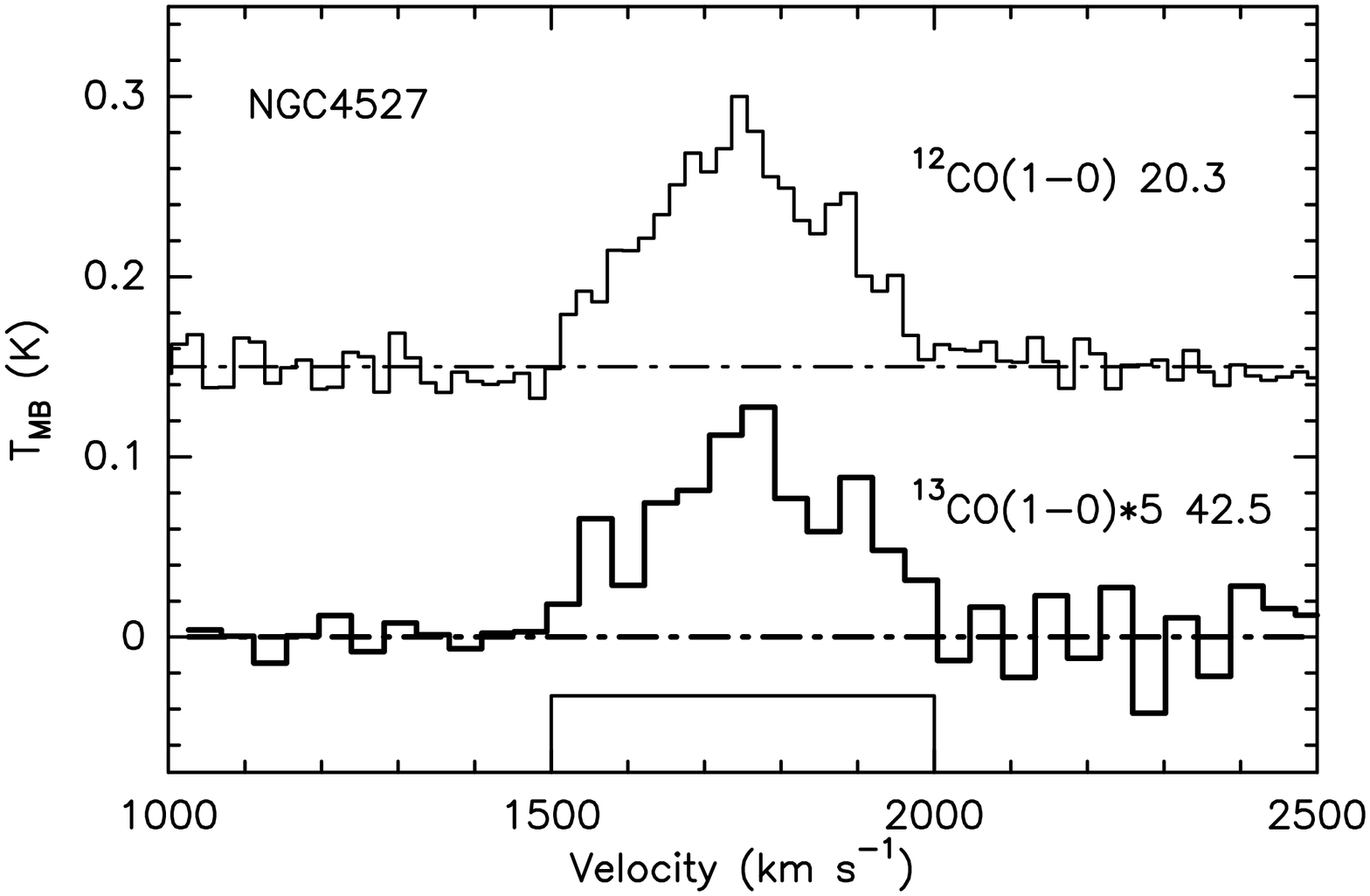}
\includegraphics[width=0.33\textwidth, angle=-90]{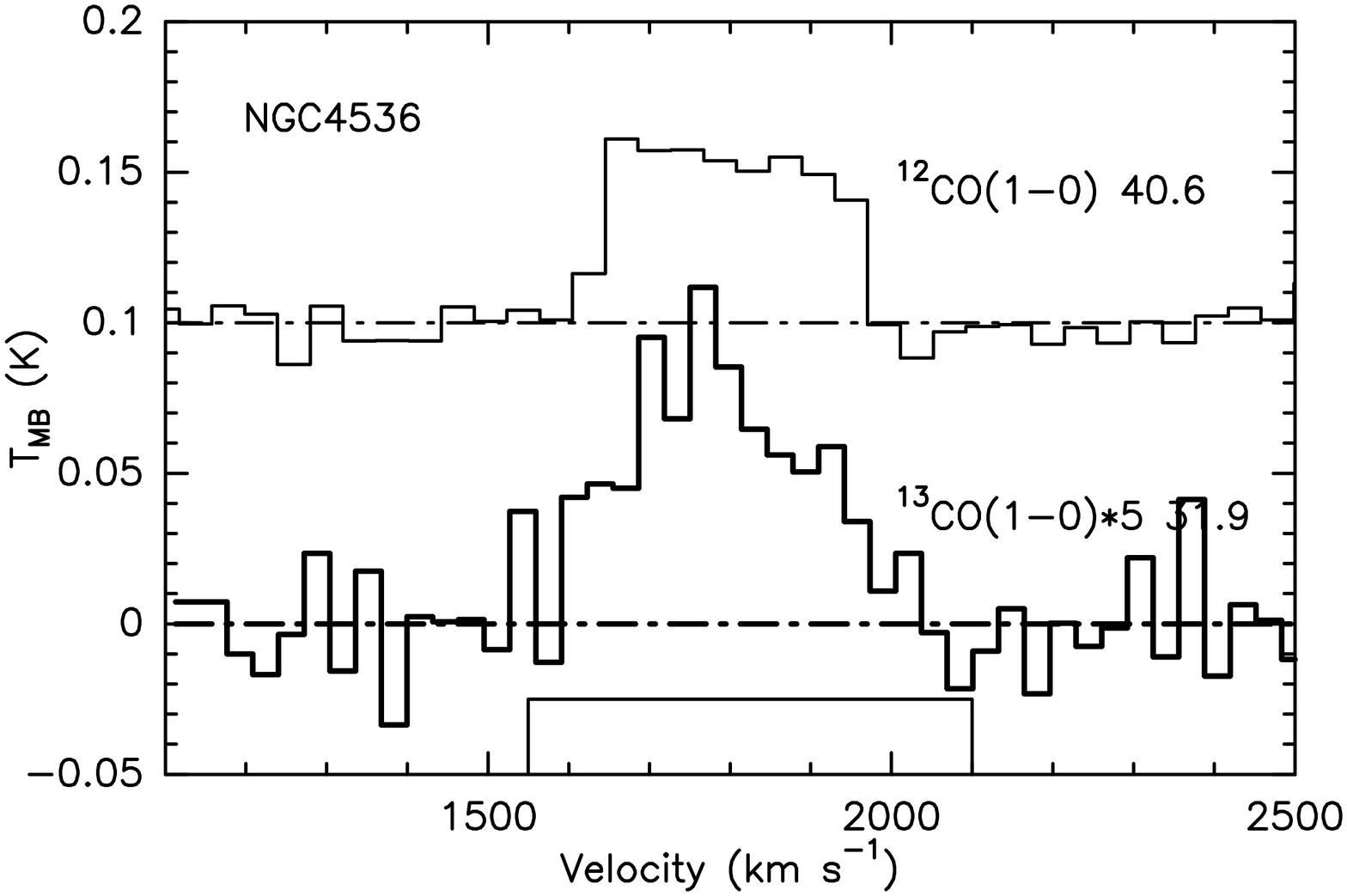}
\\[1cm]
\textbf{Fig. 1}\quad\textit{-- Continued}
\vspace{1cm}
\end{figure}

\begin{figure}[!htp]
  \centering
\includegraphics[width=0.33\textwidth, angle=-90]{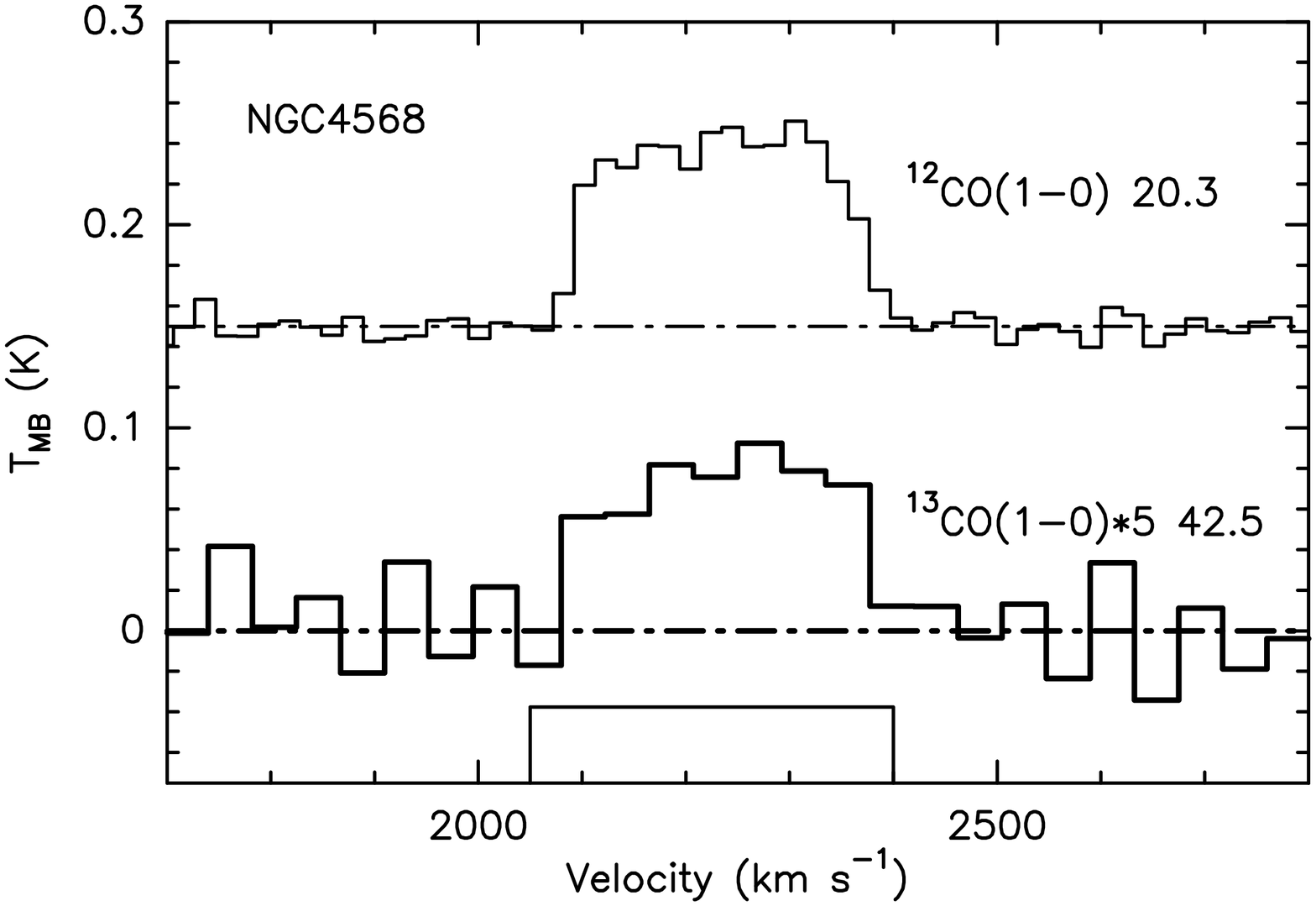}
\includegraphics[width=0.33\textwidth, angle=-90]{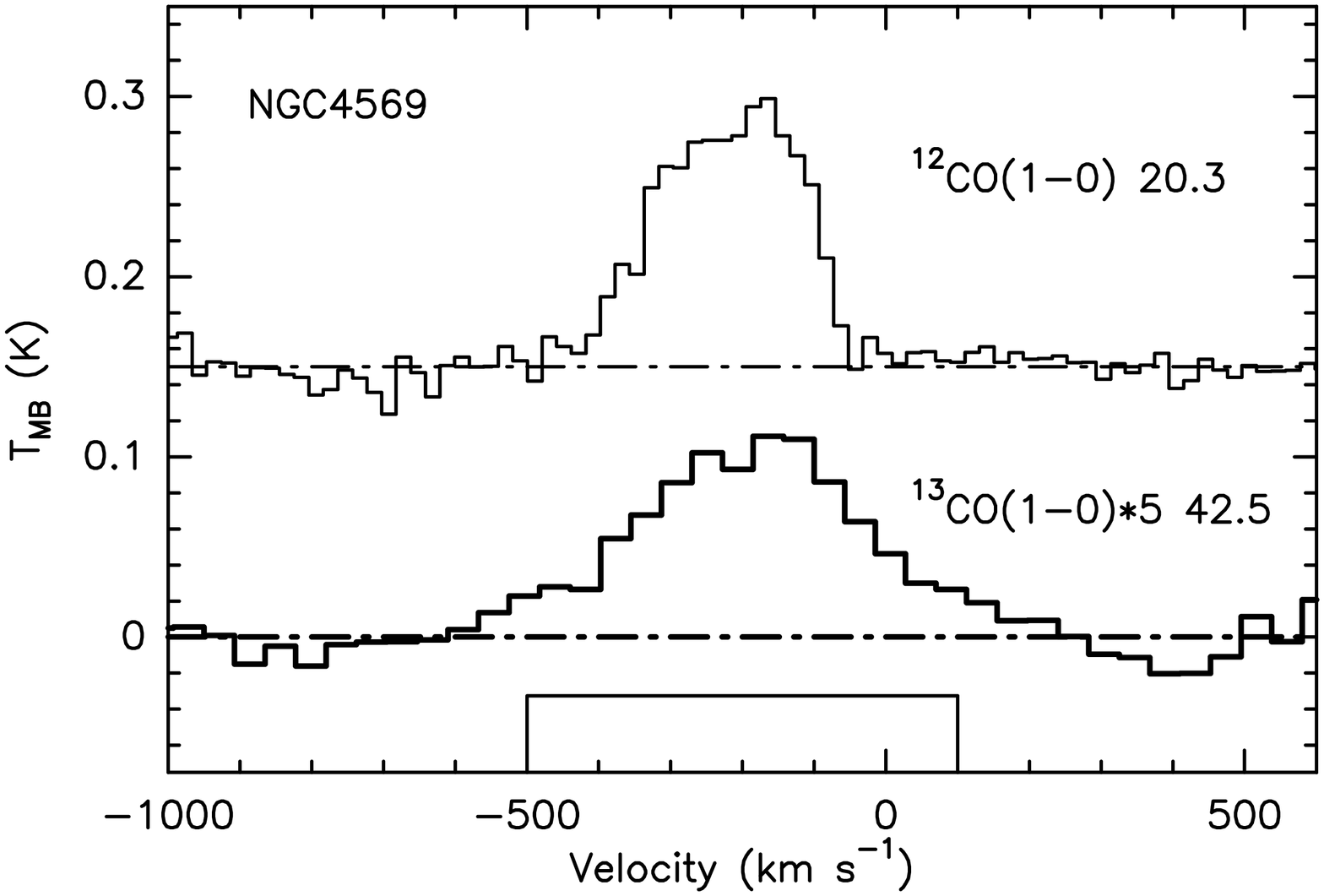}
\includegraphics[width=0.33\textwidth, angle=-90]{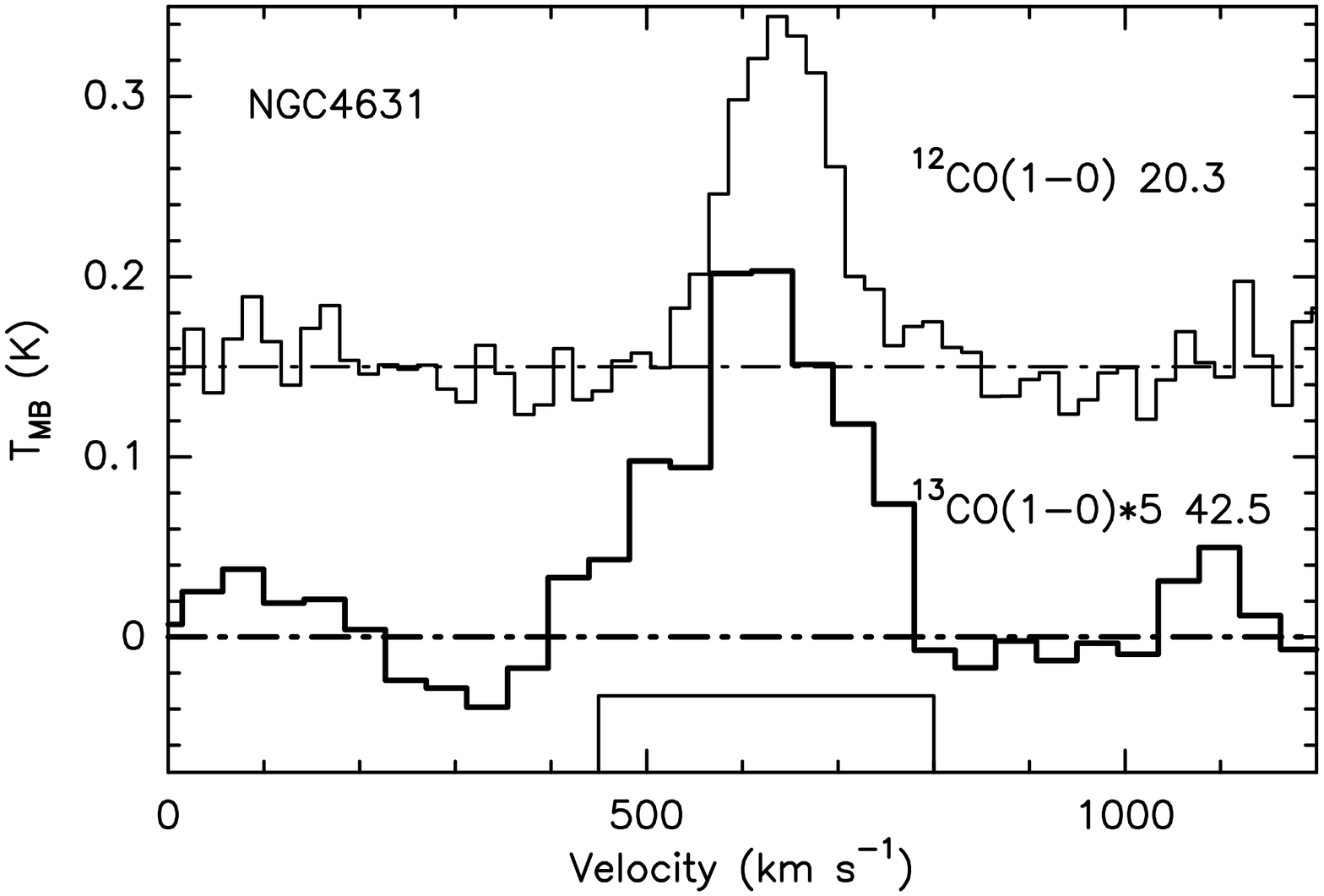}
\includegraphics[width=0.33\textwidth, angle=-90]{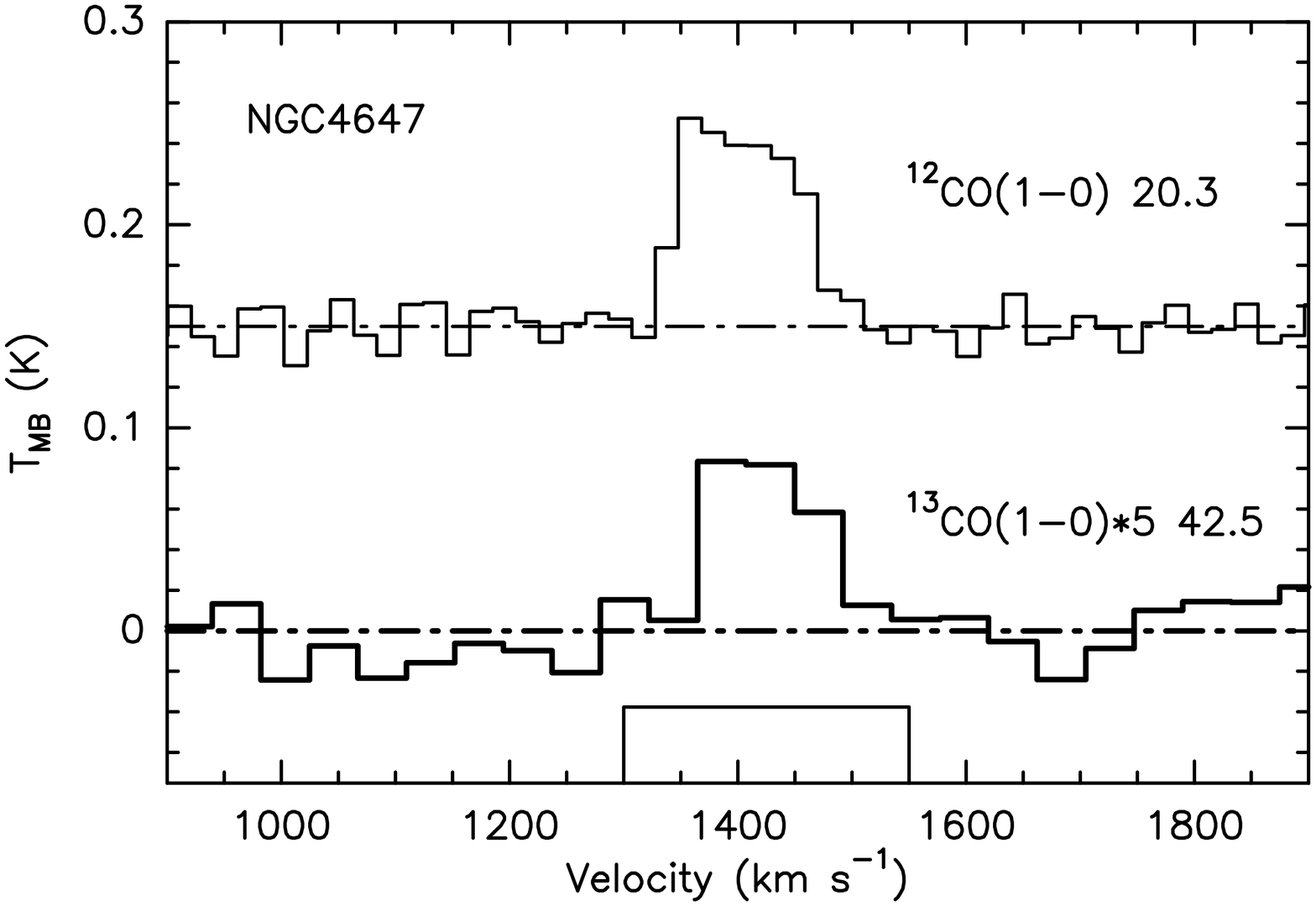}
\includegraphics[width=0.33\textwidth, angle=-90]{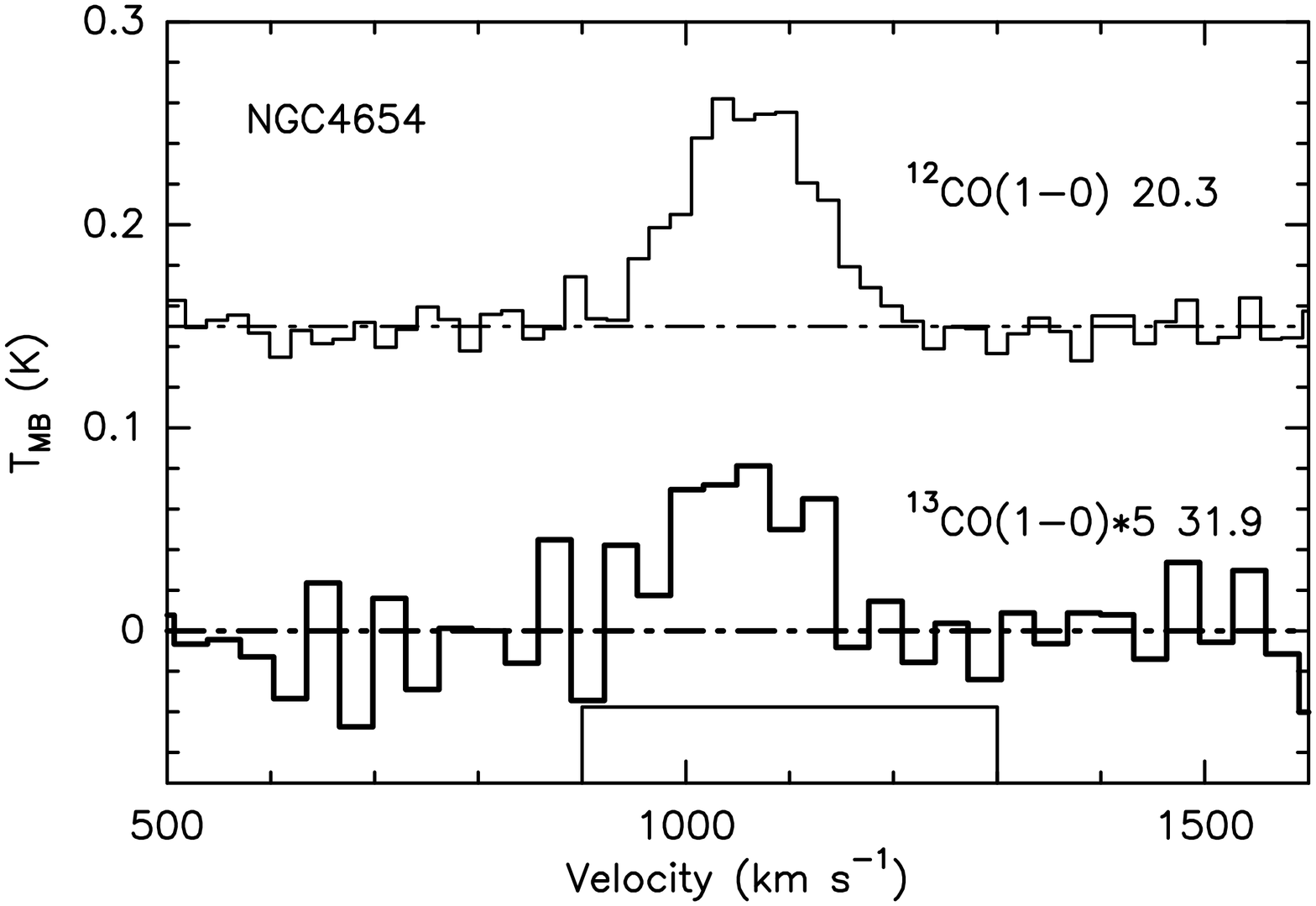}
\includegraphics[width=0.33\textwidth, angle=-90]{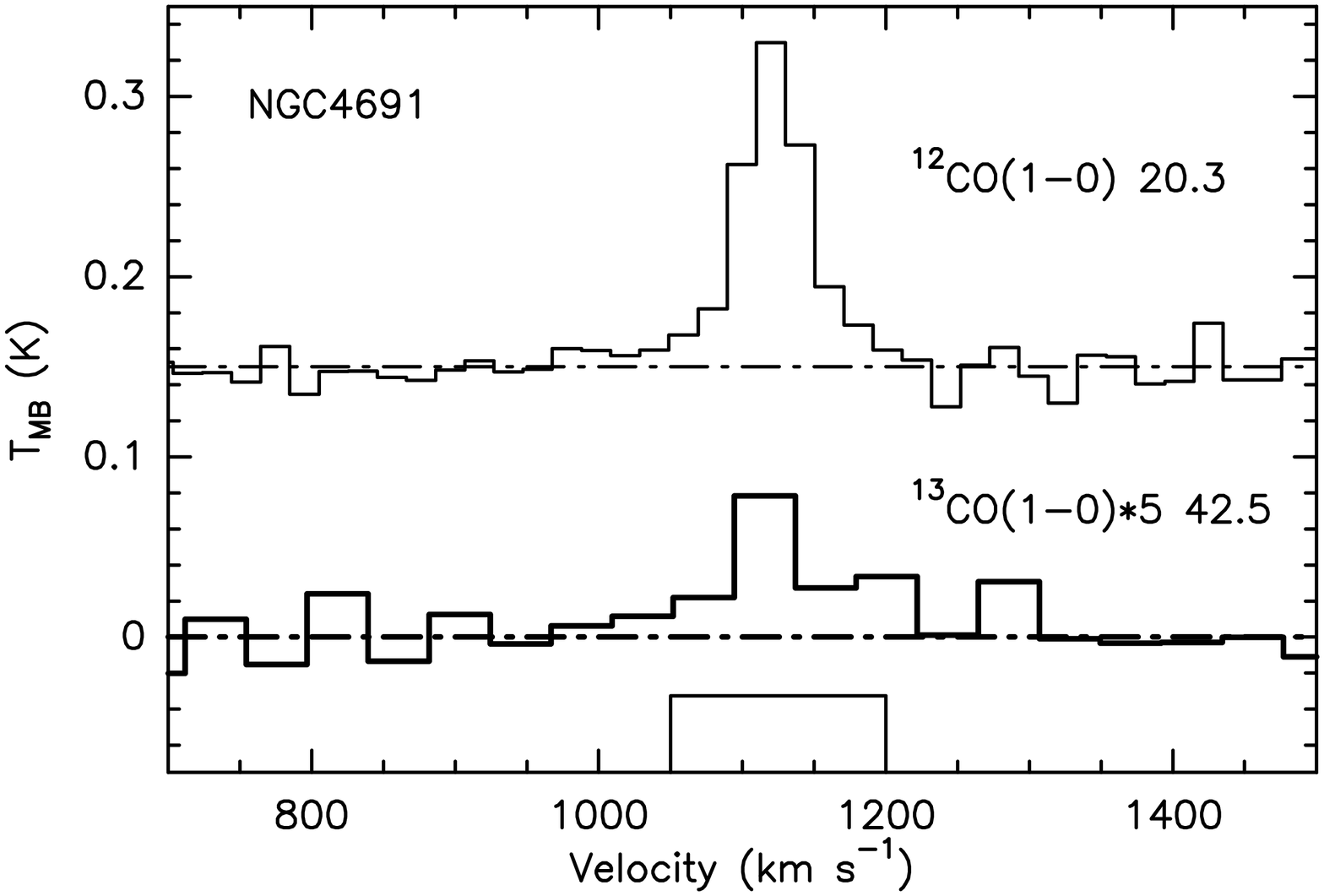}
\includegraphics[width=0.33\textwidth, angle=-90]{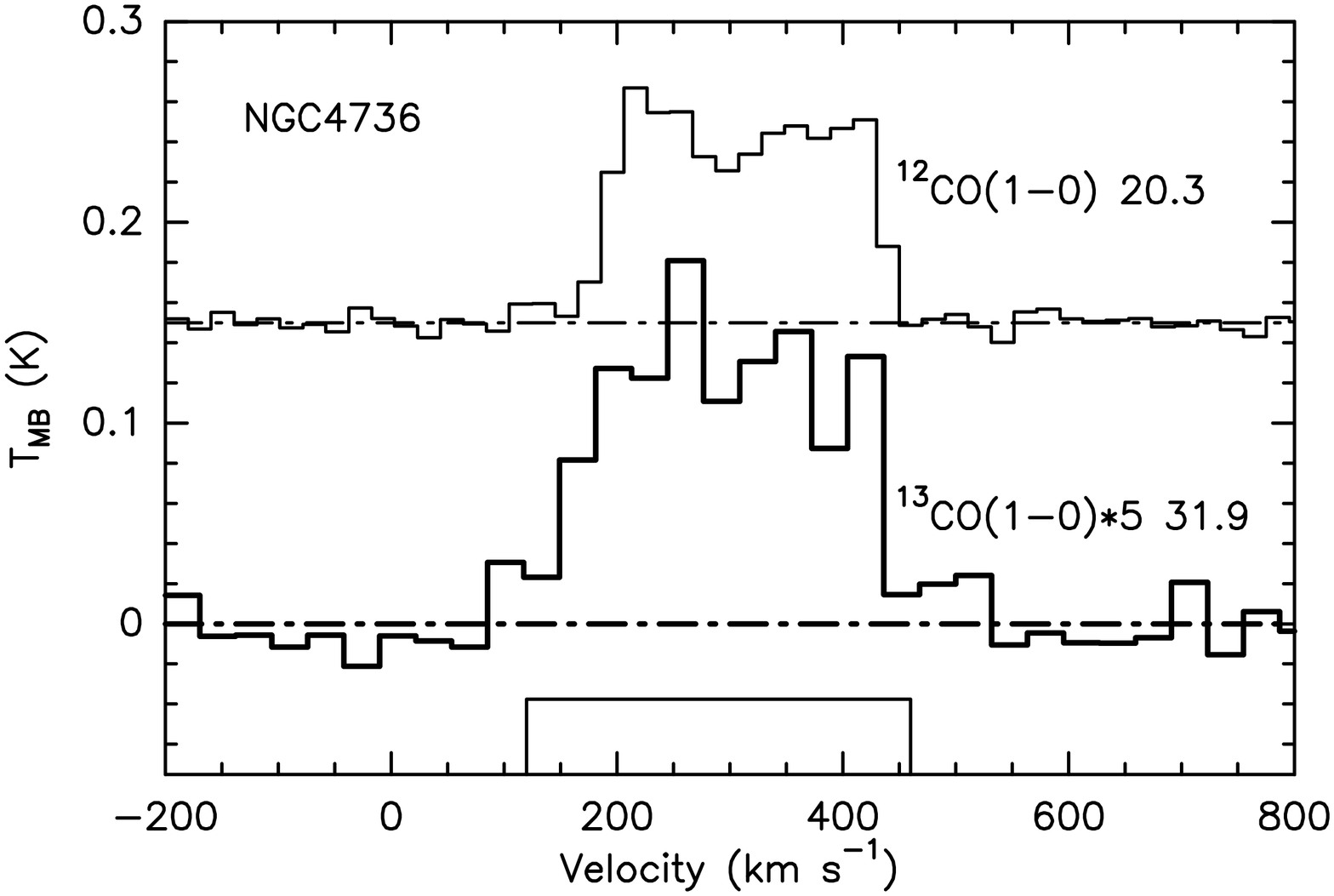}
\includegraphics[width=0.33\textwidth, angle=-90]{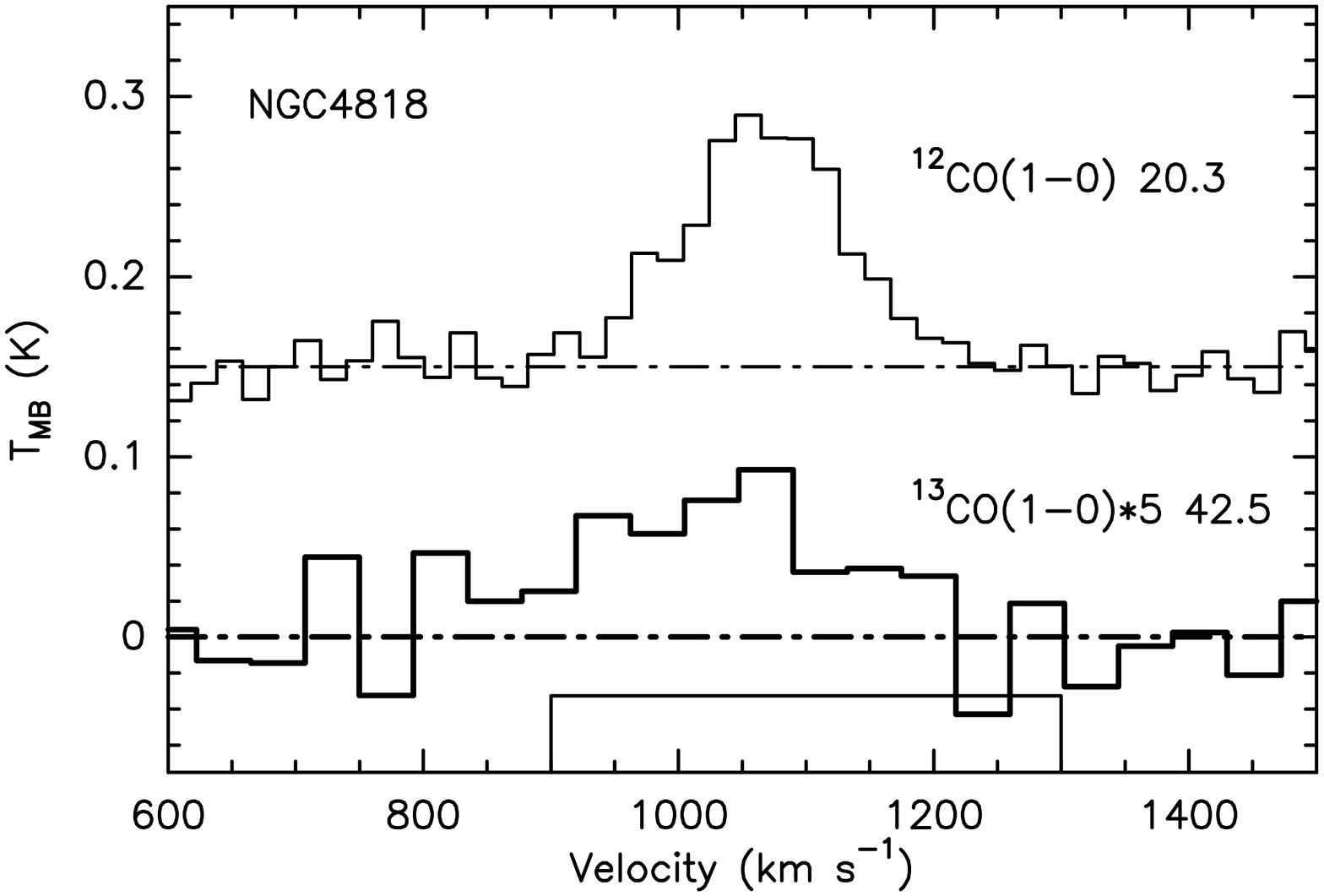}
\\[1cm]
\textbf{Fig. 1}\quad\textit{-- Continued}
\vspace{1cm}
\end{figure}

\begin{figure}[!ht]
  \centering
\includegraphics[width=0.33\textwidth, angle=-90]{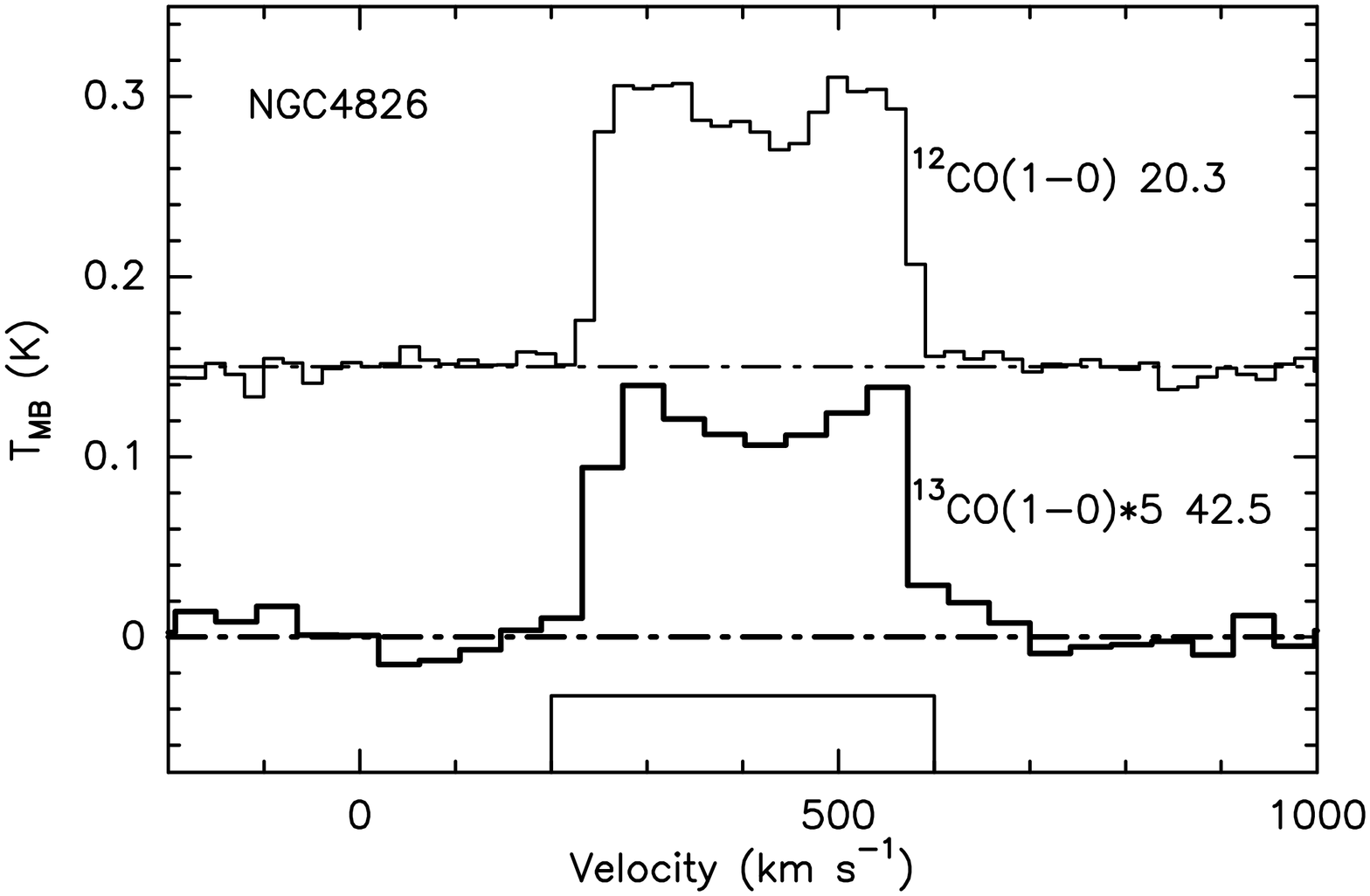}
\includegraphics[width=0.33\textwidth, angle=-90]{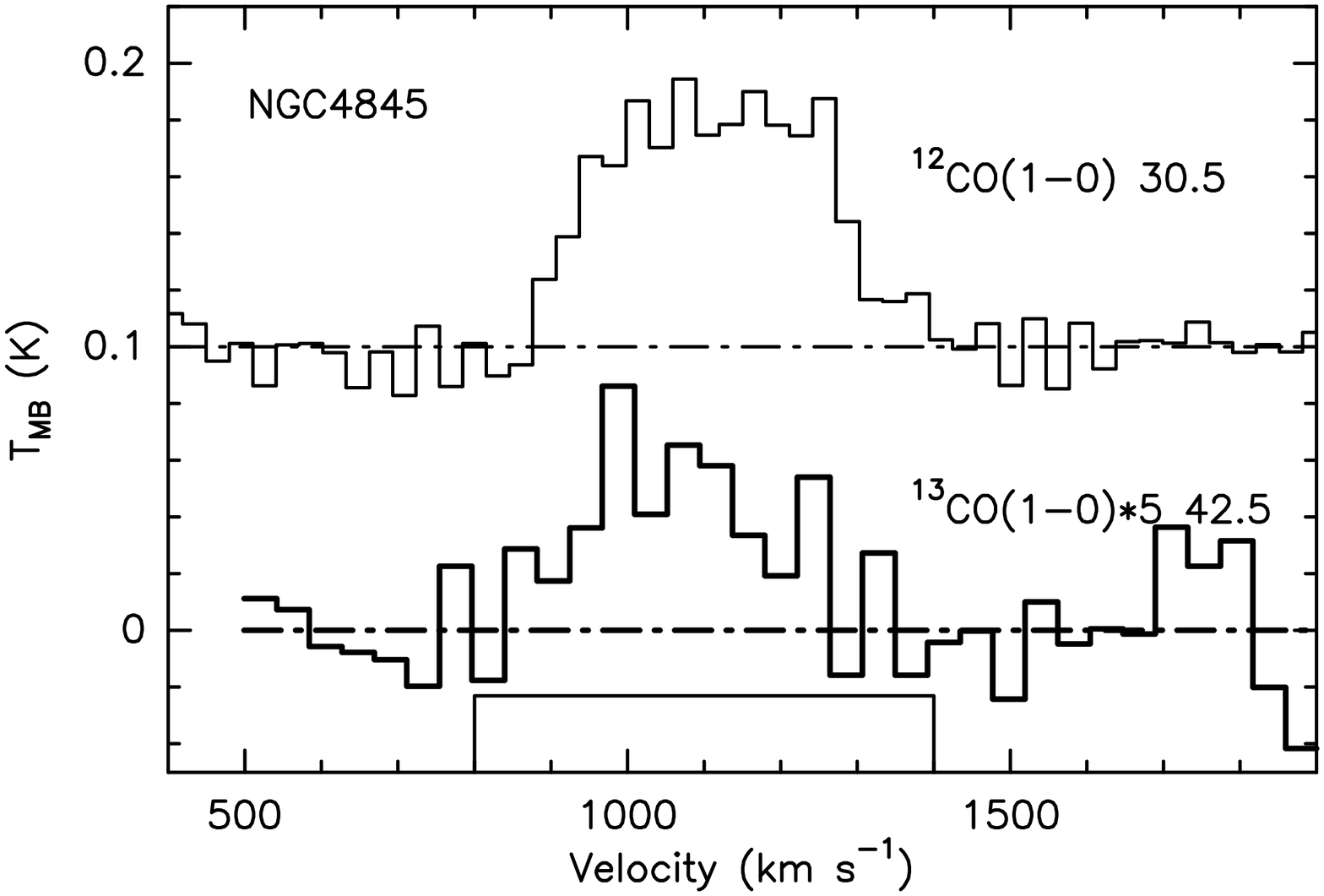}
\\[1cm]
\textbf{Fig. 1}\quad\textit{-- Continued}
\vspace{1cm}
\end{figure}

\section{Results}
\label{sect:data}

\subsection{Data reduction}
We reduced the data using CLASS, which is a part of the GILDAS \footnote{\url{http://www.iram.fr/IRAMFR/GILDAS/}} software package. The original data included individual scans. Of each spectrum, line-free channels that exhibited positive or negative spikes more than 5$\sigma$ above the rms noise were blanked or substituted with values interpolated from adjacent ones. This was only done after properly setting the velocity range limits for each spectrum and a liner baseline was subtracted from it. We then examined every spectrum and then we chose those with a relatively bad baseline and an abnormally high rms noise. Due to unstable spectrum baselines and bad weather conditions, a considerable part of the data was discarded. After converting the temperature scale to $T_{\textrm{mb}}$ from $T_\textrm{A}^*$ of the spectra and dividing by $\eta_{\textrm{mb}}$, we then averaged the converted data weighting by the inverse square of the rms noise, sigma. Note however that data from different observational seasons were treated differently. The final averaged spectra of each source were smoothed to a velocity resolution of about 20 \kms for $^{12}$CO and about 40 \kms for $^{13}$CO in order to limit the rms noise. Each averaged spectrum was fitted using the GAUSS method and the results are presented in the next subsection.

\subsection{Observational results}

We observed simultaneously $^{12}$CO, $^{13}$CO, and C$^{18}$O emissions at the centers of 58 nearby galaxies, among which 42 were detected in $^{13}$CO emission with a signal-to-noise ratio of more than 3, while C$^{18}$O emission was too weak to be detected in any galaxy with an upper limit of 2 mK (1 $\sigma$). Both $^{12}$CO and $^{13}$CO line profiles agree very well with similar centroid velocities and line widths in most cases, as shown in Fig.~\ref{Fig:spectra}. All of the spectra were smoothed to a velocity resolution of about 20 \kms for $^{12}$CO and about 40 \kms for $^{13}$CO to improve the signal-to-noise ratio, which are given in Table~\ref{Tab:obs_results}. We do not show those $^{12}$CO spectra without $^{13}$CO detection since one can easily find them in the literature.

The integrated intensity, $I_{\textrm{CO}}$, can be obtained by integrating $T_{\textrm{mb}}$ over the line emission feature,
\begin{equation}
I_{\textrm{CO}} \equiv \int T_{\textrm{mb}} dv \ [\textrm{K} \ \textrm{\kms}],
\label{eq:i}
\end{equation}
And the error of the integrated intensity is estimated through the following formula \citep{1996A&AS..115..439E},
\begin{equation}
\delta I = \sigma_{\textrm{rms}}\sqrt{\frac{\Delta v dV}{(1-\Delta v/W)}} \ [\textrm{K} \ \textrm{\kms}],
\label{eq:di}
\end{equation}
where $T_{\textrm{rms}}$ is the rms noise temperature, $\Delta v$ is the line width of the emission feature, $dV$ is the spectrum velocity resolution, and $W$ is the entire velocity range of each spectrum. These properties for each spectrum is presented in Fig.~\ref{Fig:spectra}: $W$ is taken from the visible velocity range in each plot; $\Delta v$ is the window for emission feature drawn on the bottom; $dV$ is labeled. For those without the detection of $^{13}$CO, 2$\delta I_{13}$ upper limits were given based on estimates by using the expected line width from the detected $^{12}$CO lines at exactly the same position. The peak velocities and line widths come from a Gaussian fit. The ratio of $^{12}$CO and $^{13}$CO integrated intensity is defined as
\begin{equation}
\mathcal{R} \equiv \frac{\int T_{\mathrm{mb}}(^{12}\mathrm{CO})dv}{\int T_{\mathrm{mb}}(^{13}\mathrm{CO})dv}.
\label{eq:R}
\end{equation}

\bc
\small
\tabcolsep=3pt
\renewcommand{\arraystretch}{1.2}
\begin{longtable}[c]{crrrrcrrrrr}
 \caption{Observational Results}\label{Tab:obs_results} \\
  \hline
  \hline\noalign{\smallskip}
   &  \multicolumn{4}{c}{$^{12}$CO(1$-$0)} & & \multicolumn{4}{c}{$^{13}$CO(1$-$0)} & \\
  \noalign{\smallskip}\cline{2-5}\cline{7-10}\noalign{\smallskip}
   \multicolumn{1}{c}{Galaxy} &  \multicolumn{1}{c}{$I\pm\delta I$} & \multicolumn{1}{c}{$\sigma_{\mathrm{rms}}$} & \multicolumn{1}{c}{$V$} & \multicolumn{1}{c}{$\Delta V$} & & \multicolumn{1}{c}{$I\pm\delta I$} & \multicolumn{1}{c}{$\sigma_{\mathrm{rms}}$}  & \multicolumn{1}{c}{$V$} & \multicolumn{1}{c}{$\Delta V$} & \multicolumn{1}{c}{\R}\\
   & \multicolumn{1}{c}{(K \kms)} & \multicolumn{1}{c}{(mK)} & \multicolumn{2}{c}{(\kms)} & & \multicolumn{1}{c}{(K \kms)} &\multicolumn{1}{c}{(mK)} & \multicolumn{2}{c}{(\kms)} & \\
   \multicolumn{1}{c}{(1)} & \multicolumn{1}{c}{(2)} & \multicolumn{1}{c}{(3)} & \multicolumn{1}{c}{(4)} & \multicolumn{1}{c}{(5)} & & \multicolumn{1}{c}{(6)} & \multicolumn{1}{c}{(7)} & \multicolumn{1}{c}{(8)} & \multicolumn{1}{c}{(9)} & \multicolumn{1}{c}{(10)} \\
  \noalign{\smallskip}\hline\noalign{\smallskip}
 \endfirsthead
 \caption[]{-- \textit{Continued}}\\
  \hline
  \hline\noalign{\smallskip}
   &  \multicolumn{4}{c}{$^{12}$CO(1$-$0)} & & \multicolumn{4}{c}{$^{13}$CO(1$-$0)} & \\
  \noalign{\smallskip}\cline{2-5}\cline{7-10}\noalign{\smallskip}
   \multicolumn{1}{c}{Galaxy} &  \multicolumn{1}{c}{$I\pm\delta I$} & \multicolumn{1}{c}{$\sigma_{\mathrm{rms}}$} & \multicolumn{1}{c}{$V$} & \multicolumn{1}{c}{$\Delta V$} & & \multicolumn{1}{c}{$I\pm\delta I$} & \multicolumn{1}{c}{$\sigma_{\mathrm{rms}}$}  & \multicolumn{1}{c}{$V$} & \multicolumn{1}{c}{$\Delta V$} & \multicolumn{1}{c}{\R}\\
   & \multicolumn{1}{c}{(K \kms)} & \multicolumn{1}{c}{(mK)} & \multicolumn{2}{c}{(\kms)} & & \multicolumn{1}{c}{(K \kms)} &\multicolumn{1}{c}{(mK)} & \multicolumn{2}{c}{(\kms)} & \\
   \multicolumn{1}{c}{(1)} & \multicolumn{1}{c}{(2)} & \multicolumn{1}{c}{(3)} & \multicolumn{1}{c}{(4)} & \multicolumn{1}{c}{(5)} & & \multicolumn{1}{c}{(6)} & \multicolumn{1}{c}{(7)} & \multicolumn{1}{c}{(8)} & \multicolumn{1}{c}{(9)} & \multicolumn{1}{c}{(10)} \\
  \noalign{\smallskip}\hline\noalign{\smallskip}
  \endhead
  \noalign{\smallskip}\hline\endfoot

M 066      &  33.89$\pm$0.67  &    5.54  &   719   &  233  &  &  2.75$\pm$0.22  &1.28  &   714	& 283  &  12.32$\pm$1.02  \\
M 108      &  17.94$\pm$0.67  &    8.58  &   708   &   89  &  &  2.01$\pm$0.37  &3.29  &   713	&  73  &   8.93$\pm$1.68  \\
NGC3079    &  59.17$\pm$0.96  &    6.38  &  1160   &  425  &  &  5.11$\pm$0.28  &1.17  &  1176  & 378  &  11.58$\pm$0.66  \\
NGC3169    &  31.66$\pm$1.33  &    7.69  &  1182   &  447  &  &        $<$1.50  &2.46  &    --  &  --  &        $>$21.11  \\
NGC3184    &   6.52$\pm$0.44  &    7.42  &   594   &   52  &  &  0.56$\pm$0.25  &2.89  &   601  &  42  &  11.64$\pm$5.26  \\
NGC3593    &  24.34$\pm$0.78  &    7.86  &   637   &  212  &  &  1.02$\pm$0.34  &2.37  &   616  &  42  &  23.86$\pm$7.99  \\
NGC3628    &  72.66$\pm$0.59  &    4.44  &   840   &  226  &  &  5.92$\pm$0.59  &2.73  &   843  & 252  &  12.27$\pm$1.23  \\
NGC3631    &  11.29$\pm$0.51  &    7.33  &  1147   &   76  &  &  1.17$\pm$0.27  &2.66  &  1146  &  91  &   9.65$\pm$2.27  \\
NGC3672    &  15.58$\pm$1.31  &    9.76  &  1870   &  265  &  &  3.55$\pm$0.33  &1.53  &  1769  & 186  &   4.39$\pm$0.55  \\
NGC3675    &  17.13$\pm$1.05  &    7.37  &   786   &  282  &  &  3.26$\pm$0.42  &2.04  &   814	& 312  &   5.25$\pm$0.75  \\
NGC3690    &  24.29$\pm$0.66  &    4.87  &  3093   &  260  &  &        $<$0.46  &1.36  &   --   &  --  &        $>$52.80  \\
NGC3810    &  12.42$\pm$0.47  &    5.07  &  1006   &  183  &  &  1.39$\pm$0.27  &2.03  &  1016  & 181  &   8.94$\pm$1.77  \\
NGC3893    &  13.63$\pm$0.53  &    5.08  &   963   &  238  &  &  0.83$\pm$0.20  &1.51  &   927	& 153  &  16.42$\pm$4.01  \\
NGC3938    &   7.47$\pm$0.33  &    5.42  &   813   &   66  &  &  0.85$\pm$0.23  &2.56  &   820	&  42  &   8.79$\pm$2.41  \\
NGC4030    &  21.95$\pm$1.74  &   14.29  &  1454   &  290  &  &        $<$2.24  &6.35  &   --   &  --  &         $>$9.80  \\
NGC4038    &  28.62$\pm$2.00  &   18.49  &  1632   &  127  &  &  5.74$\pm$0.76  &4.83  &  1701  & 189  &   4.99$\pm$0.75  \\
NGC4039    &  39.93$\pm$1.96  &   14.06  &  1584   &  313  &  &  4.47$\pm$0.69  &3.40  &  1734  & 146  &   8.93$\pm$1.45  \\
NGC4041    &  18.33$\pm$0.65  &    7.86  &  1234   &  152  &  &  1.39$\pm$0.46  &3.84  &  1218  & 146  &  13.19$\pm$4.39  \\
NGC4051    &  10.66$\pm$0.53  &    4.71  &   705   &  141  &  &        $<$1.02  &3.65  &    --  &  --  &        $>$10.45  \\
NGC4088    &  21.90$\pm$1.07  &    9.73  &   782   &  208  &  &  2.32$\pm$0.46  &2.89  &   758	& 263  &   9.44$\pm$1.93  \\
NGC4096    &   6.58$\pm$0.72  &    6.40  &   546   &  136  &  &        $<$0.60  &1.83  &    --  &  --  &        $>$10.97  \\
NGC4102    &  21.03$\pm$0.68  &    5.51  &   825   &  335  &  &  1.63$\pm$0.39  &2.16  &   834	& 310  &  12.90$\pm$3.12  \\
NGC4157    &  17.88$\pm$0.61  &    4.94  &   751   &  241  &  &  1.93$\pm$0.44  &2.44  &   764	& 300  &   9.26$\pm$2.14  \\
NGC4194    &   4.02$\pm$1.02  &    7.98  &  2515   &  139  &  &        $<$1.36  &3.66  &    --  &  --  &         $>$2.96  \\
NGC4212    &   5.27$\pm$1.55  &   11.72  &   -85   &  156  &  &  1.45$\pm$0.59  &3.09  &   -13	& 218  &   3.63$\pm$1.82  \\
NGC4254    &  25.55$\pm$0.49  &    5.44  &  2393   &  168  &  &  3.20$\pm$0.24  &1.82  &  2380  & 164  &   7.98$\pm$0.62  \\
NGC4258    &  52.70$\pm$1.27  &    8.80  &   423   &  348  &  &        $<$2.46  &6.80  &   --   & --   &        $>$21.42  \\
NGC4273    &  11.77$\pm$0.81  &    6.52  &  2385   &  245  &  &  2.36$\pm$0.59  &3.29  &  2405  & 178  &   4.99$\pm$1.29  \\
NGC4293    &  12.18$\pm$0.70  &    8.47  &   929   &  171  &  &        $<$0.54  &2.24  &   --   & --   &        $>$22.56  \\
NGC4298    &  12.69$\pm$0.72  &    7.74  &  1132   &  157  &  &  2.17$\pm$0.30  &2.22  &  1148  & 137  &   5.85$\pm$0.87  \\
NGC4302    &  13.25$\pm$1.67  &   18.48  &  1134   &  190  &  &        $<$1.28  &4.87  &   --   &  --  &        $>$10.35  \\
NGC4303    &  19.28$\pm$0.49  &    6.44  &  1563   &  118  &  &  1.31$\pm$0.41  &3.72  &  1557  &  73  &  14.72$\pm$4.62  \\
NGC4312    &   5.93$\pm$1.46  &    9.46  &   162   &  160  &  &  2.57$\pm$0.58  &2.58  &   184	& 224  &   2.31$\pm$0.77  \\
NGC4321    &  30.07$\pm$0.78  &    8.07  &  1585   &  162  &  &  3.60$\pm$0.56  &3.98  &  1572  & 168  &   8.35$\pm$1.32  \\
NGC4402    &  15.82$\pm$0.85  &    9.24  &   247   &  153  &  &  2.79$\pm$0.37  &2.82  &   235	& 194  &   5.67$\pm$0.81  \\
NGC4414    &  44.11$\pm$0.90  &    6.45  &   692   &  336  &  &  6.55$\pm$0.40  &2.01  &   708	& 315  &   6.73$\pm$0.43  \\
NGC4419    &  26.82$\pm$1.41  &    9.50  &  -219   &  308  &  &  5.78$\pm$0.66  &3.07  &  -135  & 333  &   4.64$\pm$0.58  \\
NGC4433    &  20.60$\pm$1.01  &    7.40  &  2946   &  273  &  &        $<$1.08  &2.74  &    --  &  --  &        $>$19.07  \\
NGC4457    &  14.51$\pm$0.62  &    6.23  &   875   &  164  &  &  1.52$\pm$0.41  &2.81  &   864	& 148  &   9.55$\pm$2.61  \\
NGC4490    &   3.18$\pm$0.43  &    6.02  &   609   &  125  &  &        $<$0.72  &3.47  &    --  &  --  &         $>$4.42  \\
NGC4501    &  29.66$\pm$0.91  &    6.86  &  2288   &  315  &  &  6.58$\pm$0.59  &3.56  &  2268  & 294  &   4.51$\pm$0.43  \\
NGC4527    &  37.20$\pm$1.21  &    9.80  &  1749   &  286  &  &  6.90$\pm$0.61  &3.44  &  1759  & 311  &   5.39$\pm$0.51  \\
NGC4535    &  21.87$\pm$1.23  &   10.60  &  1993   &  199  &  &        $<$2.28  &6.76  &   --   & --   &         $>$9.59  \\
NGC4536    &  17.33$\pm$1.13  &    5.88  &  1786   &  274  &  &  4.81$\pm$0.56  &3.32  &  1776  & 271  &   3.60$\pm$0.48  \\
NGC4567    &  12.33$\pm$0.47  &    5.58  &  2273   &  167  &  &        $<$0.58  &2.39  &   --   &  --  &        $>$12.92  \\
NGC4568    &  24.46$\pm$0.61  &    5.97  &  2236   &  230  &  &  4.33$\pm$0.68  &4.61  &  2247  & 238  &   5.65$\pm$0.90  \\
NGC4569    &  36.23$\pm$1.20  &    8.61  &  -214   &  234  &  &  7.92$\pm$0.48  &2.39  &  -186  & 346  &   4.57$\pm$0.32  \\
NGC4579    &  11.38$\pm$1.08  &    9.81  &  1492   &  272  &  &        $<$1.30  &4.09  &    --  &  --  &         $>$8.71  \\
NGC4631    &  11.73$\pm$0.74  &    9.00  &   639   &  125  &  &  2.18$\pm$0.41  &3.45  &   622	& 199  &   5.38$\pm$1.07  \\
NGC4647    &  26.42$\pm$1.87  &   18.63  &  1401   &  111  &  &  8.36$\pm$0.72  &4.96  &  1422  & 108  &   3.16$\pm$0.35  \\
NGC4654    &  16.81$\pm$0.97  &    8.54  &  1061   &  147  &  &  2.12$\pm$0.68  &4.81  &  1051  & 146  &   7.93$\pm$2.58  \\
NGC4666    &  35.61$\pm$1.16  &    8.58  &  1525   &  268  &  &        $<$3.16  &8.06  &    --  &  --  &        $>$11.26  \\
NGC4691    &  10.82$\pm$0.58  &    9.51  &  1122   &   56  &  &  1.08$\pm$0.31  &3.50  &  1122  &  97  &  10.02$\pm$2.93  \\
NGC4710    &  14.97$\pm$1.03  &    7.43  &  1086   &  290  &  &        $<$1.06  &2.61  &    --  &  --  &        $>$14.12  \\
NGC4736    &  24.31$\pm$0.43  &    4.16  &   308   &  224  &  &  7.38$\pm$0.38  &2.95  &   293	& 240  &   3.29$\pm$0.18  \\
NGC4818    &  21.54$\pm$1.41  &   11.63  &  1067   &  146  &  &  3.21$\pm$0.91  &5.21  &  1021  & 232  &   6.71$\pm$1.95  \\
NGC4826    &  48.17$\pm$0.65  &    5.88  &   409   &  298  &  &  8.40$\pm$0.33  &2.08  &   413	& 306  &   5.73$\pm$0.24  \\
NGC4845    &  30.70$\pm$1.52  &    8.69  &  1118   &  320  &  &  3.55$\pm$0.77  &3.74  &  1054  & 274  &   8.65$\pm$1.92  \\

\end{longtable}
\tablecomments{0.81\textwidth}{~Column (1):Names of the sample galaxies.~Column (2):$^{12}$CO integrated intensities and uncertainties, calculated from eq.~\eqref{eq:i} and eq.~\eqref{eq:di}.~Column (3):Baseline noises of spectra in mK.~Column (4)\&(5):Gaussian fitting results of peak velocities and FWHM line widths.~Column (6)$-$(9):Results for $^{13}$CO, for non-detections, 2-sigma upper limits are given.~Column (10): The ratios of $^{12}$CO and $^{13}$CO integrated intensities and their uncertainties.}
\ec

\subsection{Derived properties}

The derived physical parameters, such as the H$_2$ column density, the CO luminosity, and the gas mass, are presented in Table~\ref{Tab:der_resultss}. We assumed that the CO and its isotopic variants are approximately under the conditions of local thermodynamic equilibrium (LTE) when activating transitions and CO transitions are optically thick. Therefore we estimated the average optical depth of $^{13}$CO from \citep{1991A&A...247..320S}
\begin{equation}
\tau^{13}\simeq-\ln \left [1-\frac{\int T_{\textrm{R}}^*(^{13}\textrm{CO})dv}{\int T_{\textrm{R}}^*(^{12}\textrm{CO})dv} \right ],
\label{eq:tau}
\end{equation}
where $T_{\textrm{R}}^*$ should be corrected for a filling factor and therefore this is only an averaged estimation over all of the unresolved clouds in the beam.

Due to the beam dilution effect of remote molecular clouds in remote galaxies, we could hardly measure the excitation temperature, $T_{\mathrm{ex}}$, directly. We present the cold dust color temperature, $T_{\mathrm{dust}}$, calculated from IRAS far infrared data assuming a dust emissivity, $\propto \nu^1$, in Table~\ref{Tab:basic}. However, the gas and dust in the central regions of the galaxies may not couple, and therefore, we took the half value of $T_{\mathrm{dust}}$ as the gas kinetic temperature as well as the excitation temperature, $T_{\mathrm{ex}}=T_{\mathrm{k}}$.

The H$_2$ column density of galaxies was estimated from an empirical equation \citep{2001PASJ...53..757N} in the Milky Way,
\begin{equation}
N(\mathrm{H}_2) = 2 \times 10^{20}\int T_{\mathrm{mb}}(^{12}\mathrm{CO})dv ~ [\mathrm{cm}^{-2}],
\label{eq:nh2_12}
\end{equation}
where the coefficient is the standard galactic $^{12}$CO to H$_2$ conversion factor, $X$.

Using the LTE assumption, the total column density of $^{13}$CO (1-0) transition is described as \citep{2009tra..book.....W}
\begin{equation}
N(^{13}\mathrm{CO}) = 3.0 \times 10^{14} \frac{T_{\mathrm{ex}} \int \tau_{13} dv}{1-\exp[-5.3/T_{\mathrm{ex}}]}~ [\mathrm{cm}^{-2}].
\end{equation}
Here, we assumed the filling factor as 1 and $^{13}$CO optically thin so that there would be $T_{\mathrm{ex}}~\tau = T_{\mathrm{mb}}$ and $ T_{\mathrm{ex}} \int \tau_{13} dv = T_{\mathrm{mb}} \tau_{13} /(1-\mathrm{e}^{-\tau_{13}})$. We then took a relatively high ratio of $N(\mathrm{H}_2)/N(^{13}\mathrm{CO}) \sim 7.5\times10^5$ determined by the relationship between $N(^{13}\mathrm{CO})$ and visual extinction as well as $N(\mathrm{H}_2)$ and $A_{\mathrm{V}}$. Consequently, we also obtained the H$_2$ column density from the following equation,
\begin{equation}
N(\mathrm{H}_2)' = 7.5 \times N(^{13}\mathrm{CO}) = 2.25 \times 10^{20} \frac{\tau_{13}}{1-\mathrm{e}^{-\tau_{13}}} \frac{\int T_{\mathrm{mb}}(^{13}\mathrm{CO})dv}{1-\exp[-5.3/T_{\mathrm{ex}}]}~ [\mathrm{cm}^{-2}].
\label{eq:nh2_13}
\end{equation}

By equating the H$_2$ column density derived both from $^{12}$CO and $^{13}$CO, we can estimate the kinetic temperature as well as excitation temperature of the molecular gas in each galaxy.


The CO luminosity in $\mathrm{K}~\mathrm{km}~\mathrm{s}^{-1}~\mathrm{pc}^2$ can be defined as \citep{1998ApJ...507L.121T}
\begin{equation}
L_{\mathrm{CO}} \equiv area \times I(\mathrm{CO}) = \frac{\pi \theta_{\mathrm{mb}}^2 D^2}{4\ln2} \int T_{\mathrm{mb}} dv ~ [\mathrm{K}~\mathrm{km}~\mathrm{s}^{-1}~\mathrm{pc}^2],
\label{eq:l_co}
\end{equation}
where $\pi \theta_{\mathrm{mb}}^2 D^2/4\ln2$ is the total area of a Gaussian beam source in units of pc$^2$ and $\theta_{\mathrm{mb}}$ is the size of the beam in arc second. Furthermore, the CO luminosity can be equivalently expressed for a source of any size in terms of the total line flux \citep{1997ApJ...478..144S},
\begin{equation}
L_{\mathrm{CO}}  = \frac{c^2}{2k}  S(\mathrm{CO})  \nu_{\mathrm{obs}}^{-2}  D^2 (1+z)^{-3}  = 3.25 \times 10^7  S(\mathrm{CO})  \nu_{\mathrm{obs}}^{-2}  D^2 (1+z)^{-3},
\label{eq:l_co'}
\end{equation}
here $L_{\mathrm{CO}}$ is in $\mathrm{K}~\mathrm{km}~\mathrm{s}^{-1}~\mathrm{pc}^2$, $k$ is the Boltzmann constant, $\nu_{\mathrm{obs}}$ is the observational frequency we received, and
\begin{equation}
S(\mathrm{CO}) = \frac{2k\Omega\int T_{\mathrm{mb}}dv}{\lambda^2},
\end{equation}
is the flux of CO in Jy, where $\Omega$ is the solid angle of the source beam.

Taking the H$_2$ column density derived from the $^{13}$CO emission, we can calibrate the $X$ factor.

\bc
\small
\tabcolsep=5pt
\renewcommand{\arraystretch}{1.32}
 \begin{longtable}[c]{cc*{8}{c}r}
 \caption{Derived Parameters of $^{13}$CO Detected Galaxies}\label{Tab:der_resultss} \\
  \hline
  \hline\noalign{\smallskip}
   \multicolumn{1}{c}{Galaxy} & $\tau_{13}$ &  $N(\mathrm{H}_2)$ & $N(\mathrm{H}_2)'$  & log$L_{^{12}\mathrm{CO}}$ & log$L_{^{13}\mathrm{CO}}$& $T_{\mathrm{ex}}$ & $X$  \\
   &  & \multicolumn{2}{c}{$(10^{21}~\mathrm{cm}^{-2})$} &  \multicolumn{2}{c}{ $(\mathrm{K}~\mathrm{km}~\mathrm{s}^{-1}~\mathrm{pc}^2)$} & (K) &  \\
   \multicolumn{1}{c}{(1)} & (2) & (3) & (4) & (5) & (6) & (7) & (8)  \\
  \noalign{\smallskip}\hline\noalign{\smallskip}
 \endfirsthead
 \caption[]{-- \textit{Continued}} \\
  \hline
  \hline\noalign{\smallskip}
   \multicolumn{1}{c}{Galaxy} & $\tau_{13}$ &  $N(\mathrm{H}_2)$ & $N(\mathrm{H}_2)'$  & log$L_{^{12}\mathrm{CO}}$ & log$L_{^{13}\mathrm{CO}}$& $T_{\mathrm{ex}}$ & $X$  \\
   &  & \multicolumn{2}{c}{$(10^{21}~\mathrm{cm}^{-2})$} &  \multicolumn{2}{c}{ $(\mathrm{K}~\mathrm{km}~\mathrm{s}^{-1}~\mathrm{pc}^2)$} & (K) &  \\
   \multicolumn{1}{c}{(1)} & (2) & (3) & (4) & (5) & (6) & (7) & (8)  \\
  \noalign{\smallskip}\hline\noalign{\smallskip}
  \endhead
  \noalign{\smallskip}\hline\endfoot

M 066      &  0.08    &  6.78 (0.13)   &   2.46(0.20)     &  $8.39^{+0.01}_{-0.01}$  &	$7.35^{+0.03}_{-0.04}$   &   52.88    & 0.72(0.06)   \\
M 108      &  0.12    &  3.59 (0.13)   &   1.75(0.32)     &  $8.39^{+0.02}_{-0.02}$  &	$7.49^{+0.07}_{-0.09}$   &   36.86    & 0.98(0.18)   \\
NGC3079    &  0.09    & 11.83 (0.19)   &   4.57(0.25)     &  $9.15^{+0.01}_{-0.01}$  &	$8.13^{+0.02}_{-0.02}$   &   49.37    & 0.77(0.04)   \\
NGC3184    &  0.09    &  1.30 (0.09)   &   0.44(0.20)     &  $7.87^{+0.03}_{-0.03}$  &	$6.85^{+0.16}_{-0.26}$   &   49.67    & 0.67(0.30)   \\
NGC3593    &  0.04    &  4.87 (0.16)   &   0.90(0.30)     &  $7.65^{+0.01}_{-0.01}$  &	$6.32^{+0.12}_{-0.18}$   &   107.17   & 0.37(0.12)   \\
NGC3628    &  0.08    & 14.53 (0.12)   &   5.40(0.54)     &  $8.72^{+0.00}_{-0.00}$  &	$7.68^{+0.04}_{-0.05}$   &   52.64    & 0.74(0.07)   \\
NGC3631    &  0.11    &  2.26 (0.10)   &   0.96(0.22)     &  $8.58^{+0.02}_{-0.02}$  &	$7.64^{+0.09}_{-0.11}$   &   40.28    & 0.85(0.20)   \\
NGC3672    &  0.26    &  3.12 (0.26)   &   3.16(0.29)     &  $8.93^{+0.04}_{-0.04}$  &	$8.34^{+0.04}_{-0.04}$   &   15.39    & 2.03(0.25)   \\
NGC3675    &  0.21    &  3.43 (0.21)   &   2.67(0.34)     &  $8.30^{+0.03}_{-0.03}$  &	$7.63^{+0.05}_{-0.06}$   &   19.51    & 1.56(0.22)   \\
NGC3810    &  0.12    &  2.48 (0.09)   &   1.18(0.23)     &  $8.32^{+0.02}_{-0.02}$  &	$7.42^{+0.08}_{-0.09}$   &   36.91    & 0.95(0.19)   \\
NGC3893    &  0.06    &  2.73 (0.11)   &   0.70(0.17)     &  $8.42^{+0.02}_{-0.02}$  &	$7.25^{+0.09}_{-0.12}$   &   72.17    & 0.52(0.13)   \\
NGC3938    &  0.12    &  1.49 (0.07)   &   0.69(0.19)     &  $8.07^{+0.02}_{-0.02}$  &	$7.17^{+0.10}_{-0.14}$   &   36.22    & 0.93(0.25)   \\
NGC4038    &  0.22    &  5.72 (0.40)   &   5.60(0.74)     &  $8.98^{+0.03}_{-0.03}$  &	$8.33^{+0.05}_{-0.06}$   &   18.23    & 1.96(0.29)   \\
NGC4039    &  0.12    &  7.99 (0.39)   &   4.14(0.64)     &  $9.13^{+0.02}_{-0.02}$  &	$8.22^{+0.06}_{-0.07}$   &   36.90    & 1.04(0.17)   \\
NGC4041    &  0.08    &  3.67 (0.13)   &   1.21(0.40)     &  $8.84^{+0.02}_{-0.02}$  &	$7.76^{+0.12}_{-0.17}$   &   56.94    & 0.66(0.22)   \\
NGC4088    &  0.11    &  4.38 (0.21)   &   2.03(0.40)     &  $8.45^{+0.02}_{-0.02}$  &	$7.52^{+0.08}_{-0.10}$   &   39.29    & 0.93(0.19)   \\
NGC4102    &  0.08    &  4.21 (0.14)   &   1.61(0.39)     &  $8.64^{+0.01}_{-0.01}$  &	$7.57^{+0.09}_{-0.12}$   &   55.60    & 0.77(0.19)   \\
NGC4157    &  0.11    &  3.58 (0.12)   &   1.59(0.36)     &  $8.36^{+0.01}_{-0.02}$  &	$7.44^{+0.09}_{-0.11}$   &   38.46    & 0.89(0.20)   \\
NGC4212    &  0.32    &  1.05 (0.31)   &   1.36(0.55)     &  $7.95^{+0.11}_{-0.15}$  &	$7.44^{+0.15}_{-0.23}$   &   11.77    & 2.58(1.30)   \\
NGC4254    &  0.13    &  5.11 (0.10)   &   2.78(0.21)     &  $8.63^{+0.01}_{-0.01}$  &	$7.78^{+0.03}_{-0.03}$   &   32.42    & 1.09(0.08)   \\
NGC4273    &  0.22    &  2.35 (0.16)   &   2.18(0.54)     &  $8.30^{+0.03}_{-0.03}$  &	$7.65^{+0.10}_{-0.12}$   &   18.24    & 1.85(0.48)   \\
NGC4298    &  0.19    &  2.54 (0.14)   &   1.74(0.24)     &  $8.33^{+0.02}_{-0.03}$  &	$7.61^{+0.06}_{-0.06}$   &   22.32    & 1.37(0.20)   \\
NGC4303    &  0.07    &  3.86 (0.10)   &   1.15(0.36)     &  $8.51^{+0.01}_{-0.01}$  &	$7.39^{+0.12}_{-0.16}$   &   64.15    & 0.60(0.19)   \\
NGC4312    &  0.57    &  1.19 (0.29)   &   2.60(0.59)     &  $8.07^{+0.10}_{-0.12}$  &	$7.75^{+0.09}_{-0.11}$   &   5.19     & 4.39(1.47)   \\
NGC4321    &  0.13    &  6.01 (0.16)   &   3.06(0.48)     &  $8.70^{+0.01}_{-0.01}$  &	$7.83^{+0.06}_{-0.07}$   &   34.16    & 1.02(0.16)   \\
NGC4402    &  0.19    &  3.16 (0.17)   &   2.33(0.31)     &  $8.43^{+0.02}_{-0.02}$  &	$7.72^{+0.05}_{-0.06}$   &   21.48    & 1.47(0.21)   \\
NGC4414    &  0.16    &  8.82 (0.18)   &   5.80(0.35)     &  $9.00^{+0.01}_{-0.01}$  &	$8.22^{+0.03}_{-0.03}$   &   26.52    & 1.32(0.08)   \\
NGC4419    &  0.24    &  5.36 (0.28)   &   5.51(0.63)     &  $8.65^{+0.02}_{-0.02}$  &	$8.04^{+0.05}_{-0.05}$   &   16.59    & 2.06(0.26)   \\
NGC4457    &  0.11    &  2.90 (0.12)   &   1.38(0.37)     &  $8.21^{+0.02}_{-0.02}$  &	$7.27^{+0.10}_{-0.14}$   &   39.79    & 0.95(0.26)   \\
NGC4501    &  0.25    &  5.93 (0.18)   &   5.62(0.50)     &  $8.70^{+0.01}_{-0.01}$  &	$8.09^{+0.04}_{-0.04}$   &   15.96    & 1.90(0.18)   \\
NGC4527    &  0.21    &  7.44 (0.24)   &   6.50(0.57)     &  $8.80^{+0.01}_{-0.01}$  &	$8.11^{+0.04}_{-0.04}$   &   20.16    & 1.75(0.16)   \\
NGC4536    &  0.33    &  3.47 (0.23)   &   5.40(0.63)     &  $8.44^{+0.03}_{-0.03}$  &	$7.94^{+0.05}_{-0.05}$   &   11.62    & 3.11(0.42)   \\
NGC4568    &  0.19    &  4.89 (0.12)   &   3.76(0.59)     &  $8.61^{+0.01}_{-0.01}$  &	$7.91^{+0.06}_{-0.07}$   &   21.38    & 1.54(0.24)   \\
NGC4569    &  0.25    &  7.25 (0.24)   &   7.06(0.43)     &  $8.79^{+0.01}_{-0.01}$  &	$8.17^{+0.03}_{-0.03}$   &   16.27    & 1.95(0.13)   \\
NGC4631    &  0.21    &  2.35 (0.15)   &   2.13(0.40)     &  $7.70^{+0.03}_{-0.03}$  &	$7.02^{+0.07}_{-0.09}$   &   20.11    & 1.81(0.36)   \\
NGC4647    &  0.38    &  5.28 (0.37)   &   7.81(0.67)     &  $8.65^{+0.03}_{-0.03}$  &	$8.20^{+0.04}_{-0.04}$   &   9.47     & 2.96(0.33)   \\
NGC4654    &  0.13    &  3.36 (0.19)   &   1.77(0.57)     &  $8.45^{+0.02}_{-0.03}$  &	$7.60^{+0.12}_{-0.17}$   &   32.16    & 1.05(0.34)   \\
NGC4691    &  0.11    &  2.16 (0.12)   &   1.06(0.30)     &  $8.52^{+0.02}_{-0.02}$  &	$7.57^{+0.11}_{-0.15}$   &   42.02    & 0.98(0.29)   \\
NGC4736    &  0.36    &  4.86 (0.09)   &   8.03(0.41)     &  $7.61^{+0.01}_{-0.01}$  &	$7.14^{+0.02}_{-0.02}$   &   10.12    & 3.30(0.18)   \\
NGC4818    &  0.16    &  4.31 (0.28)   &   3.46(0.98)     &  $8.13^{+0.03}_{-0.03}$  &	$7.36^{+0.11}_{-0.14}$   &   26.41    & 1.61(0.47)   \\
NGC4826    &  0.19    &  9.63 (0.13)   &   7.72(0.30)     &  $7.52^{+0.01}_{-0.01}$  &	$6.81^{+0.02}_{-0.02}$   &   21.79    & 1.60(0.07)   \\
NGC4845    &  0.12    &  6.14 (0.30)   &   3.09(0.67)     &  $8.70^{+0.02}_{-0.02}$  &	$7.81^{+0.09}_{-0.11}$   &   35.55    & 1.01(0.22)   \\

\end{longtable}
\tablecomments{0.84\textwidth}{~Column (1): Galaxy names of those $^{13}$CO detected.~Column (2): The optical depth of $^{13}$CO emission derived from eq.~\eqref{eq:tau}.~Column (3)-(4): The H$_2$ column density derived from $^{12}$CO and $^{13}$CO, respectively, from eq.~\eqref{eq:nh2_12} and eq.~\eqref{eq:nh2_13}. ~Column (5)-(6): $^{12}$CO and $^{13}$CO luminosities derived from eq.~\eqref{eq:l_co} or eq.~\eqref{eq:l_co'} respectively.~Column (7): Excitation temperature calculated by equating the H$_2$ column density derived both from $^{12}$CO and $^{13}$CO.~Column (8): The X factor, calculated by dividing H$_2$ column density and CO integrated intensity, in unit of $10^{20}\mathrm{cm}^{-2} [ \mathrm{K}~\mathrm{km}~\mathrm{s}^{-1} ]^{-1}$.}
\ec

\begin{figure}[!ht]
 \centering
\includegraphics[width=10cm]{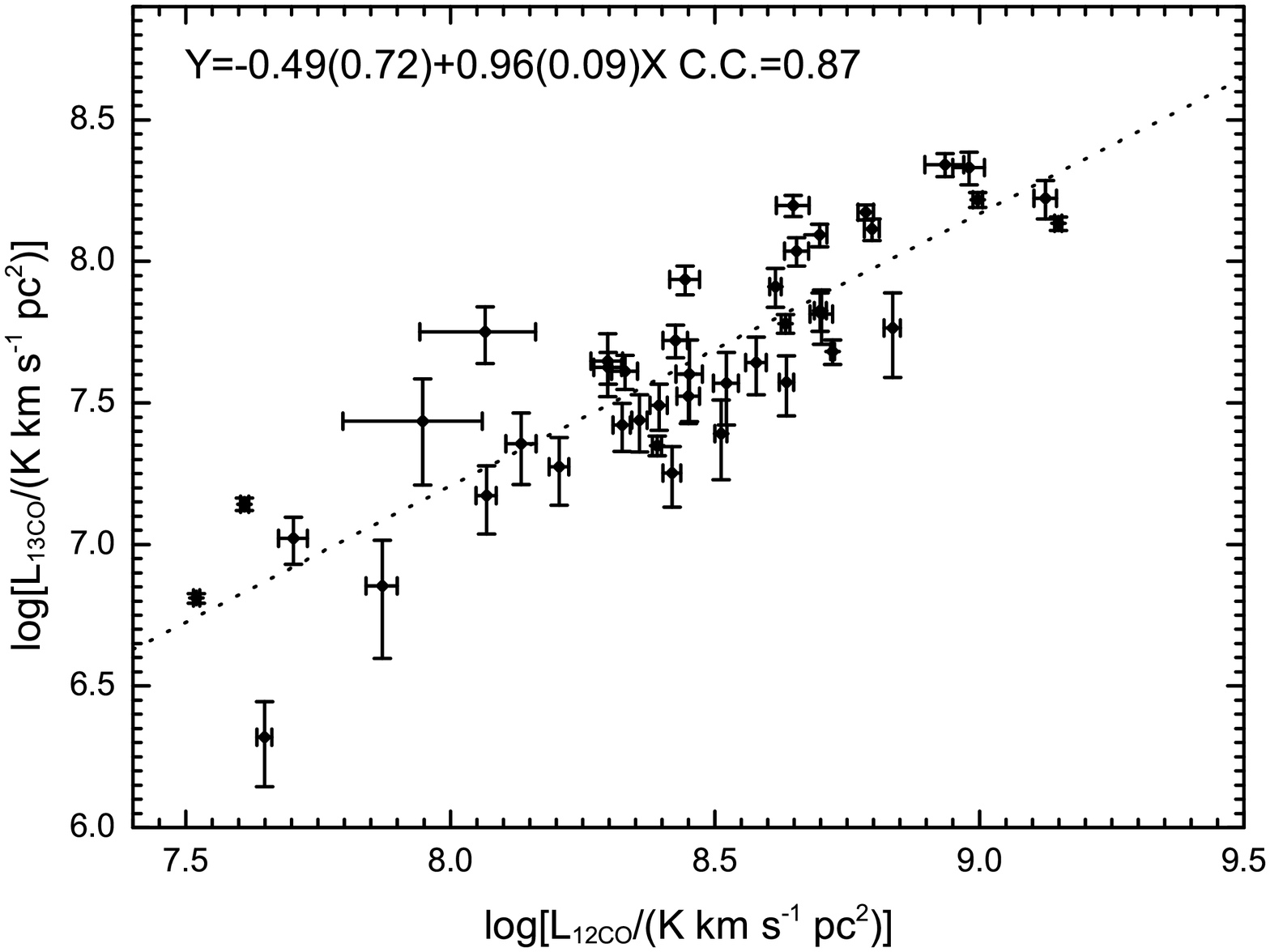}
\caption{The comparison between the $^{12}$CO and $^{13}$CO luminosity. A correlation is validated with C.C. value of 0.87. The dotted line corresponds to the average ratio, \R.}
\label{Fig:L12-L13}
\end{figure}

\section{Discussions}
\label{sect:analysis}

We confirmed the detection of $^{13}$CO emissions in 42 galaxies. However, this does not mean that other galaxies do not have $^{13}$CO emissions. The detection limitation restricted by wavelike baseline reaches beyond the signal of those with relatively weak sources. Of those both $^{12}$CO and $^{13}$CO were detected in 42. We present the central intensity ratio, \R, with an average value of 8.14$\pm$4.21, ranging mostly from 5 to 13. NGC 4212, NGC 4312, NGC4536, NGC 4631, and NGC 4736 have very low ratios less than 4. This is probably due to systematic uncertainties or a higher optical depth of the gas in the central positions of the galaxies. The uncertainty of \R may not merely indicate the accuracy of our measurement but also reflect pointing errors. The average ratio is slightly lower than previous estimations of 11$\pm$3 \citep{1991A&A...249..323A}, 9.3$\pm$3.6 \citep{1991A&A...247..320S} and 11.3$\pm$3.3 for normal galaxies. \citet{1986ApJ...302..680Y} found no evidence for systematic variation in \R with radius, and \citet{1991A&A...247..320S} did not find clear evidence either. The average of all off-center points is somewhat less than the average of the centers. It is suggested that galaxies which display variations in \R have varying large-scale properties of their molecular cloud distributions. Thus, our lower estimation of the average ratio, \R, may not result merely from the different main beam efficiency between telescopes, but also due to the different sample of galaxies with regions larger than the central that are covered in a one beam-size field in our observations.

We compared $^{12}$CO luminosity with $^{13}$CO luminosity in Fig.~\ref{Fig:L12-L13}. Of course a tight correlation can be found because of the same distance and the deviation accounts for variants of \R. We could not determine why \citet{1998ApJ...507L.121T} claimed that more luminous galaxies have lower $^{13}$CO luminosity with respect to $^{12}$CO. Perhaps a wider range of CO luminosity data of other galaxies is needed.

\begin{figure}[!ht]
 \centering
\includegraphics[width=10cm]{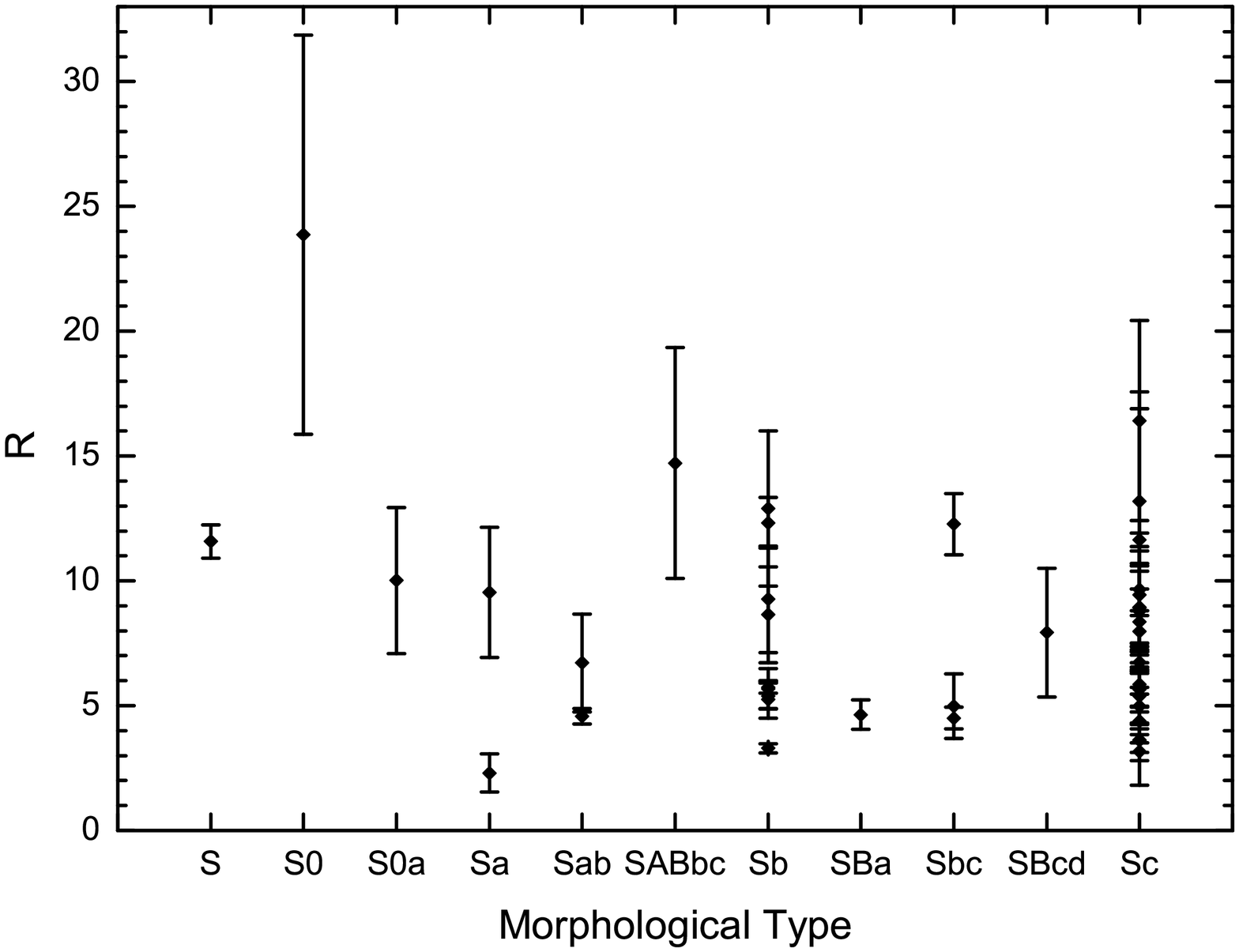}
\caption{The intensity ratio of $^{12}$CO/$^{13}$CO, \R, is independent of the morphological type. There is no clear relationship between \R and the morphological type evolution in subclasses of spirals.
}
\label{Fig:R-Type}
\end{figure}

\begin{figure}[!ht]
 \centering
\includegraphics[width=10cm]{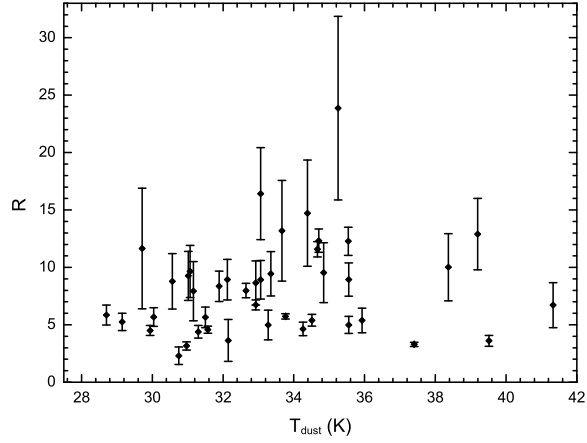}
\caption{The correlation between \R and dust temperature. There seems to be little relationship between them below a temperature of 40 K.}
\label{Fig:R-T_dust}
\end{figure}

Generally, it is suggested that the intensity ratio, \R, is a measure of the cloud environment in galaxies \citet{1991A&A...249..323A}; \citet{1995A&A...300..369A}. High ratio values (\R$>20$) might originate in turbulent, high-pressure gas in the centers of luminous interactive galaxies or mergers, intermediate values ($10\leq$\R$\leq15$) refer to normal starbursts, and low values (\R$\simeq6$) represent the disk population of clouds. The deficiency of $^{13}$CO due to isotope-selective photo-dissociation may alternatively account for a high \R.

\citet{1991A&A...247..320S} found that the intensity ratio of $^{12}$CO/$^{13}$CO \R is independent of the morphological type. Our result in Fig.~\ref{Fig:R-Type} is similar to \citet{1989ApJS...70..699Y} and \citet{1989ApJ...342L..15S}: there is no clear relationship between the intensity ratio, \R, and the morphological type evolution in subclasses of spirals.

Fig.~\ref{Fig:R-T_dust} illustrates the relationship between $T_{\mathrm{dust}}$ and \R. It has been claimed that high CO luminosity in luminous far-infrared galaxies are due to a greater excitation temperature of CO gas rather than a higher mass quantity \citet{1988ApJ...325..389M}; \citet{1991ApJ...373..423S}. However, we concluded the same results as \citet{1991A&A...247..320S}: there seems to be no clear relationship between them below a temperature of 40 K. Unfortunately, we were unable to test whether there exists a trend of \R and $T_{\mathrm{dust}}$ beyond 40 K.

The CO-H$_{2}$ conversion factor gives us a direct way to estimate H$_{2}$ gas in molecular clouds through CO. In the Milky Way, we usually take a universal value of $X=2\times10^{20}~\mathrm{cm}^{-2}~(\mathrm{K}~\mathrm{km}~\mathrm{s}^{-1})^{-1}$ as an estimation. For an estimation of extragalaxies, we found an average value of $1.44\pm0.84\times10^{20}~ \mathrm{cm}^{-2}~(\mathrm{K}~\mathrm{km}~\mathrm{s}^{-1})^{-1}$, which is slightly lower than the standard value in the Milky Way.

\section{Summary}
\label{sect:summary}

Using the PMO 13.7-m millimeter-wave telescope, we simultaneously observed the $^{12}$CO, $^{13}$CO, and C$^{18}$O J=1$-$0 rotational transitions in the centers of 58 nearby galaxies with relatively strong $^{12}$CO emissions. We detected $^{13}$CO emissions in 42 out of the 58 galaxies, but a null detection of C$^{18}$O emission with a sigma upper limit of 2 mK. The main two results are summarized as follows:
\begin{itemize}
\item[(1)] We presented results of spectra, integrated intensity of both $^{12}$CO and $^{13}$CO emissions in each galaxy. Central beam ratios, \R, of $^{12}$CO and $^{13}$CO range mostly from 5 to 13, with an average value of 8.14$\pm$4.21, which is slightly lower than previous estimates for normal galaxies.
\item[(2)] We calculated $^{12}$CO and $^{13}$CO luminosities and clear correlations are validated. We computed the column density of H$_{2}$ gas from $I(^{13}\mathrm{CO})$ and then calibrated the X-factor, finding an average value of $1.44\pm0.84\times10^{20}~ \mathrm{cm}^{-2}~(\mathrm{K}~\mathrm{km}~\mathrm{s}^{-1})^{-1}$, which is slightly lower than the standard value in the Milky Way.
\end{itemize}

\begin{acknowledgements}
Special thanks to the PMO Qinghai Station staff for their help.
\end{acknowledgements}




\bibliographystyle{raa}
\bibliography{ref}


\end{document}